\documentclass[aps,prc,showpacs,superscriptaddress,twocolumn,10pt,floatfix,nofootinbib]{revtex4-2}

\usepackage{natorb}

\begin{document}
\ifproofpre{}{\count\footins = 1000}  

\title{Natural orbitals for the \textit{ab initio} no-core configuration interaction approach}

\author{Patrick J.~Fasano}
\affiliation{Department of Physics and Astronomy, University of Notre Dame, Notre Dame, Indiana 46556-5670, USA}

\author{Chrysovalantis Constantinou}
\altaffiliation[Present address: ]{Computation-Based Science and Technology Research Center, The Cyprus Institute, 2121 Aglantzia, Nicosia, Cyprus}
\affiliation{Department of Physics and Astronomy, University of Notre Dame, Notre Dame, Indiana 46556-5670, USA}
\affiliation{Center for Theoretical Physics, Sloane Physics Laboratory,
  Yale University, New Haven, Connecticut 06520-8120, USA}

\author{Mark A.~Caprio}
\affiliation{Department of Physics and Astronomy, University of Notre Dame, Notre Dame, Indiana 46556-5670, USA}

\author{Pieter Maris}
\affiliation{Department of Physics and Astronomy, Iowa State University, Ames, Iowa 50011-3160, USA}

\author{James P.~Vary}
\affiliation{Department of Physics and Astronomy, Iowa State University, Ames, Iowa 50011-3160, USA}

\date{March 22, 2022}

\begin{abstract}
\textit{Ab initio} no-core configuration interaction (NCCI) calculations for the
nuclear many-body problem have traditionally relied upon an antisymmetrized
product (Slater determinant) basis built from harmonic oscillator orbitals.
The accuracy of such calculations is limited by the finite dimensions which are
computationally feasible for the truncated many-body space.  We therefore
seek to improve the accuracy obtained for a given basis size by optimizing
the choice of single-particle orbitals.
Natural orbitals, which diagonalize the one-body density matrix, provide a
basis which maximizes the occupation of low-lying orbitals, thus accelerating
convergence in a configuration-interaction basis, while also possibly providing physical
insight into the single-particle structure of the many-body wave function.
We describe the implementation of natural orbitals in the NCCI framework, and
examine the nature of the natural orbitals thus obtained, the properties of the
resulting many-body wave functions, and the convergence of observables.
After taking $\isotope[3]{He}$ as an illustrative testbed, we explore aspects of
NCCI calculations with natural orbitals for the ground state of the $p$-shell neutron halo
nucleus $\isotope[6]{He}$.
 \end{abstract}

\maketitle

\section{Introduction}
\label{sec-intro}

The goal of \textit{ab initio} nuclear
theory~\cite{navratil2000:12c-ab-initio,pieper2004:gfmc-a6-8,neff2004:cluster-fmd,hagen2007:coupled-cluster-benchmark,quaglioni2009:ncsm-rgm,bacca2012:6he-hyperspherical,shimizu2012:mcsm,dytrych2013:su3ncsm,barrett2013:ncsm,baroni2013:7he-ncsmc}
is to predict the behavior of the nuclear many-body system starting
from underlying internucleon
interactions~\cite{wiringa1995:nn-av18,entem2003:chiral-nn-potl,shirokov2007:nn-jisp16,epelbaum2009:nuclear-forces}.
However, the nuclear many-body problem lives in an infinite-dimensional
space.  Thus, in practical numerical computations, the problem must be replaced by an
approximate, truncated representation, and, given finite computational resources,
can only be solved with finite accuracy.

This accuracy may be expected to depend critically upon the choice of many-body
basis used to define the truncated space for the problem.  The many-body basis
is in turn generated from some underlying set of single-particle states.  More
specifically, given the rotational invariance of the nuclear problem, we
consider some underlying set of single-particle \textit{orbitals}, of definite
angular momentum $j$.  While the choice of orbitals has been a central concern
in quantum many-body calculations for the electron structure of atoms and
molecules~\cite{helgaker2000:electron-structure}, it has been largely neglected
in \textit{ab initio} nuclear many-body calculations.

In the no-core configuration interaction (NCCI), or no-core shell model (NCSM),
approach~\cite{navratil2000:12c-ab-initio,barrett2013:ncsm}, the many-body basis
consists of antisymmetrized products (Slater determinants) of single-particle
states.  The many-body problem is then recast as a Hamiltonian matrix
eigenproblem in terms of this basis.  Harmonic oscillator
orbitals~\cite{moshinsky1996:oscillator} have traditionally been used to define
the basis for NCCI calculations.  This choice is motivated in part by technical
convenience.  Namely, two-body matrix elements of translationally invariant
operators such as the Hamiltonian are conveniently evaluated in the oscillator
basis, via the Moshinsky transformation~\cite{moshinsky1996:oscillator} from the
relative oscillator basis.  Furthermore, an exact separation of the
center-of-mass motion is obtained with an oscillator basis truncated according
to the $\Nmax$ scheme, \textit{i.e.}, by total number of oscillator
excitations~\cite{elliott1955:com-shell,caprio2020:intrinsic}.

The calculated results for energies, electromagnetic observables, \textit{etc.},
from an \textit{ab initio} NCCI calculation depend on the truncated space in
which this calculation is carried out.  As $\Nmax$ is increased towards
infinity, the calculated results in principle converge towards those which would
be obtained in the full, untruncated space for the nuclear many-body problem.
However, a rapid growth in dimension of the many-body space, with increasing
$\Nmax$ and number of nucleons, limits the accuracy which can be obtained.

Despite the computationally-convenient properties of the harmonic-oscillator
orbitals, within a many-body calculation, there is no reason to presume that
they are ``optimal'' as the underlying single-particle basis for expanding the
many-body wave function.  Moreover, in at least one way, they are qualitatively
mismatched to the problem.  Notably, as the solutions to the infinitely-bound
harmonic oscillator problem, the oscillator functions fall off at large distance
with Gaussian asymptotics, \textit{i.e.}, $\sim e^{-r^2/(2b^2)}$.  However, the
nuclear attraction is of finite-range.  Consequently, the single-particle wave
functions arising in mean-field descriptions of the nucleus instead fall off
exponentially, \textit{i.e.}, $\sim e^{-\kappa r}$.  While a suitable fall-off
can be recovered, out to any finite distance of relevance, by taking a
superposition of oscillator functions, doing so may require a large number of
oscillator functions (see, \textit{e.g.}, Fig.~4 of
Ref.~\cite{davies1966:hartree-fock}).

We are therefore motivated to look beyond the traditional harmonic oscillator
many-body basis, to increase the accuracy attainable for a given NCCI problem
dimension.  In the present work, we explore the improvement which may be
obtained by optimizing the choice of underlying orbitals used to construct the
basis configurations.  While we might simply prescribe a set of orbitals of some
analytic form (\textit{e.g.},
Refs.~\cite{caprio2012:csbasis,caprio2014:cshalo}), in the hopes that these
might provide some benefit relative to the harmonic oscillator orbitals, a more
informed choice can be obtained by first carrying out some preliminary many-body
calculation, and using the resulting information on the many-body wave function
for guidance in constructing the orbitals.

In this spirit, the \textit{natural
  orbitals}~\cite{loewdin1955:natural-orbitals-part1,shull1955:natural-orbitals-helium,loewdin1956:natural-orbital,davidson1972:natural-orbital,mahaux1991:single-particle}
have been used extensively in atomic and molecular electron-structure
theory~\cite{davidson1972:natural-orbital,helgaker2000:electron-structure}, and
have also found application in the nuclear
problem~\cite{lalazissis1992:natural-orbital-shell,stoitsov1993:natural-orbital-nuclei,stoitsov1993:natural-orbital-correlation,shin2017:6li-gsm-ncfc,jaganathen2017:gamow-shell-psd-quantified}.  They are constructed in a way intended to reduce the number of
antisymmetrized product states required for an accurate representation of the
many-body wave function, thereby accelerating the convergence of its description
in a configuration interaction
basis~\cite{loewdin1960:natural-orbital,kobe1969:natural-orbital-variational}.

Natural orbitals are defined with reference to some many-body state
$\tket{\Psi}$~--- not necessarily a single Slater determinant, but a
general, correlated many-body state.  The corresponding set of natural orbitals
is obtained by diagonalizing the one-body density matrix of $\tket{\Psi}$.  The
eigenvectors define the natural orbitals, and the corresponding eigenvalues
represent the mean occupations of these orbitals within the reference many-body
state $\tket{\Psi}$.

In order to find the \textit{true} natural orbitals for a given nuclear state, say, the
ground state, we would have to have first solved the full many-body problem for
this state, thence obtaining the densities.  However, even from an approximate
initial solution for the many-body wave function, which yields approximate
densities, we may still obtain \textit{approximate} natural orbitals.  It is these which
we may attempt to use in constructing an improved basis for the many-body
calculation.

Here we explore the use of natural orbitals in NCCI calculations.  The initial
many-body calculation, providing the densities used to define the natural
orbitals, is a traditional $\Nmax$-truncated oscillator-basis calculation.  The
natural orbitals for a subsequent many-body calculation are thus obtained as a
unitary transformation on the original oscillator orbitals.  In addition to
illustrating the convergence properties of the resulting NCCI calculations, we
attempt to illuminate the properties of the natural orbitals and probe
some of the implications for center-of-mass motion.

Preliminary results of the present work
were reported in
Refs.~\cite{constantinou2017:natorb-natowitz16,constantinou2017:diss}.
Complementary approaches have since also been explored where natural orbitals
for use in \textit{ab initio} nuclear many-body calculations are obtained from
solutions of a spatially-localized two-body (deuteron)
problem~\cite{puddu2018:deuteron-natural-orbitals} or from many-body
perturbation theory for closed-shell
nuclei~\cite{tichai2019:natorb-mbpt,hoppe2021:natorb-mbpt-imsrg}.  The
implications of natural orbitals for wave function entanglement in NCCI
calculations have also been examined~\cite{robin2021:6he-entanglement}.

Whereas the preliminary results presented in
Refs.~\cite{constantinou2017:natorb-natowitz16,constantinou2017:diss}
were based on the earlier JISP16 interaction~\cite{shirokov2007:nn-jisp16},
the present examples are based on NCCI calculations using the Daejeon16
internucleon interaction~\cite{shirokov2016:nn-daejeon16}.
Relative to JISP16, Daejeon16 has the advantage of providing both faster convergence
of calculated observables and improved agreement with experimental
binding and excitation energies~\cite{maris2019:daejeon16-lenpic-pshell-ntse18}.

We first review the framework for calculations with natural orbitals: defining
how symmetry-adapted natural orbitals (of definite angular momentum and parity)
are extracted from the density matrix (Sec.~\ref{sec:methods:no}) and outlining
how these are obtained and used within the NCCI framework
(Sec.~\ref{sec:methods:ncci}).  Then, to see how the formalism
is reflected in actual NCCI calculations, we take $\isotope[3]{He}$ as the
simplest nontrivial example: examining the convergence of energy and radius
observables for $\isotope[3]{He}$ (Sec.~\ref{sec:results-3he:orbitals}),
inspecting the radial wave functions of the natural orbitals themselves
(Sec.~\ref{sec:results-3he:orbitals}), and diagnosing the center-of-mass motion
of the many-body wave function (Sec.~\ref{sec:results-3he:Ncm}). After
establishing this baseline, we explore aspects of NCCI calculations for the
neutron halo properties of $\isotope[6]{He}$ (Sec.~\ref{sec:results-6he}).

\section{Natural orbitals}
\label{sec:methods}

\subsection{Natural orbitals and rotational symmetry}
\label{sec:methods:no}

Recall that we seek orbitals which will provide rapid convergence in a finite
basis of antisymmetrized product states. Our many-body basis is built out of an
ordered set of single-particle orbitals where we favor ``lower-lying'' orbitals (and
disfavor ``higher-lying'' orbitals) when deciding which orbitals to use in
constructing basis states.  We would therefore be best served by a set of
orbitals such that the ``lower-lying'' orbitals contribute disproportionately to the
most important antisymmetrized products.

It is therefore natural to construct orbitals in a way that maximizes the mean
occupation of the lowest-lying orbitals~--- and, correspondingly, minimizes the mean
occupation of the higher-lying orbitals~--- in the many-body state.  In a particle-hole
picture, this may be thought of as minimizing the depletion of the Fermi sea.
Natural orbitals, in a well-defined sense, accomplish this goal.

Suppose we are interested in finding single-particle states in which to
efficiently represent a particular many-body state $\tket{\Psi}$. The
single-particle properties of $\tket{\Psi}$ are described by its
\textit{one-body density operator}
$\hat{\rho}_\Psi$~\cite{ring1980:nuclear-many-body}\footnote{Such a one-body
  density operator, derived from a pure state of a many-body system, is properly
  known as a \textit{one-body reduced density
    operator}~\cite{coleman2000:reduced-density-matrices}.}, which is an
operator on the single-particle space.  \textit{Natural orbitals} are, quite simply,
obtained as its eigenstates.

Taken in this traditional
sense~\cite{loewdin1955:natural-orbitals-part1,shull1955:natural-orbitals-helium,loewdin1956:natural-orbital},
the natural ``orbitals'' are not orbitals \textit{per se}, as usually construed
in nuclear physics.  They are, rather, simply a set of independently-defined
single-particle states, unrelated to each other by any explicit symmetry
constraint.  However, one may refine the definition of the natural orbitals, so
as to manifestly respect the symmetries of the
system~\cite{mcweeny1968:symmetry-natorb-part1-adapted,davidson1972:natural-orbital}.
In the case of the rotationally-invariant nuclear problem, the resulting
\textit{symmetry-adapted natural orbitals} become orbitals in the usual
nuclear-physics sense, of $nlj$~orbitals~\cite{suhonen2007:nucleons-nucleus}.
In the following, we first review the formulation of natural orbitals in the
traditional sense, \textit{i.e.}, without explicitly embedding the nuclear
symmetries, then establish the symmetry-adapted natural orbitals appropriate to
nuclear NCCI calculations.

Although the definition of $\hat{\rho}_\Psi$ as an operator is independent of
the choice of basis for the single-particle space, this operator may be
expressed in terms of any discrete single-particle basis as
\begin{equation}
  \label{eqn:obdo-defn}
  \hat{\rho}_\Psi = \sum_{\alpha \beta}\tket{\alpha} \, \tme{\Psi}{a^\dagger_\beta a_\alpha}{\Psi} \, \tbra{\beta}.
\end{equation}
Here, the labels $\alpha$ and $\beta$ specify the single-particle basis states,
\textit{e.g.}, for the nuclear problem, they may represent the magnetic
substates $\alpha=(n_a l_a j_a m_\alpha)$ of
$nlj$~orbitals~\cite{suhonen2007:nucleons-nucleus}, while $a^\dagger_\alpha$ and
$a_\alpha$ represent the creation and annihilation operators, respectively, for
a nucleon in state $\tket{\alpha}$.

The eigenstates $\tket{\phi_i}$ of
$\hat{\rho}_\Psi$ are what we take as the natural orbitals for the many-body
reference state $\tket{\Psi}$.  In terms of this eigenbasis, the
expression~(\ref{eqn:obdo-defn}) for $\hat\rho_\Psi$ reduces to the familiar
canonical form for a density operator as a real linear combination of projection
operators (\textit{e.g.}, Ref.~\cite{sakurai1994:qm}),
\begin{equation}
  \label{eqn:obdo-expansion}
  \hat\rho_\Psi = \sum_i \lambda_i \tdyad{\phi_i}{\phi_i},
\end{equation}
where the $\lambda_i$ are the corresponding real eigenvalues for the
$\tket{\phi_i}$ (we rely here on the observation that $\hat\rho_\Psi$ is a
self-adjoint operator).

If we work in terms of a discrete basis for the single-particle space,
$\hat\rho_\Psi$ is represented as the \textit{one-body density matrix} $\rho$,
with matrix elements $\rho_{\alpha\beta} =
\tme{\alpha}{\hat{\rho}_\Psi}{\beta}$, which may be read off
from~(\ref{eqn:obdo-defn}) as
\begin{equation}
  \label{eqn:obdm-defn}
  \rho_{\alpha\beta}
  =
  \tme{\Psi}{a^\dagger_\beta a_\alpha}{\Psi}.
\end{equation}
The operator
eigenproblem for $\hat\rho_\Psi$ reduces to the matrix eigenproblem for $\rho$.
The
eigenvectors then express the natural orbitals $\tket{\phi_i}$ in terms of the
underlying basis.
Changing to a natural orbital
basis for the single-particle space makes the density matrix diagonal, with
entries $\lambda_i$, as is apparent from~(\ref{eqn:obdo-expansion}).

A diagonal matrix element $n_\alpha=\rho_{\alpha\alpha}$ is simply the
expectation value of the number operator $\hat{N}_\alpha=a_\alpha^\dagger
a_\alpha$, and thus represents the mean occupation of the single-particle state
$\alpha$ in the many-body state $\tket{\Psi}$.  Thus, the eigenvalues $\lambda_i$ for the natural orbitals $\tket{\phi_i}$
of a reference state $\tket{\Psi}$ represent their mean occupations in
this reference state, \textit{i.e.}, $n_{\phi_i}=\lambda_i$.  Consequently, these
eigenvalues must satisfy the properties expected for mean occupations: $0 \leq
\lambda_i \leq 1$ and $\sum_i \lambda_i =A$, where $A$ is the number of nucleons
in the system.

To see the relevance of the natural orbitals to the problem of identifying an
optimal basis of antisymmetrized product states, first consider the case where
$\tket{\Psi}$ is itself a single antisymmetrized product, specifically, of the
first $A$ single-particle states taken from some particular disrete basis
$\lbrace \tket{\alpha_i} \rbrace$, \textit{i.e.}, $\tket{\Psi}=\tket{\alpha_1,
  \alpha_2,\cdots,\alpha_A}$.  The one-body density matrix taken in this basis
is already diagonal, with occupation numbers $n_{\alpha_i}=1$ for occupied
states or $0$ for unoccupied
states~\cite{ring1980:nuclear-many-body,dirac1930:thomas-atom-exchange}.

If we were instead working in terms of some other single-particle basis $\lbrace
\tket{\beta_i} \rbrace$, $\tket{\Psi}$ would not manifestly be represented as a
simple antisymmetrized product state.  However, evaluating the density matrix in
this basis $\lbrace \tket{\beta_i} \rbrace$, and diagonalizing the resulting
matrix, will recover the $\lbrace \tket{\alpha_i} \rbrace$ basis as the natural
orbital basis, thereby revealing $\tket{\Psi}$ as a single antisymmetrized
product state.  (More properly, it will recover the $\lbrace \tket{\alpha_i}
\rbrace$ basis to within an arbitrary freedom of choice of basis within the
spaces spanned by the occupied and unoccupied orbitals separately, as each of
these represents a degenerate eigenspace of the density operator, with
eigenvalues $1$ and $0$, respectively.)  Such a transformation back to a single
antisymmetrized product state is possible if and only if the density matrix has
eigenvalues which are all either $0$ or
$1$~\cite{coleman2000:reduced-density-matrices}.

Of course, we are more generally interested in many-body states which
incorporate correlations.  There is no single-particle basis in which such a
state can be represented as a single antisymmetrized product, and the
eigenvalues of the one-body density operator are no longer simply $0$ and $1$.

However, the transformation to the natural orbital basis still generates
single-particle states for which the mean occupations ``fall as quickly as
possible'', in a very particular sense.  Namely, we order the natural orbitals
$\tket{\phi_i}$ by decreasing eigenvalue ($\lambda_1 \geq \lambda_2 \geq \dots
$), that is, in order of decreasing mean occupation $n_{\phi_i}=\lambda_i$.  The
total mean occupation of the first $q$ single-particle states in any basis is
$n_{\leq q}= \sum_{i=1}^{q} \rho_{ii}$, and the total mean occupation of the
first $q$ natural orbitals, in particular, is $n_{\leq q}'=\sum_{i=1}^q
\lambda_i$.  By a general property of traces of Hermitian
matrices~\cite{fan1949:transformation-eigenvalues-part1}, the partial trace (sum
of the first $q$ diagonal entries) in any basis is bounded from above by the
partial trace in the eigenbasis (sum of the first $q$ eigenvalues).  Thus,
\begin{equation}
  \label{eqn:fan-inequality}
  n_{\leq q}= \sum_{i=1}^{q} \rho_{ii} \leq \sum_{i=1}^{q} \lambda_i =n_{\leq q}'.
\end{equation}
That is, for any $q$, the natural orbitals constitute the basis which maximizes the
total mean occupation of the first $q$ single-particle states~\cite{loewdin1960:natural-orbital}.

The ``naive'' or generic natural orbitals as defined above, by simply
diagonalizing $\rho$ without further precautions, fail to take into account the
symmetry properties of the system.  Despite their name, these natural orbitals
are simply an independent set of single-particle states, without well-specified
quantum numbers, rather than orbitals \textit{per se}, in the sense that
``orbitals'' are usually meant in rotationally-invariant problems, as we now
elaborate.

Consider, in particular, the symmetries present in nuclear
configuration-interaction calculations.  To ensure that the many-body space
supports states of definite angular momentum and parity, the single-particle
states used to build the basis configurations are not arbitrary, but must form
orbitals in the traditional shell-model sense.  An orbital is a set of magnetic
substates $\tket{nljm}$ ($m=-j,\ldots,+j$), which together form an angular
momentum multiplet of definite $j$ and definite parity $P=(-)^l$.  (Since $l$
and $j$ can differ only by $1/2$, the condition of definite parity is sufficient
to also enforce definite $l$.)  The different magnetic substates of the orbital
are related to each other by angular momentum laddering and share the same radial wave
function $R_{nlj}$.

The many-body reference state $\tket{\Psi_{JM}^P}$ for a nuclear
configuration-interaction calculation will have definite angular momentum $J$,
projection $M$ (assuming the problem is formulated in the $M$
scheme~\cite{whitehead1977:shell-methods}), and parity $P$.  These properties of
the reference state serve to impose some, but not all, of the requisite
properties for the natural orbitals to constitute true $nlj$~orbitals (or the
$m$-substates thereof).  By inspection of~(\ref{eqn:obdm-defn}), and the
additive nature of the $m$ quantum number, it is clear that the density matrix
for a reference state $\tket{\Psi_{JM}^P}$ cannot connect single-particle states
$\tket{nljm}$ with different $m$.  Similarly, by the multiplicative nature of
parity, it cannot connect single-particle states of different parity.

However, the density matrix will in general connect single-particle states
$\tket{nljm}$ with different $j$, leading to natural orbitals without definite
angular momentum.\footnote{Only for the special case of a reference state with
  $J=0$ do spherical tensor selection rules prevent the density matrix from
  connecting and thus mixing single-particle states of different $j$.  Even
  here, caution would be necessary in diagonalizing $\rho$, as it would contain
  redundant $ljm$ blocks, one for each $m=-j,\ldots,j$.  Diagonalizing these blocks
  together would lead to degeneracies and thus ambiguity (and, in general, undesirable $m$-mixing) in the choice of
  eigenstates within each degenerate eigenspace, while diagonalizing each block
  of definite $m$ independently would still fail to enforce consistent phase
  relations between the $m$-substates of an $nlj$ orbital.}
Due to such considerations, calculations involving natural orbitals are
instead commonly based on symmetry-adapted natural
  orbitals~\cite{mcweeny1968:symmetry-natorb-part1-adapted,davidson1972:natural-orbital}.
These are obtained by diagonalizing only that part of the one-body density
matrix which is invariant under the action of the symmetry group, namely, for
the present problem, angular momentum (and parity).

We construct a rotational scalar one-body density matrix $\bar\rho$ in terms of
the spherical tensor scalar coupled
product~\cite{suhonen2007:nucleons-nucleus,edmonds1960:am} of the creation and
annihilation operators for an orbital.\footnote{\label{fn:a-dagger-a-tilde}The
  creation and annihilation operators for the magnetic substates $\alpha=(n_a
  l_a j_a m_\alpha)$ of an orbital $a=(n_a l_a j_a)$ together constitute
  spherical tensors $a_a^\dagger$ and $\tilde{a}_a$ with components
\begin{math}
  (a^\dagger_{n_al_aj_a})_{m_\alpha} = a_{n_al_aj_a m_\alpha}^\dagger
\end{math}
and
\begin{math}
(\tilde{a}_{n_al_aj_a})_{m_\alpha} = (-)^{j_a+m_\alpha}a_{n_al_aj_a, -m_\alpha},
\end{math}
respectively~\cite{suhonen2007:nucleons-nucleus}.}
This rotational scalar one-body density matrix has elements
\begin{equation}
  \label{eqn:obrdm-scalar}
 \bar\rho_{ab} =
  \tme{\Psi_{JM}^P}{[a_b^\dagger \tilde{a}_a]_{00}}{\Psi_{JM}^P},
\end{equation}
or, equivalently, in terms of the original, uncoupled one-body density matrix elements defined in~(\ref{eqn:obdm-defn}),
\begin{math}
 \bar\rho_{ab} =
  \delta_{j_aj_b}\hat{\jmath}_a^{-1}\sum_m \rho_{(n_a l_a j_a m)(n_b l_b j_b m)},
\end{math}
where we adopt the notation $\hat{\jmath}\equiv(2j+1)^{1/2}$.

This scalar density matrix $\bar{\rho}$ is now simply a matrix with respect to orbitals (labeled by $a$),
rather than their magnetic substates (labeled by $\alpha$).  The matrix elements $\bar\rho_{ab}$ must be independent of the magnetic substate $M$ of the reference
state, since they are given in~(\ref{eqn:obrdm-scalar}) as the matrix element of a scalar
operator in the many-body space.\footnote{Alternatively, the vestigial reference to the $M$
quantum number in~(\ref{eqn:obrdm-scalar}) can be eliminated by recourse to the
Wigner-Eckart theorem~\cite{edmonds1960:am}, which gives
\begin{math}
  \label{eqn:rhobar-rme}
\bar\rho_{ab} = \hat{J}^{-1}
  \trme{\Psi_{J}^P}{[a_b^\dagger \tilde{a}_a]_{0}}{\Psi_{J}^P},
\end{math}.}

Nonzero matrix elements $\bar\rho_{ab}$ only arise between orbitals of the same
angular momentum ($j_a=j_b$), parity, and thus (as argued above)
orbital angular momentum ($l_a=l_b$).
That is, the scalar one-body density
matrix is block diagonal in $lj$.
Symmetry-adapted natural orbitals, obtained as eigenvectors of
$\bar\rho$, may thus be found by diagonalizing independently within each $lj$
subspace.  The resulting natural orbitals are related to the underlying orbitals simply by a unitary transformation
\begin{equation}
\tket{\phi_{n'ljm}}=\sum_{n}A^{lj}_{nn'}\tket{nljm}
\end{equation}
on the radial wave functions $R_{nlj}$ within each $lj$ space separately.

The total number operator, summed over all magnetic substates of an orbital, is
\begin{math}
  \hat{N}_a=\hat{\jmath}_a[a_a^\dagger \tilde{a}_a]_{00}.
\end{math}
Thus, the diagonal matrix elements
$\bar{\rho}_{aa}$ of the scalar density matrix are proportional to the mean occupation of the orbital $a$,
  \begin{equation}
    \label{eqn:n-orbital}
  n_a =
  \hat{\jmath}_a \bar{\rho}_{aa},
\end{equation}
which ranges from $0$ to the degeneracy $2j_a+1$ of the orbital.  For the
symmetry-adapted natural orbital $\phi_a$, the corresponding eigenvalue
$\lambda_a$ is then proportional to the mean occupation of the
orbital.\footnote{If the spherical tensor annihilation operator in
  footnote~\ref{fn:a-dagger-a-tilde} is instead defined with the common
  alternative phase convention $(\tilde{a}_a)_{j_a,m_\alpha} =
  (-)^{j_a-m_\alpha}a_{(n_a, l_a, j_a, -m_\alpha)}$~\cite{rowe2010:rowanwood},
  which differs by an overall sign, then we instead have
  $\hat{N}_a=-\hat{\jmath}_a[a_a^\dagger \tilde{a}_a]_{00}$, and $n_a =
  -\hat{\jmath}_a \bar{\rho}_{aa}$.}

Ordering the natural orbitals by decreasing eigenvalue, separately within each
$lj$ subspace, serves to define a radial $n$ quantum number, which is now
simply a counting index with no strict relation to the number of radial nodes.
Ordering by decreasing eigenvalue or, equivalently, decreasing mean occupation,
again serves to maximize the mean occupation of the ``lower-lying'' orbitals, as
in~(\ref{eqn:fan-inequality}), but now only within each $lj$ subspace.

For the rotationally-invariant many-body problem with symmetry-adapted natural
orbitals, in contrast to the situation above for ``naive'' natural orbitals, we
would not in general expect the transformation to natural orbitals to reveal a
many-body reference state to be a single antisymmetrized product state.  Even
for a pure shell-model \textit{configuration}, \textit{i.e.}, defined by a
specific distribution of nucleons over $nlj$ orbitals, a state of definite $J$
is in general obtained as a linear combination of many such antisymmetrized
product states, involving different choices of occupied $m$-substates for each
orbital, as required to couple the angular momenta of the individual nucleons to
yield resultant total angular momentum $J$~\cite{whitehead1977:shell-methods}.\footnote{The notable exception is a
  closed-shell configuration, for which the resulting $J=0$ state is indeed
  simply an antisymmetrized product state.}  Transformation to the
symmetry-adapted natural orbitals serves to reveal if a reference state can be
represented, not as a single antisymmetrized product state, but rather as a pure
shell-model configuration, for some choice of basis orbitals.  More generally,
it serves to allow the expansion of the many-body wave function in terms of
fewer low-lying configurations.

\subsection{Natural orbitals in the NCCI framework}
\label{sec:methods:ncci}

In NCCI calculations~\cite{navratil2000:12c-ab-initio,barrett2013:ncsm}, the
many-body basis consists of antisymmetrized product states built from some
underlying orbitals, usually those of the three-dimensional isotropic harmonic
oscillator.  The nuclear Hamiltonian
\begin{equation}
\label{eqn:H}
H=\Tintr+V+a\Ncm,
\end{equation}
is then represented as a matrix in terms of this basis.  Here $\Tintr$ is the
two-body intrinsic kinetic energy
operator~\cite{bethe1937:nuclear-kinetic-com,brussaard1977:shell,caprio2020:intrinsic},
$V$ represents the internucleon interaction (typically limited to two-body or
three-body contributions), and the final Lawson
term~\cite{gloeckner1974:spurious-com,whitehead1977:shell-methods,lawson1980:shell},
proportional to the number operator $\Ncm$ for center-of-mass oscillator quanta,
optionally serves to control the center-of-mass motion (as discussed further
below).

The NCCI many-body basis states are defined as antisymmetrized products of
single-particle states described by quantum numbers $nljm$, where $n$ is the
radial quantum number, $l$ the orbital angular momentum, $j$ the resultant
angular momentum after coupling to spin, and $m$ its projection.  Each product
state thus has definite total angular momentum projection $M=\sum_{i=1}^A m_i$
and parity $P=\Pi_{i=1}^A(-)^{l_i}$.  In a typical $M$-scheme
calculation~\cite{whitehead1977:shell-methods}, the basis is restricted to fixed
$M$ and $P$.  The individual basis states do not have definite angular momentum,
but, since the Hamiltonian is rotationally invariant,\footnote{For states of
  definite total angular momentum to emerge from the diagonalization, the
  many-body space spanned by this basis should also be ``complete'' for this
  purpose, \textit{i.e.}, invariant under rotations.  Such is guaranteed in the
  standard construction procedure for an $M$-scheme basis, where all
  $m$-substates of a given orbital are treated on an equal footing.  But this
  assumption would in general be violated if we were to treat $m$-substates
  unequally in the basis truncation, as might happen if we were to work with
  ``naive'' natural orbitals (Sec.~\ref{sec:methods:no}).} states of definite
total angular momentum $J\geq\abs{M}$ emerge from the diagonalization.

In the usual case where we adopt oscillator orbitals, each orbital is
furthermore identified by its oscillator major shell, or number of
oscillator quanta, $N=2n+l$~\cite{suhonen2007:nucleons-nucleus}.  An
antisymmetrized product state then has $N = \sum_{i=1}^{A} N_{i}$
oscillator quanta overall, where $N_{i}=2n_i+l_i$ represents the number of
oscillator quanta contributed by the $i$th particle. The total number of quanta
may be reexpressed as $N=N_{0} + \Nex$, where $N_0$ is the number of
quanta in the lowest filling of oscillator shells permitted by the Pauli
principle for the given nucleus, so that $\Nex$ then represents the number of
excitation quanta relative to this lowest filling.

The $\Nmax$ truncation scheme restricts the basis configurations to those with
$\Nex\leq\Nmax$, that is, limiting the total number of excitation quanta.  Thus,
$\Nmax=0$ yields a traditional ``$0\hw$'' shell model space, in which all
nucleons are restricted to the valence shell (and an inert core).  Since the parity of a harmonic
oscillator configuration is $P=(-)^{N_0+\Nex}$, a basis consisting of
configurations with $\Nex=0,2,\ldots,\Nmax$ (with $\Nmax$ even) yields a
truncated space of the same parity as the lowest oscillator configuration
(normal parity), while a basis consisting of configurations with
$\Nex=1,3,\ldots,\Nmax$ (with $\Nmax$ odd) yields a truncated space of the
opposite parity (nonnormal parity)~\cite{lane1960:reduced-widths}.  The growth
in dimension of the nuclear many-body space with increasing $\Nmax$ is
illustrated in Fig.~\ref{fig:dimension}, for selected nuclides through the lower
$sd$ shell.

\begin{figure}[tp]
\begin{center}
\includegraphics[width=\ifproofpre{0.9}{0.5}\hsize]{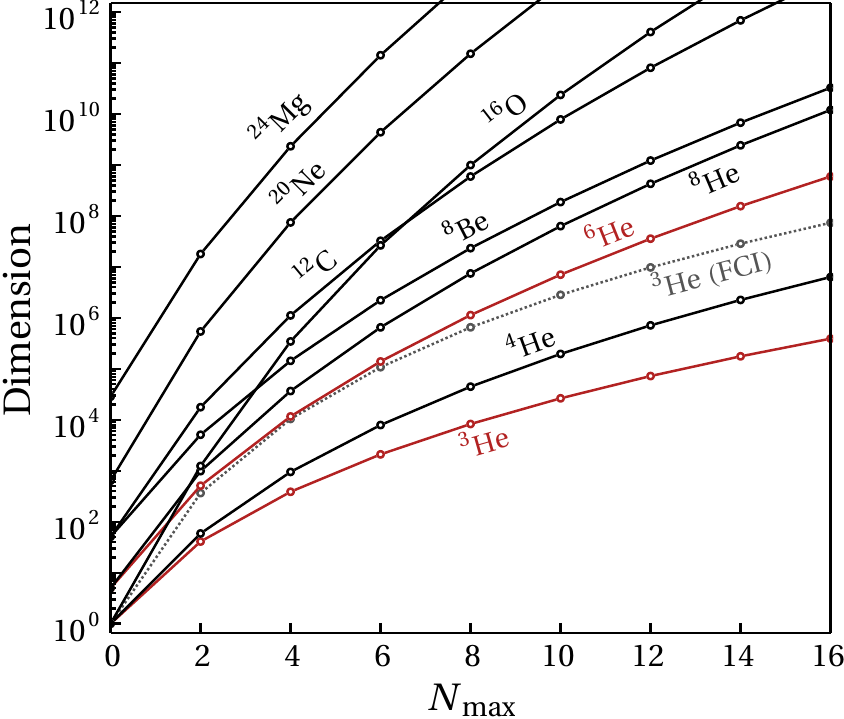}
\end{center}
\caption{Dimension of the NCCI many-body space as a function of the number of
  oscillator excitations $\Nmax$ included in the basis, including for
  $\isotope[3,6]{He}$~(highlighted).  The dimension of the FCI space constructed
  from the same orbitals is also shown for $\isotope[3]{He}$ (dotted gray
  line). Dimensions are those obtained with $M$-scheme bases ($M=0$ for
  even-mass nuclei, or $M=1/2$ for odd-mass nuclei) for the normal-parity space.
}
\label{fig:dimension}
\end{figure}

The truncated space spanned by such a basis, and thus the results of an NCCI
calculation, depend on both the many-body basis truncation parameter $\Nmax$ and
the oscillator length $b$ of the underlying oscillator single-particle
basis.  This length scale is commonly quoted as an oscillator energy $\hw$, in
terms of which $b=(\hbar c)/[(m_Nc^2)(\hw)]^{1/2}$, where $m_N$ is the mean
nucleon mass ($m_Nc^2\approx938.92\,\mathrm{MeV}$).  See, \textit{e.g.},
Refs.~\cite{bogner2008:ncsm-converg-2N,maris2013:ncsm-pshell,caprio2015:berotor-ijmpe,caprio2021:emratio}
for illustrations of convergence of observables with respect to these basis
parameters.

The $\Nmax$ truncation for the oscillator basis holds a special place
in NCCI calculations due to its properties regarding center-of-mass motion.  The
physically-relevant degrees of freedom for describing nuclear structure and
excitations reside in the motion of the nucleons relative to their common center
of mass, rather than in the motion of this center of mass relative to the
laboratory frame.  However, given that the NCCI appproach is formulated in terms
of antisymmetrized products of single-particle states defined with respect to
the laboratory frame, the center-of-mass coordinate cannot be strictly
eliminated as a degree of freedom in the many-body wave function.  Nonetheless,
this motion can at least be brought into a known, controlled form.

Namely, the $\Nmax$ truncation, in particular, ensures that nuclear eigenstates can be
obtained exhibiting an exact separation between a pure oscillator $0s$
wave function for the center of mass coordinate (\textit{i.e.}, the
center-of-mass degree of freedom is frozen into its zero point motion) and an
intrinsic wave function for the motion of the nucleons relative to each other
(see Sec.~II\,B of Ref.~\cite{caprio2020:intrinsic} for a detailed explanation
of the reasoning).  The Lawson term in~(\ref{eqn:H}) selects such eigenstates
with $0s$ center-of-mass motion, by shifting any remaining states involving
center-of-mass excitation out of the low-lying spectrum.  Thus, states involving
excitation of the intrinsic wave function are cleanly separated from what would
otherwise be a thicket of spurious excitations in the calculated spectrum (see
Fig.~8 of Ref.~\cite{caprio2012:csbasis} for an illustration of the effect on
the spectrum).  Moreover, such factorization greatly simplifies the calculation
of certain observables, including the r.m.s.\ radius, electric monopole ($E0$),
magnetic dipole ($M1$), and electric quadrupole ($E2$)
observables~\cite{caprio2020:intrinsic}.

Here it is important to note that the factorized $0s$ center-of-mass wave
function thus obtained has an oscillator parameter $\hwcm=\hw$ which is
determined by the oscillator parameter of the underlying single-particle basis.
Equivalently, in terms of oscillator lengths, the $0s$ wave function in the
center-of-mass coordinate has an $\hw$-dependent oscillator length
$\bcm=A^{-1/2} b$ (see Sec.~F.3 of Ref.~\cite{caprio2020:intrinsic}).  Thus,
many-body calculations carried out in $\Nmax$-truncated oscillator bases of
different $\hw$ result in different ``spectator'' center-of-mass motions.  That
is, the many-body eigenstates obtained using these different bases may converge
towards the same intrinsic structure with increasing $\Nmax$, but not the same
center-of-mass wave function.  This will be important to keep in mind when
interpreting the $\hw$-dependence of the natural orbitals thus obtained (as in
Sec.~\ref{sec:results-3he:orbitals} below).

If we move beyond the traditional oscillator basis in $\Nmax$
truncation, as we must to make use of natural orbitals, we forsake the formal
comfort of having a guaranteed exact center-of-mass factorization.  However, in
practice, an approximate factorization may still be
obtained~\cite{hagen2009:coupled-cluster-com,*hagen2010:coupled-cluster,roth2009:it-ncsm-cc-cm-truncated,caprio2012:csbasis,hergert2016:imsrg},
either since it naturally emerges in the calculation (as explored for the
natural orbital basis in Sec.~\ref{sec:results-3he:Ncm} below) or with some help
from a Lawson term.  Furthermore, the impact upon observables of any spurious
contribution may be mitigated through judicious use of translationally-invariant
intrinsic operators~\cite{caprio2020:intrinsic}.

Indeed, alternate choices both for orbitals and for truncation have already been
applied in NCCI calculations.  For instance, orbitals defined in terms of the
Laguerre
functions~\cite{shull1955-continuum,weniger1985:fourier-plane-wave,mccoy2016:lgalg},
a standard set of basis functions in electron-structure
theory~\cite{helgaker2000:electron-structure}, have been
explored~\cite{caprio2012:csbasis,caprio2014:cshalo}.

For the many-body truncation, calculations have also been performed using the so-called
full configuration interaction (FCI)
truncation~\cite{helgaker2000:electron-structure}, which simply retains all
configurations built by distributing nucleons over the given set of orbitals
(this is simply the traditional fermionic many-body space obtained from a given
set of single-particle states~\cite{negele1988:many-particle}). In the context
of NCCI calculations, the FCI basis is taken as all configurations involving a
given set of oscillator shells.  However, convergence with respect to the many-body basis
size is found to be much slower than for traditional $\Nmax$
calculations~\cite{abe2012:fci-mcsm-ncfc}.  More general many-body truncation
schemes\footnote{Here we specifically have in mind truncation schemes for a
  traditional configuration interaction basis of antisymmetrized product states.
  It should be noted that symmetry-adapted coupling schemes for NCCI
  calculations, based on $\grpsu{3}$~\cite{dytrych2013:su3ncsm} or
  $\grpsptr$~\cite{dytrych2008:sp-ncsm} symmetry groups, are subject to
  truncation schemes of a different nature, as these schemes involve a change of
  basis, before truncation, to correlated many-body basis states.}  are also
feasible, \textit{e.g.}, in which orbitals are weighted by measures other than
the number of oscillator quanta~\cite{vary2018:gentrunc-ostuka17} or in which
the basis configurations are selected through more sophisticated importance
criteria~\cite{roth2007:it-ncsm-40ca}.

Regardless of basis choice, the essential inputs into the construction of the
Hamiltonian matrix in the NCCI basis are the two-body matrix elements of this
Hamiltonian (assuming the internucleon interaction $V$ is two-body, or
three-body matrix elements, if the interaction is three-body, \textit{etc.}).
These must be obtained for the given choice of orbitals.  The rest of the
Hamiltonian construction follows from the standard treatment of $n$-body
operators in second quantization~\cite{negele1988:many-particle}.  The
eigenproblem is thus cast as a large, sparse matrix diagonalization problem,
which is solved numerically using, \textit{e.g.}, the Lanczos
algorithm~\cite{lanczos1950:algorithm,whitehead1977:shell-methods}.

One-body densities are readily extracted from the resulting wave functions.
These densities are commonly used for the computation of one-body observables,
such as matrix elements of electromagnetic operators for moments and
transitions~\cite{suhonen2007:nucleons-nucleus}, and as inputs to reaction
calculations~\cite{goldberger1964:collision}.  More precisely, while the electromagnetic operators, taken
properly in the center-of-mass frame, involve two-body or higher contributions,
they may effectively be replaced by one-body operators when the center-of-mass
motion has the harmonic-oscillator $0s$ form noted
above~\cite{caprio2020:intrinsic,navratil2021:trinv-obme}.  The scalar
densities~(\ref{eqn:obrdm-scalar}), in particular, are also the necessary
ingredient for deducing natural orbitals appropriate to the NCCI framework
(Sec.~\ref{sec:methods:no}).

Our procedure is thus to carry out an initial NCCI calculation in a traditional
$\Nmax$ truncated oscillator basis.  One of the calculated eigenstates, say, the
ground state, is taken as the reference state for generating natural orbitals,
and the relevant scalar densities are extracted.

To see which oscillator orbitals contribute to the resulting natural
orbitals, note that, in an $\Nmax$-truncated NCCI basis, the configurations
involve nucleons reaching orbitals with $\Nmax$ quanta above the valence shell.
The active orbitals thus have $N\leq N_v + \Nmax$, where $N_v$ is the number of
oscillator quanta for the valence shell (\textit{e.g.}, $N_v=0$ for an
``$s$-shell'' nucleus, or $N_v=1$ for a ``$p$-shell'' nucleus).  The calculated
scalar densities reflect only these active orbitals, and the natural orbitals
resulting from diagonalizing the resulting density matrix represent mixtures of
only these orbitals, that is, oscillator orbitals of the same $(lj)$
and with $N\leq N_v + \Nmax$.

The resulting natural orbitals are again labeled by quantum numbers $nlj$,
where now the radial quantum number $n$ no longer necessarily reflects the
number of nodes in the radial wave function but simply reflects the chosen
ordering of natural orbitals by decreasing eigenvalue (\textit{i.e.}, decreasing
mean occupation) as discussed above (Sec.~\ref{sec:methods:no}).  For example,
consider an $\Nmax=4$ calculation for the $s$-shell nucleus $\isotope[4]{He}$.
Within the $s_{1/2}$, or $(l,j)=(0,1/2)$, subspace, the resulting scalar
densities connect the $0s_{1/2}$, $1s_{1/2}$, and $2s_{1/2}$ orbitals
($N=0,2,4$, respectively), and diagonalizing the scalar density matrix thus mixes these
orbitals to define natural orbitals $0s_{1/2}'$, $1s_{1/2}'$, and $2s_{1/2}'$.

It is now straightforward to carry out an NCCI calculation in a new basis,
formed from antisymmetrized products of natural orbitals.  The same many-body
machinery is used as in the original oscillator-basis calculation.  It is merely
necessary to carry out a change of basis~\cite{hagen2006:gdm-realistic} on the
two-body matrix elements of the Hamiltonian~(\ref{eqn:H}) (see Sec.~III\,C of
Ref.~\cite{caprio2012:csbasis}).  Only a finite set of two-body matrix elements
in the oscillator basis are required as input to the transformation,
since, as just noted, each natural orbital is obtained from a finite set of
underlying oscillator orbitals.  Then, evaluation and diagonalization of
the many-body Hamiltonian matrix proceed as before.

However, in defining an NCCI calculation in terms of natural orbitals, a
fundamental question arises as to how to truncate the many-body basis.  The
choice may be expected to profoundly affect the results and, in particular,
determine how the accuracy obtained from the many-body calculation relates to
basis size.\footnote{Admittedly, this same comment applies to the choice of truncation
scheme for NCCI calculations defined in terms of oscillator orbitals as
well, discussed above, where the freedom of choice is commonly ignored.}

The transformation from oscillator orbitals to natural orbitals is
simply a unitary transformation on the single-particle space.  More
specifically, this transformation is restricted to the low-lying subspace
spanned by oscillator states with $N\leq N_v + \Nmax$.  Since the
many-body basis consists of antisymmetrized products of the single-particle
orbitals, a change of basis on the single-particle space inherently induces a
change of basis on the many-body product space.

However, if all antisymmetrized products are retained, as in the FCI truncation,
then, while the basis itself may change, the many-body space spanned by this basis is
invariant under such a rearrangement of the single-particle space.  Thus, an FCI
calculation based on the original oscillator orbitals, or on natural
orbitals obtained by a unitary transformation of these, yield identical results.
No benefit in convergence is achieved.  The truncated many-body spaces obtained
before and after transformation to natural orbitals only differ when the
set of antisymmetrized product states constituting the many-body basis is
truncated in a nontrivial fashion, that is, to a proper subspace of the FCI
space (as compared in Sec.~\ref{sec:results-3he:obs} below).
The dimension of the $\Nmax$ truncated space and the enveloping FCI space
  involving the same orbitals (dotted line) for $\isotope[3]{He}$ may be compared in Fig.~\ref{fig:dimension}.

An obvious, though not necessarily optimal, choice of many-body truncation
scheme, as adopted here, is to simply carry over the formal structure of the
$\Nmax$ truncation.  The natural orbitals are already identified by $nlj$
labels, where, again, $n$ reflects the chosen ordering by decreasing occupation
in the reference state.  For each of these orbitals, we may simply define a
weighting label $N=2n+l$ (as in
Refs.~\cite{caprio2012:csbasis,tichai2019:natorb-mbpt}), although this label no
longer has any direct meaning in terms of oscillator quanta.  We then proceed as
before, by treating this label as an additive quantity, thereby defining $N =
\sum_{i=1}^{A} N_{i}$ for a many-body configuration, and imposing a nominal
$\Nmax$ truncation on the configurations.  This truncation no longer has any
direct connection to the oscillator excitation quanta in the system,
nor does it guarantee exact center-of-mass separability.  However, conveniently
for purposes of comparison, the dimension of the problem is exactly as it was
for the original $\Nmax$-truncated oscillator basis
(Fig.~\ref{fig:dimension}).
 
\section{Illustration of natural orbitals in NCCI calculations: \boldmath$\isotope[3]{He}$}
\label{sec:results-3he}

\subsection{Convergence of observables}
\label{sec:results-3he:obs}

\begin{figure}[tp]
    \begin{center}
        \includegraphics[width=\ifproofpre{0.9}{0.5}\hsize]{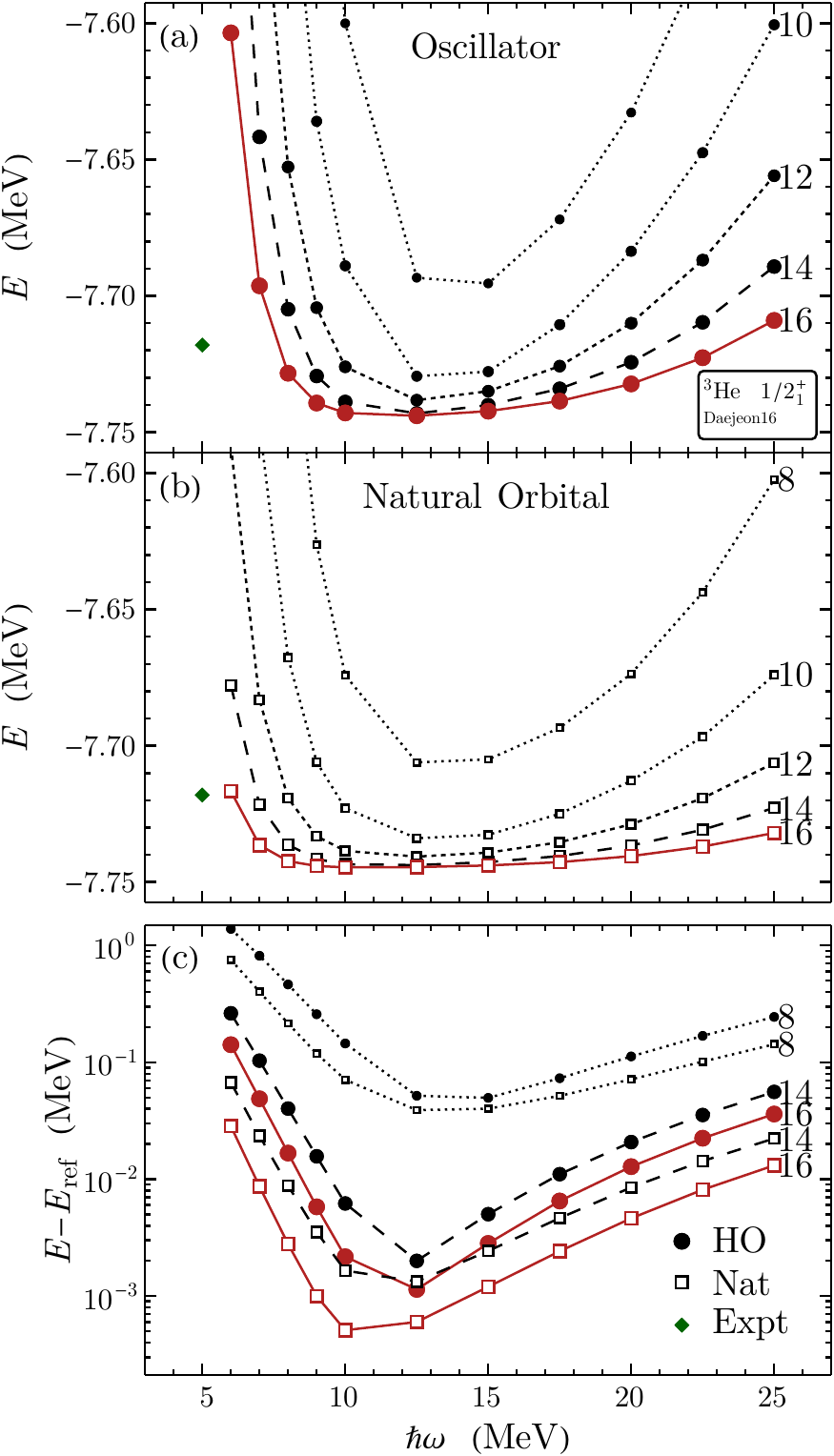}\\
    \end{center}
    \caption{Convergence of $\isotope[3]{He}$ ground-state energy, as calculated
      in (a)~oscillator (solid circles) and (b)~natural-orbital (open
      squares) bases, shown also (c)~on a logarithmic scale as the residual $E-\Eref$ with
      respect to the true ``full-space'' value.  Calculated values are shown as
      functions of the basis parameter $\hw$, for successive even value of
      $\Nmax$, from $\Nmax=8$ (dotted lines) to $16$ (solid lines, highlighted).
      The experimental binding energy (solid diamond)~\cite{wang2021:ame2020} is also shown.
      }
    \label{fig:3he-convergence-energy}
\end{figure}

To see how the formalism just elaborated (Sec.~\ref{sec:methods}) is reflected
in actual NCCI calculations, let us now examine the convergence of
observables in illustrative NCCI calculations, making use of symmetry-adapted
natural orbitals.  Here we take $\isotope[3]{He}$ as the simplest
nontrivial example.  The comparatively slow growth of dimension with $\Nmax$ for
this nuclide (Fig.~\ref{fig:dimension}) means that essentially converged results
can be obtained, as a reference against which the convergence of lower-$\Nmax$
results can be compared.

Results for the ground state energy eigenvalue of $\isotope[3]{He}$ are shown in
Fig.~\ref{fig:3he-convergence-energy}, first as obtained in the oscillator basis
[Fig.~\ref{fig:3he-convergence-energy}(a)], then as obtained in the natural
orbital basis [Fig.~\ref{fig:3he-convergence-energy}(b)].
For these illustrations, we take the Daejeon16 internucleon
interaction~\cite{shirokov2016:nn-daejeon16}, which is based on the two-body
part of the Entem-Machleidt (EM) \nthreelo{} chiral EFT
interaction~\cite{entem2003:chiral-nn-potl}, subsequently softened via a
similarity renormalization group (SRG)
transformation~\cite{bogner2007:srg-nucleon} to enhance convergence and then
adjusted via a phase-shift equivalent transformation to provide better
description of nuclei with $A\leq16$.
Calculations are obtained using
the many-body code
MFDn~\cite{aktulga2013:mfdn-scalability,shao2018:ncci-preconditioned}, along
with codes for the transformation of two-body matrix elements from the oscillator
basis to the natural-orbital basis~\cite{code-shell}, and no Lawson term
[see~(\ref{eqn:H})] is included in the Hamiltonian for the calculations in the
natural-orbital basis.
We also show the experimental binding energy~\cite{wang2021:ame2020} for comparison.

The eigenvalues obtained in the
oscillator-basis calculations [Fig.~\ref{fig:3he-convergence-energy}(a)] follow
a familiar convergence pattern (\textit{e.g.},
Refs.~\cite{bogner2007:srg-nucleon,maris2013:ncsm-pshell}).  Each curve
represents calculations at fixed $\Nmax$ (from $8$ to $16$), for varying $\hw$,
and has a variational minimum with respect to $\hw$, which arises in the
vicinity of $\hw=12.5\,\MeV$ for this particular nuclide, state, and
interaction.  Increasing $\Nmax$, at given $\hw$, strictly expands the space in
which the calculation is carried out, and is thus guaranteed by the variational
principle to monotonically lower the ground state energy.  Convergence towards
the true eigenvalue, as would be obtained in the full, untruncated many-body
space, is signalled by insensitivity to the basis truncation $\Nmax$
(compression of successive curves), as well as local insensitivity to the
oscillator parameter $\hw$ (flattening of the curves).  For the ground state
energy, this manifests as compression of the curves against a variational floor.

For each of these oscillator-basis calculations, the resulting one-body
densitites yield a set of approximate natural orbitals, which define the natural
orbital basis, which we then use for a subsequent many-body calculation, as
outlined in Sec.~\ref{sec:methods:ncci}.  For the resulting energies
[Fig.~\ref{fig:3he-convergence-energy}(b)], each curve again represents
calculations at fixed $\Nmax$, now in the sense of the nominal $\Nmax$ trucation
scheme for natural orbitals (Sec.~\ref{sec:methods:ncci}).

Comparing the overall shapes of the curves, of $E$ \textit{vs.} $\hw$, in
Fig.~\ref{fig:3he-convergence-energy}, we may observe that the natural-orbital
basis provides an overall flattening, or reduced dependence on $\hw$, in the
vicinity of the variational minimum.  However, for a more direct quantitative
comparison of the results obtained with the two bases, the approximately
exponential nature of the convergence with
$\Nmax$~\cite{forssen2008:ncsm-sequences,bogner2007:srg-nucleon,maris2009:ncfc}
means that comparison can be carried out more readily on a logarithmic scale.
To provide a meaningful zero point for the logarithmic scale, we must take the
residual with respect to a ``converged'' reference value $\Eref$, which we obtain from
higher-$\Nmax$ calculations (for $\Nmax\approx 24$, the energy in the vicinity
of the variational minimum is converged to the $\keV$ scale).  The energies,
thus recast as residuals, are shown on a logarithmic scale in
Fig.~\ref{fig:3he-convergence-energy}(c), for the results obtained both with the
oscillator (filled circles) and natural-orbital (open squares) bases.  To
provide clear separation in the plot, only the $\Nmax=8$, $14$, and $16$ results
are shown.

At lower $\Nmax$, as exemplified by the $\Nmax=8$ results (dotted lines) in
Fig.~\ref{fig:3he-convergence-energy}(c), there is little distinction between
the results obtained in oscillator and natural-orbital bases.  This is
perhaps to be expected.  In the limit of $\Nmax=0$, the bases for the oscillator
and subsequent natural-orbital calculations are strictly identical.  More generally, a
low-$\Nmax$ underlying oscillator calculation provides little opportunity for
high-$N$ orbitals to appear in the densities and thus natural orbitals.

At higher $\Nmax$, as exemplified by the $\Nmax=14$ and $16$ results (dashed and
solid lines, respectively) in Fig.~\ref{fig:3he-convergence-energy}(c), one way
of comparing the results is to measure the advance obtained by the transformation
to natural orbitals in terms of the equivalent increase in $\Nmax$ required with
a traditional oscillator basis to achieve the same advance.  In this
sense, for calculations in the vicinity of the variational minimum, the energies
obtained with natural orbitals are approximately ``one step'' in $\Nmax$ ahead
of those obtained with oscillator orbitals.  Away from the variational minimum,
the advantage provided by the natural orbitals is more marked, reflecting the
comparative $\hw$-independence already noted for these calculation in the
natural-orbital basis.

Alternatively, we may assess the results of the change of basis in terms of the
fraction by which it reduces the residual, \textit{i.e.}, how far it brings us
towards the true value which would be obtained in the full, untruncated space.
On a logarithimic scale, a given downward vertical shift represents a given
fractional reduction.  Comparing the $\Nmax=16$ results obtained in the two
bases, we may observe an approximately uniform downward shift, across the range
of $\hw$, representing a reduction in the residual by a factor of $\sim3$ (a
somewhat greater reduction is attained with the natural-orbital basis for
$\hw\approx10\,\MeV$).

\begin{figure}[tp]
    \begin{center}
        \includegraphics[width=\ifproofpre{0.9}{0.5}\hsize]{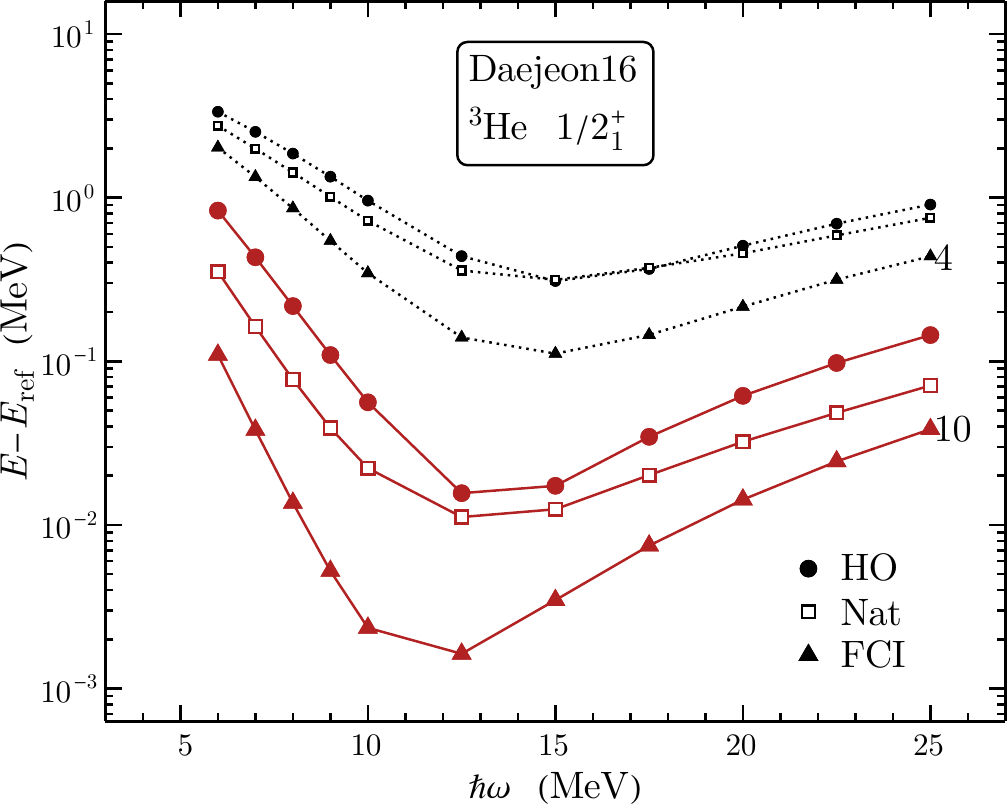}\\
    \end{center}
    \caption{Comparison of $\isotope[3]{He}$ ground-state energies as calculated
      in spaces defined by $\Nmax$-truncated bases~--- oscillator
      (solid circles) or natural-orbital (open squares)~--- and the
      corresponding enveloping FCI space (solid triangles).
      Energies are shown as residuals, as in Fig.~\ref{fig:3he-convergence-energy}.
      Calculated values
      are shown as functions of the basis parameter $\hw$, for $\Nmax=4$ (dotted
      lines) and $10$ (solid lines, highlighted).
      }
    \label{fig:3he-e-hw-scan-fci}
\end{figure}

However, there is an obvious bound on the improvement which may be expected from
the transformation to the natural-orbital basis derived from an
$\Nmax$-truncated oscillator basis calculation.  Recall that the active orbitals
in the oscillator-basis calculation and the subsequent natural orbitals span the
same single-particle space.  Both the $\Nmax$-truncated oscillator basis and the
nominally $\Nmax$-truncated constructed from the ensuing natural orbitals span
subspaces of the same enveloping FCI space defined by those orbitals
(Sec.~\ref{sec:methods:ncci}).  This FCI space is, in general, of much higher
dimension~\cite{abe2012:fci-mcsm-ncfc}.  \textit{E.g.}, for an $\Nmax=10$
calculation for $\isotope[3]{He}$, which has dimension $2.6\times10^4$, the FCI
space consists of all product states involving orbitals through the $N=10$
oscillator shell, which has the substantially larger dimension $2.8\times10^6$
(Fig.~\ref{fig:dimension}).  We might hope that the $\Nmax$-truncated natural
orbital basis might allow us to reach comparable accuracy in a much smaller
space, but it cannot access any components of the true wave function which lie
outside of the FCI truncated space.

For the ground state energy, in particular, the result in the FCI space provides
a variational lower bound on the results in the subspaces.  Thus, it is
informative to compare the improvement obtained with natural orbitals to the
maximum improvement which could be obtained in the enveloping FCI space.  The
calculated $\isotope[3]{He}$ ground state energies obtained in the oscillator
basis (filled circles) and natural-orbital basis (open squares) are compared
with the variational bound provided by the enveloping FCI space (filled
trianges) in Fig.~\ref{fig:3he-e-hw-scan-fci}.  Here again, as in
Fig.~\ref{fig:3he-convergence-energy}(c), values are shown as residuals relative
to the true energy, on a logarithmic scale.

At low $\Nmax$, as exemplified by the $\Nmax=4$ results (dotted lines), a factor
of $\sim3$ improvement is possible within the FCI space, near the variational
minimum and over most of the $\hw$ range shown.  Yet, as already noted, the
transformation to natural orbitals conveys negligible benefit, at least within
the nominal $\Nmax$ truncation scheme.

At higher $\Nmax$, as exemplified by the $\Nmax=10$ results (solid lines), the
improvement possible within the FCI space ranges from a factor of $\sim4$, at
the extreme $\hw$ shown, to an order of magnitude, near the variational minimum.
Near the variational minimum, the improvement attained in the natural-orbital
basis, which reduces the residual by less than a factor of $2$, is by this
measure perhaps disappointing.  Further from the variational minimum, however,
the improvement afforded by the $\Nmax$-truncated natural-orbital basis becomes
an appreciable fraction of that possible within the FCI space.  A natural
question is whether the improvement possible within the FCI space could be more
fully achieved, still with a reduction in dimension comparable to that afforded
by the $\Nmax$ truncation scheme, but under a more physically-informed
truncation scheme, \textit{e.g.}, one which makes use of the information on
expected occupations of the orbitals provided by the eigenvalues of the density
matrix.

\begin{figure}[tp]
    \begin{center}
        \includegraphics[width=\ifproofpre{0.9}{0.5}\hsize]{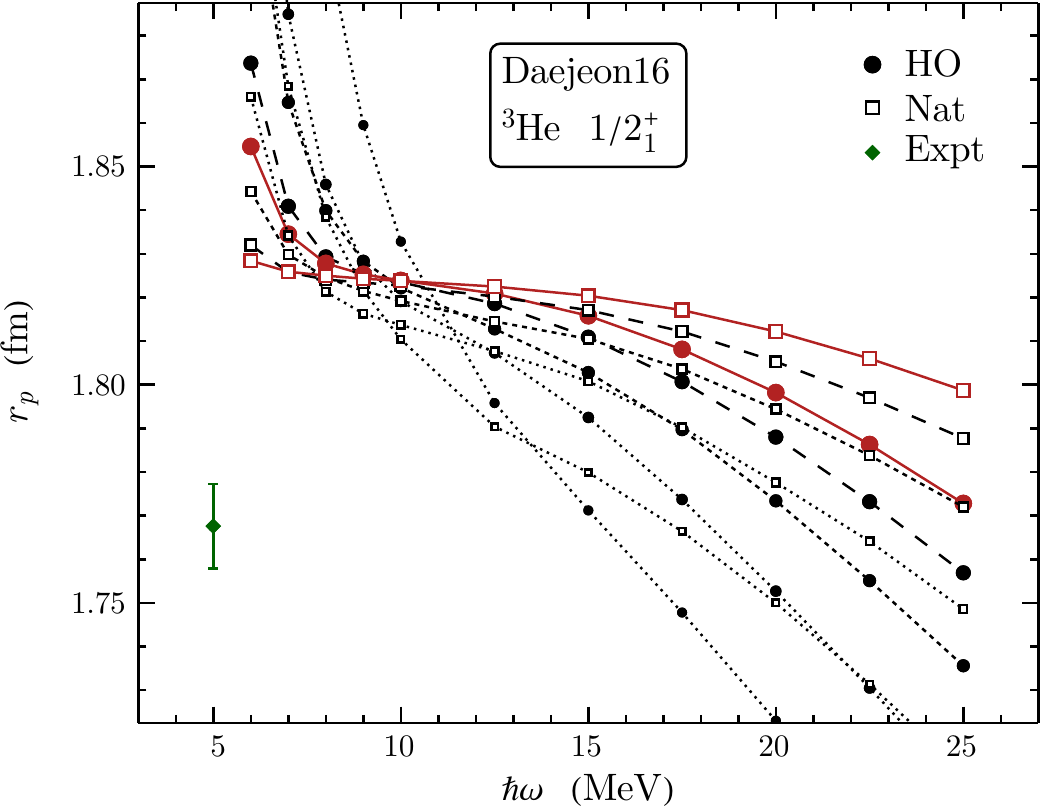}
    \end{center}
    \caption{Convergence of $\isotope[3]{He}$ ground-state point-proton
      r.m.s.\ radius, as calculated in oscillator (solid circles)
      and natural-orbital (open squares) bases.  Calculated values are shown as functions of the basis parameter
      $\hw$, for successive even value of $\Nmax$, from $\Nmax=8$ (dotted
      lines) to $16$ (solid lines, highlighted).
      The value deduced from the experimental charge radius~\cite{angeli2013:charge-radii} is also shown (filled diamond).
      }
    \label{fig:3he-convergence-hw-radius}
\end{figure}

As an initial illustrative example of the convergence obtained for an observable
other than the energy, we consider the point-proton root-mean-square (r.m.s.)
radius $r_p$ of the $\isotope[3]{He}$ ground state.  (The point-proton radius is
simply related to the physically-accessible charge
radius $r_c$~\cite{angeli2013:charge-radii}, after hadronic physics
corrections~\cite{friar1997:charge-radius-correction,lu2013:laser-neutron-rich}.)  The r.m.s.\ radius,
like electric quadrupole ($E2$) observables, is sensitive to the large-radius
behavior of the wave function, as the $r^2$ operator more heavily weights the
tails of the wave functions.  The convergence of such observables is therefore
notably troublesome in NCCI calculations in an oscillator
basis~\cite{bogner2008:ncsm-converg-2N,maris2013:ncsm-pshell,caprio2021:emratio}.
However, improved asymptotic behavior of the single-particle basis, as one
anticipates with the natural orbitals (and as illustrated below in
Sec.~\ref{sec:results-3he:orbitals}), might therefore be expected to
particularly impact the convergence of such observables.

The calculated results for $r_p$ are shown in
Fig.~\ref{fig:3he-convergence-hw-radius}, where the values obtained with
the oscillator (filled circles) and natural-orbital (open squares) bases are overlaid.
An approach to convergence is signaled by the ``shouldering'' of the curves, to
form a region of local $\hw$-independence (flattening) and compression of curves
for successive $\Nmax$ against each other.
The value for $r_p$ deduced from the experimental $r_c$~\cite{angeli2013:charge-radii}
is shown for comparison (filled diamond).

The oscillator-basis calculations for the radius are already atypically
well-converged for $\isotope[3]{He}$ (compare, \textit{e.g.}, Sec.~\ref{sec:results-6he} below).  Note the highly expanded vertical scale in
Fig.~\ref{fig:3he-convergence-hw-radius} (on the scale of $0.1\,\fm$ overall).
For the underlying oscillator calculations (filled circles), the
various curves for different $\Nmax$ (from $10$ to $16$) cross in the vicinity
of $\hw=10\,\MeV$.  (Such crossings have been suggested, purely heuristically,
as a means of estimating the true radius as it would be obtained in the full,
untruncated
space~\cite{nogga2006:7li-ncsm-chiral,bogner2008:ncsm-converg-2N,cockrell2012:li-ncfc},
though in practice this prescription must be treated with
caution~\cite{caprio2014:cshalo}.)

The subsequent calculations in the natural-orbital basis (open squares) do not
share such a sharply-defined crossing point.  Rather, they more clearly
demonstrate the traditional hallmarks of convergence, namely, flattening and
compression of the curves.  For instance, the $\Nmax=16$ curve varies by
$\lesssim0.04\,\fm$ over the range of $\hw$ from $10\,\MeV$ to $20\,\MeV$, while
the $\Nmax=14$ and $16$ curves differ by less than $\lesssim 0.01\,\fm$ over this
same range.  As a consequence of this flattening, by the high end of the $\hw$
range shown ($\hw=25\,\MeV$), the calculations in the natural-orbital basis lie
two steps in $\Nmax$ ``ahead'' of the calculation in the oscillator basis.  The
question, of course, is how this difference in convergence behavior actually
aids in the problem of direct interest in less well-converged cases, which is to
accurately estimate the true value of the observable, as it would be found in
the full, untruncated space.

\subsection{Natural orbitals}
\label{sec:results-3he:orbitals}
\begin{figure}[tp]
    \begin{center}
        \includegraphics[width=\ifproofpre{0.9}{0.5}\hsize]{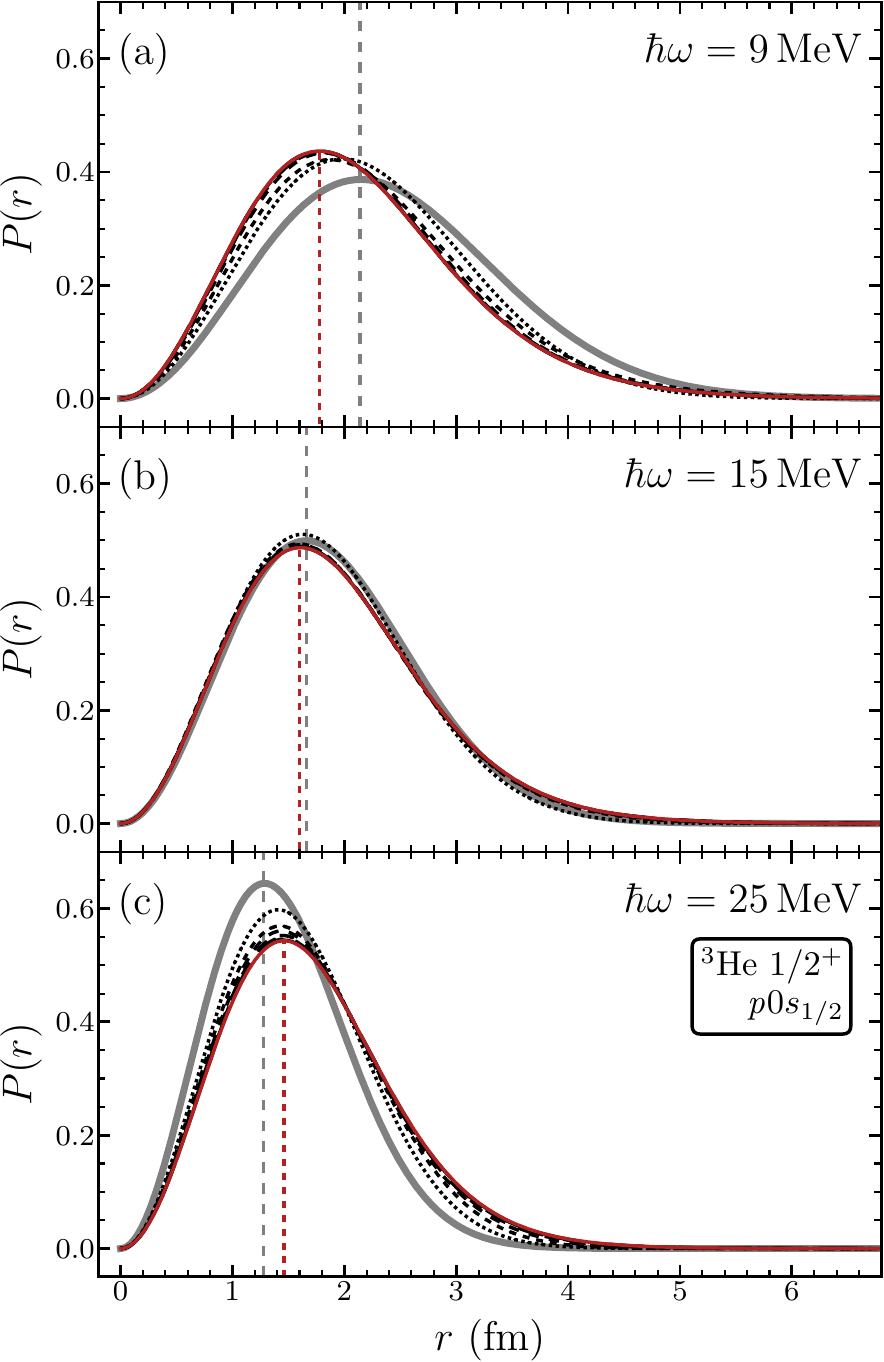}\\
    \end{center}
    \caption{Radial wave functions obtained for the $\isotope[3]{He}$ proton
      $0s_{1/2}$ natural orbital, from different underlying oscillator-basis
      calculations, plotted as the radial probability density $P(r)=r^2
      \abs{\psi(r)}^2$.  Results are shown as obtained from underlying
      oscillator-basis calculations with (a)~$\hw=9\,\MeV$, (b)~$\hw=15\,\MeV$,
      and (c)~$\hw=25\,\MeV$.  Radial wave functions are shown for $\Nmax=2$
      (dotted lines) through $\Nmax=16$ (solid lines, highlighted), with the
      oscillator $0s$ function for the given $\hw$ (thick gray lines) shown for
      comparison. The locations of the peaks of the underlying
      harmonic-oscillator orbital and $\Nmax=16$ natural orbital are marked with
      dashed vertical lines.  }
    \label{fig:3he-orbitals-Nmax-evolution-Pr}
\end{figure}
\begin{figure}[tp]
    \begin{center}
        \includegraphics[width=\ifproofpre{0.9}{0.5}\hsize]{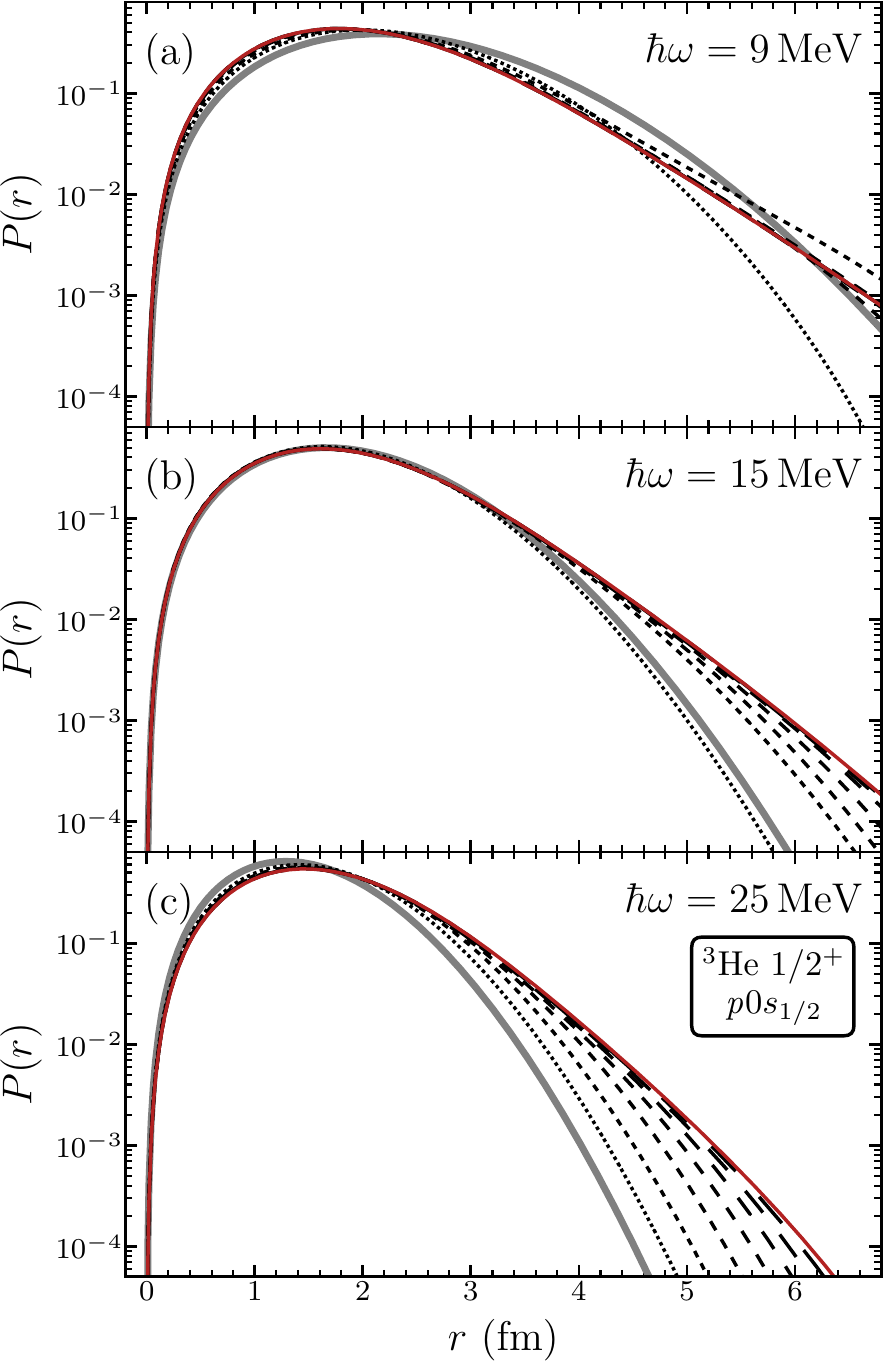}\\
    \end{center}
    \caption{Radial wave functions obtained for the $\isotope[3]{He}$ proton
      $0s_{1/2}$ natural orbital, from different underlying oscillator-basis
      calculations, plotted as the radial probability density $P(r)$, as in
      Fig.~\ref{fig:3he-orbitals-Nmax-evolution-Pr}, but now on a logarithmic
      scale.
    }
    \label{fig:3he-orbitals-Nmax-evolution-logPr}
\end{figure}
\begin{figure*}[tp]
    \begin{center}
        \includegraphics[width=\ifproofpre{0.9}{0.9}\hsize]{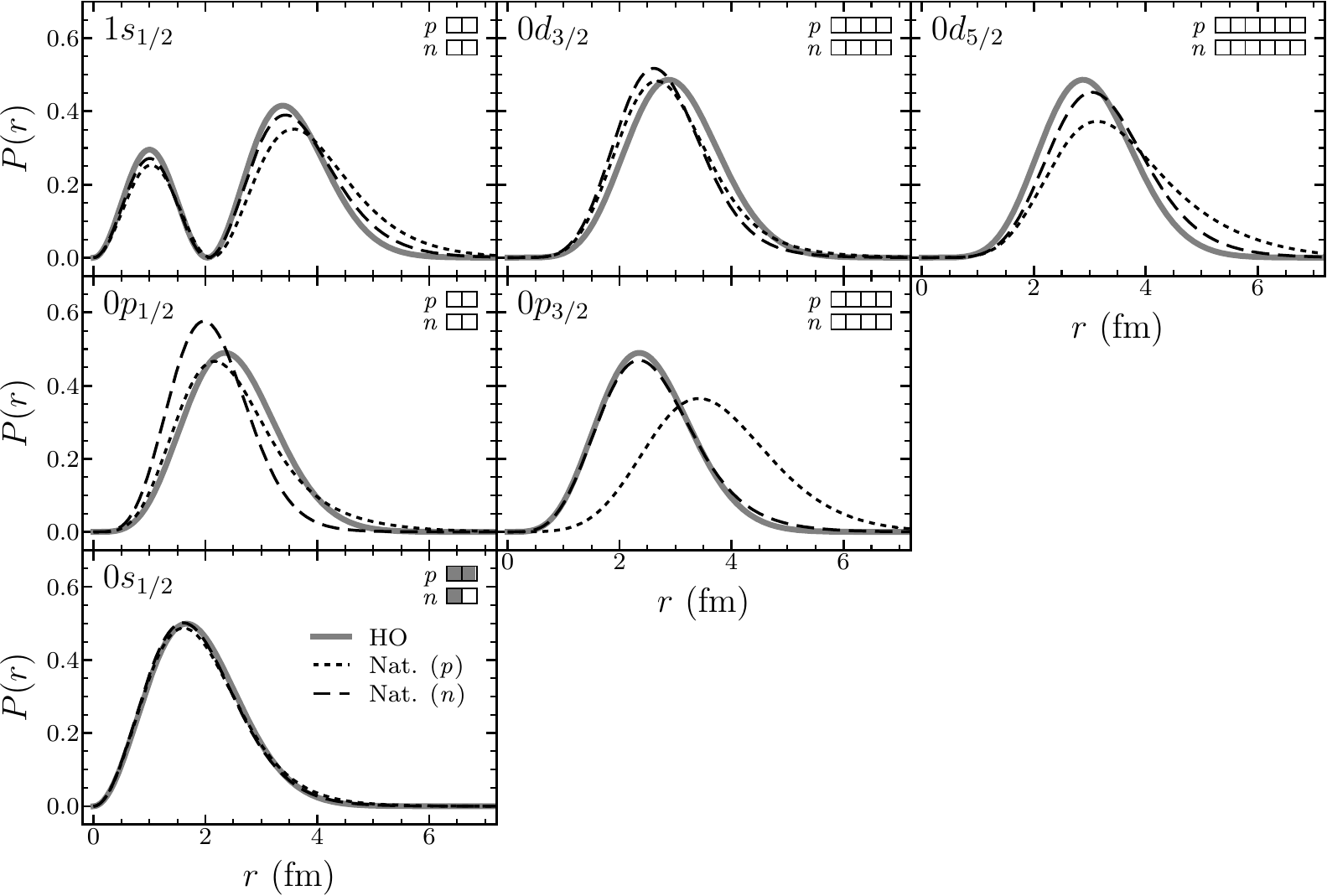}\\
    \end{center}
    \caption{Radial wave functions for the $\isotope[3]{He}$ $s$-, $p$-, and
      $sd$-shell natural orbitals, for both protons (short dashed lines) and
      neutrons (long dashed lines), plotted as the radial probability density
      $P(r)$.  These are obtained from the underlying oscillator-basis
      calculation near the variational minimum ($\hw=15\,\MeV$) and at high
      $\Nmax$ ($\Nmax=16$).  The corresponding oscillator radial
      functions for $\hw=15\,\MeV$ (thick gray lines) are shown for comparison.
      The mean occupation $n_a$ for each natural orbital, from the corresponding
      eigenvalue of the scalar density matrix, is indicated by the filling of
      the bar at top right (upper bar for protons, lower bar for neutrons).
    }
    \label{fig:3he-orbitals-Pr}
\end{figure*}

Let us now examine the natural orbitals obtained (and subsequently used) in the
present $\isotope[3]{He}$ calculations, with the aim of understanding their
dependence on the underlying oscillator calculation and thereby also of gaining
some insight into their influence on the convergence of observables in the
many-body calculation.  Recall that the natural orbitals in these many-body
calculations are approximations to the ``true'' natural orbitals for the
$\isotope[3]{He}$ ground state, since they are deduced from the approximate
$\isotope[3]{He}$ ground state densities obtained in finite, truncated
oscillator-basis NCCI calculations (Sec.~\ref{sec:methods:ncci}).  The
densities, and thus the resulting natural orbitals, depend upon both the $\Nmax$
and $\hw$ of the underlying oscillator-basis calculation.

Recall, furthermore, that the symmetry-adapted natural orbitals
(Sec.~\ref{sec:methods:no}) appropriate to NCCI calculations preserve the $l$ and
$j$ quantum numbers, changing only the radial wave function, by ``mixing''
underlying oscillator orbitals of different $n$ within an $lj$ space.  We
focus first on the $0s_{1/2}$ orbital, as this is the notionally ``occupied''
orbital in a simple shell-model picture, and is indeed still the most
heavily-occupied orbital in the actual NCCI calculations.  We then explore the
properties of the notionally ``unoccupied'' excited orbitals.
While the occupations of these excited (or notionally unoccupied) orbitals are
comparatively small, it is these orbitals which drive the convergence of the
many-body calculation in a natural-orbital basis.

The radial wave function for the $0s_{1/2}$ natural orbital for protons, in
particular, is shown in Fig.~\ref{fig:3he-orbitals-Nmax-evolution-Pr}, where its
dependence on the $\Nmax$ and $\hw$ of the underlying oscillator calculation is
mapped out.  (The behavior for the neutron $0s_{1/2}$ orbital is qualitatively
similar.)  Here, the radial wave function is plotted as the radial probability
density $P(r)=r^2 \abs{\psi(r)}^2$, from $\Nmax=2$ (dotted lines) to $\Nmax=16$
(solid lines), separately for $\hw=9\,\MeV$
[Fig.~\ref{fig:3he-orbitals-Nmax-evolution-Pr}(a)], $15\,\MeV$
[Fig.~\ref{fig:3he-orbitals-Nmax-evolution-Pr}(b)], and $25\,\MeV$
[Fig.~\ref{fig:3he-orbitals-Nmax-evolution-Pr}(c)].  The $0s$ radial function
for the underlying oscillator basis is also shown for comparison (thick gray
line).  Note that the natural orbital obtained from an $\Nmax=0$ oscillator
calculation is still simply this oscillator function, as the resulting densities
do not mix the fully-occupied $s$-shell orbitals with the fully-unoccupied
higher orbitals.

The densities, and thus the resulting natural orbitals, are expected to
eventually converge with increasing $\Nmax$.  Such is indeed seen in
Fig.~\ref{fig:3he-orbitals-Nmax-evolution-Pr}, if we examine the curves within a
given panel, \textit{i.e.}, obtained for different $\Nmax$ but at a given choice
of $\hw$.  On this scale, the shape of the radial wave function appears to
change comparatively little for $\Nmax$ beyond about $4$ or $6$.

The $\hw$ dependence is more subtle.  All observables (energies, electromagnetic
matrix elements, radii, \textit{etc.}) obtained from the densities retain some
$\hw$ dependence at finite $\Nmax$ due to their sensitivity to the $\Nmax$- and
$\hw$-dependent \textit{intrinsic} structure of the approximate
$\isotope[3]{He}$ ground state obtained in a truncated oscillator calculation.
At finite $\Nmax$, some $\hw$ dependence of the natural orbitals may similarly
be expected to arise from such sensitivity to the $\Nmax$- and $\hw$-dependence
of the calculted intrinsic structure.  This dependence is expected to ultimately
disappear with increasing $\Nmax$, as the intrinsic structure converges.

However, recall (Sec.~\ref{sec:methods:ncci}) that even in the large $\Nmax$
limit the natural orbitals for the NCCI problem are not uniquely
defined. Rather, they may be expected to have an inherent $\hw$-dependence
arising from the \textit{center-of-mass} zero-point motion of the reference
many-body state, which varies with the $\hw$ of the underlying oscillator basis.
Thus, it should not be surprising that, even at high $\Nmax$, the natural
orbitals obtained from underlying oscillator-basis calculations with different
$\hw$ do not coincide.  Compare the solid curves in the different panels of
Fig.~\ref{fig:3he-orbitals-Nmax-evolution-Pr}.  These clearly do not coincide,
with the location of the maximum moving to smaller radius with increasing $\hw$.

To characterize how the radial wave functions for the natural orbitals at high
$\Nmax$ (solid lines) differ qualitatively from those of the underlying
oscillator functions (thick gray lines), in
Fig.~\ref{fig:3he-orbitals-Nmax-evolution-Pr}, we shall find it convenient to
separately consider the central region of the wave function and its large-radius
tail (porous though this distinction may be).  Let us first consider the central
region, that is, around the peak in the wave function.

For the $0s_{1/2}$ natural orbital obtained from the reference wave function
calculated in an $\hw=15\,\MeV$ oscillator basis
[Fig.~\ref{fig:3he-orbitals-Nmax-evolution-Pr}(b)], there is little apparent
change going from the underlying oscillator function to the natural orbital.
For the natural orbital obtained in an $\hw=9\,\MeV$ oscillator basis
[Fig.~\ref{fig:3he-orbitals-Nmax-evolution-Pr}(a)], which has a longer
oscillator length [recall $b\propto(\hw)^{-1/2}$], the peak shifts inward, to
lower radius, relative to the underlying oscillator function, though not all the
way to the peak location for $\hw=15\,\MeV$
[Fig.~\ref{fig:3he-orbitals-Nmax-evolution-Pr}(b)].  Alternatively, for the
natural orbital obtained in an $\hw=25\,\MeV$ oscillator basis, which has a
shorter oscillator length [Fig.~\ref{fig:3he-orbitals-Nmax-evolution-Pr}(c)],
the peak shifts outward, to larger radius, relative to the underlying oscillator
function, though again not all the way to the peak location for
$\hw=15\,\MeV$.

Either way, a portion of the effect of transforming from the underlying
oscillator basis to natural orbitals is to ``dilate'' the radial function to
more closely resemble a $0s$ oscillator function of $\hw\approx15\,\MeV$.  The
effect is to moderate the change in characteristic length scale for the natural
orbitals, as the $\hw$ for the underlying oscillator basis is varied, as
compared to the change in oscillator length for the underlying oscillator
orbitals themselves.  This reduced $\hw$ dependence of the orbitals (at least in
the central region) presumably contribues to the reduction in $\hw$ dependence
found for the observables in the calculations based on the natural-orbital basis
(Sec.~\ref{sec:results-3he:obs}).

A simple and intuitive explanation for this behavior of the orbitals is that the
natural orbitals are the result of a compromise between the intrinsic structure
and center-of-mass motion embodied within the reference wave function.  The
intrinsic structure is described well by nucleons occupying orbitals resembling
an $\hw=15\,\MeV$ $0s_{1/2}$ oscillator orbital, regardless of underlying
oscillator basis.  But the center-of-mass motion of the reference wave function
for $\hw=9\,\MeV$ is well described by nucleons in an $\hw=9\,\MeV$ $0s_{1/2}$
oscillator orbital.  The resulting $0s_{1/2}$ natural orbital lies somewhere
inbetween.  Similarly, the center-of-mass motion of the reference wave function
for $\hw=25\,\MeV$ is well described by nucleons in an $\hw=25\,\MeV$ $0s_{1/2}$
oscillator orbital, and the resulting $0s_{1/2}$ natural orbital lies somewhere
inbetween.

Turning now to the tail region of the orbital, the natural question is the
extent to which the natural orbitals take on the exponential asymptotics
anticipated from the mean-field description of the nucleus.  Recall that these asymptotics
are expected to be particularly important for the convergence of long-distance
observables (Sec.~\ref{sec:results-3he:obs}).

The asymptotic behavior is more readily apparent if we replot the radial
probability densities on a logarithimic scale, as in
Fig.~\ref{fig:3he-orbitals-Nmax-evolution-logPr}.  A tail with exponential
asymptotics appears as a straight line on such a plot, while a tail with the
Gaussian asymptotics characteristic of the oscillator functions appears as
downward-curving parabola, as seen for the underlying oscillator radial
functions (grey lines).  We may observe that the tail ``grows in'', with the
inclusion of additional oscillator functions, so that exponential asymptotics
(\textit{i.e.}, straight-line falloff on the log plot) are gradually
established, extending to larger radii with increasing $\Nmax$. (One may compare
to Fig.~4 of Ref.~\cite{davies1966:hartree-fock}, for a classic illustration of
an exponential tail growing in for a Hartree-Fock orbital, or to Fig.~1 of
Ref.~\cite{caprio2014:cshalo}, for the schematic example of a Woods-Saxon
orbital expanded in an oscillator basis~\cite{suhonen2007:nucleons-nucleus}.)
The emergence of exponential asymptotics is most clearly visible for the
$\hw=15\,\MeV$ natural orbitals
[Fig.~\ref{fig:3he-orbitals-Nmax-evolution-logPr}(b)], where the progression
from the underlying oscillator orbital to the true, high-$\Nmax$ natural orbital
is not complicated by a signficant radial shift in the peak location.

NCCI calculations for $\isotope[3]{He}$ in a natural-orbital basis involve, of
course, not just the notionally occupied $0s_{1/2}$ orbital, but also basis
configurations incorporating the higher, notionally unoccupied, natural
orbitals, as well.  Some of the low-lying natural orbitals are shown in
Fig.~\ref{fig:3he-orbitals-Pr}, for both protons (short dashed lines) and
neutrons (long dashed lines).  Here we follow the analogy to an oscillator
basis, by focusing on natural orbitals with $nlj$ quantum numbers corresponding
to the traditional $N=0$ ($s$), $1$ ($p$), and $2$ ($sd$) oscillator shells.
We focus on the natural orbitals obtained from the $\hw=15\,\MeV$
oscillator-basis calculation at $\Nmax=16$, so that the proton $0s_{1/2}$
orbital here corresponds to the highest-$\Nmax$ case shown in
Fig.~\ref{fig:3he-orbitals-Nmax-evolution-Pr}(b).  Again, the underlying
oscillator orbital is shown for comparison (thick gray lines).

Let us first consider the ``occupations''~(\ref{eqn:n-orbital}) of these
orbitals in the reference wave function, which we know from the corresponding
eigenvalues of the scalar density matrix (Sec.~\ref{sec:methods:no}).  (Such
occupations provide only an estimate of the occupation in any subsequent
many-body calculation using the natural-orbital basis.)  The occupations are
shown graphically at the top of each panel in Fig.~\ref{fig:3he-orbitals-Pr},
but at this scale are indistinguishable from those of the traditional shell
model description (in which $n_{0s_{1/2}}=1$ for the neutrons, $n_{0s_{1/2}}=2$
for the protons, and all other orbitals are unoccupied).  More precisely, for
the present illustrative calculation, we have $n_{0s_{1/2}}\approx0.96$ for the
neutrons and $n_{0s_{1/2}}\approx1.92$ for the protons.  The next most occupied
orbitals are the $p$-shell orbitals and the $1s_{1/2}$ orbital of the $sd$
shell, with mean occupations of $\sim10^{-2}$, while occupations fall off
towards $\sim10^{-3}$ and below for higher orbitals.

Overall, in Fig.~\ref{fig:3he-orbitals-Pr}, the general impression is that the
natural orbitals simply ``tweak'' the oscillator radial functions, with modest
shifts to the peak location and overall shape (again, a linear scale does not do
justice to changes in the asymptotics).  The difference in proton and neutron
structure in the reference many-body calculation for $\isotope[3]{He}$ is
manifest in the differences between corresponding proton and neutron natural orbitals.  The
distinction is most striking for the proton $0p_{3/2}$ orbital, which is shifted
to markedly larger radii than the corresponding neutron orbital (which remains
close to the underlying oscillator function).  In general, the proton radial
functions develop more pronounced tails than the neutron orbitals, visible even
on a linear scale, suggestive of Coulomb repulsion effects.

In atomic and molecular electron structure theory, it is recognized that an
important characteristic of the natural orbitals, including the unoccupied
orbitals, is their tendency to remain localized in the region of high particle
density~\cite{davidson1972:natural-orbital}.  This is to be contrasted with the
unoccupied (virtual) Hartree-Fock orbitals, which instead provide an expansion
of the continuum.

It is thus worth elaborating on an essential difference between natural orbitals
and Hartree-Fock orbitals (\textit{e.g.},
Ref.~\cite{ring1980:nuclear-many-body}).  The unoccupied natural orbitals are
well-defined, from the densities of the reference many-body calculation.  In
contrast, the basic variational condition for the Hartree-Fock ground state
focuses entirely on optimizing the occupied orbitals, so as to minimize the
energy in a single Slater determinant.  The unoccupied orbitals are entirely
unconstrained by this variational condition (except insofar as they must span an
orthogonal complement to the occupied orbitals).  The iterative calculational
procedure for obtaining Hartree-Fock orbitals introduces a single-particle
eigenproblem (involving Hartree and exchange potentials), intended to yield the
occupied orbitals.  While the set of solutions can be extended to provide a
definition (one particular choice) for the unoccupied Hartree-Fock orbitals, it
is not at all obvious that these unoccupied Hartree-Fock orbitals should be
particularly well-suited for efficiently expanding the many-body wave functions
in a configuration-interaction basis.

\subsection{Center-of-mass factorization}
\label{sec:results-3he:Ncm}

\begin{figure}[tp]
    \begin{center}
        \includegraphics[width=\ifproofpre{0.9}{0.5}\hsize]{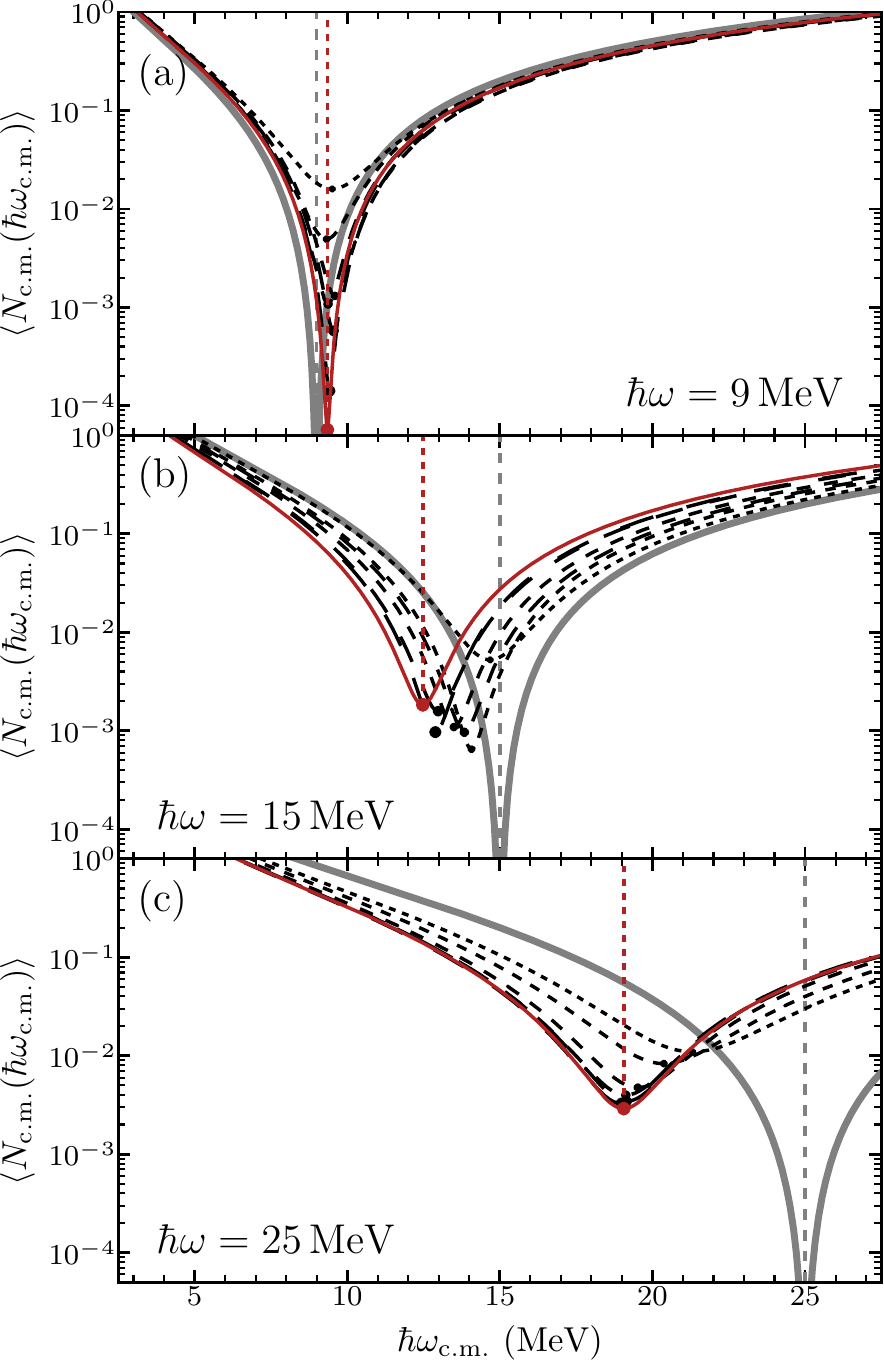}\\
    \end{center}
    \caption{Dependence of $\tbracket{\Ncm}$ on $\hwcm$, for $\isotope[3]{He}$
      ground state wave functions obtained in calculations with a
      natural-orbital basis, derived from underlying oscillator-basis
      calculations with (a)~$\hw=9\,\MeV$, (b)~$\hw=15\,\MeV$, and
      (c)~$\hw=25\,\MeV$.  Results are shown for calculations with $\Nmax=4$
      (short-dashed lines) through $\Nmax=16$ (solid lines, highlighted), with
      the curve obtained for an oscillator $0s$ wave function with $\hwcm=\hw$
      (thick gray lines)~--- or, equivalently, the calculation in an $\Nmax=0$
      natural-orbital basis~--- shown for comparison.  The underlying oscillator
      basis $\hw$ is indicated (dotted vertical line), as are the minimal
      $\hwcmtilde$ and $\Ncmtilde$ for each curve (dots, with dotted vertical
      line at highest $\Nmax$).
    }
    \label{fig:3he-Ncm-hwcm-functional-dependence}
\end{figure}
\begin{figure}[tp]
    \begin{center}
        \includegraphics[width=\ifproofpre{0.9}{0.5}\hsize]{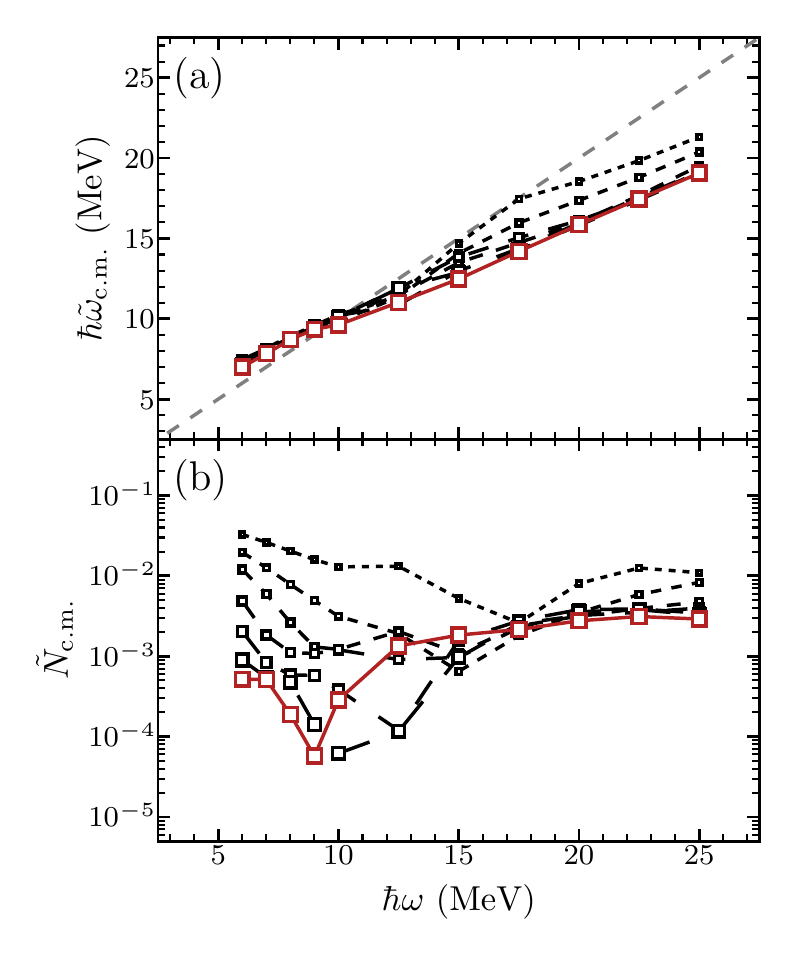}\\
    \end{center}
    \caption{Dependence of the approximate $0s$ center-of-mass motion of the
      calculated $\isotope[3]{He}$ ground state (and its degree of
      contamination) on the $\hw$ of the underlying oscillator basis, in
      calculations with a natural-orbital basis, as measured by (a)~$\hwcmtilde$
      and (b)~$\Ncmtilde$.  Results are shown for calculations with $\Nmax=4$
      (dotted lines) through $\Nmax=16$ (solid lines, highlighted).
      }
\label{fig:3he-minNcm-hw-scan}
\end{figure}

A factorized and well-controlled center-of-mass motion is important, as
discussed in Sec.~\ref{sec:methods:ncci}, if the results of the many-body
calculation are to be of practical use, beyond limited calculations for
ground-state observables.  Recall that the $\Nmax$-truncated oscillator basis is
special, in that the many-body wave functions resulting from NCCI calculations
with this basis factorize into intrinsic and center-of-mass parts, and the
center-of-mass part can be selected to have pure oscillator $0s$ zero-point
motion in the center-of-mass coordinate.  Such exact factorization is no longer
guaranteed, and no longer to be expected, if we move away from the
$\Nmax$-truncated oscillator basis.  However, approximate factorization may
arise, with or without the persuasion of a Lawson term in the Hamiltonian.  Let
us therefore diagnose the center-of-mass motion which arises in our present
calculations with the natural-orbital basis, and how it depends upon the choice
of underlying oscillator basis.

For many purposes, we might be satisfied by factorization involving an arbitrary
center-of-mass wave function.  For instance, angular momentum selection rules
which allow the intrinsic electromagnetic operators to be replaced with one-body
operators in practical calculations~\cite{caprio2020:intrinsic} require
factorization with an $s$-wave center-of-mass wave function, as
\begin{math}
  \tket{\Psi_J}=[\tket{\Psi^{\text{intr}}_J}\times\tket{\Psi^{\text{c.m.}}_{L_{\text{c.m.}}=0}}]_J,
\end{math}
but not specifically an oscillator $0s$ wave function.  However, in
practice, we do not have a good way to measure how well a many-body wave
function factorizes, unless the factorization specifically involves $0s$
harmonic-oscillator motion.

Specifically, the expectation value of the center-of-mass number operator $\Ncm$
allows us to measure deviations from pure $0s$ center-of-mass
motion~\cite{hagen2009:coupled-cluster-com,*hagen2010:coupled-cluster,roth2009:it-ncsm-cc-cm-truncated,caprio2012:csbasis,hergert2016:imsrg}.
Such $0s$ center-of-mass motion then incidentally implies factorization as
\begin{math}
  \tket{\Psi_J}=[\tket{\Psi^{\text{intr}}_J}\times\tket{\Psi^{\text{c.m.}}_{0s}}]_J.
\end{math}
The definition of a center-of-mass harmonic-oscillator number operator depends
upon the oscillator parameter $\hwcm$ taken for the center-of-mass motion:
\begin{equation}
  \label{eqn:Ncm-hwcm}
  \begin{aligned}
    \Ncm(\hwcm)
    &= \cveccm^\dagger \cdot \cveccm
    \\
    &=
    \frac12
    (\hwcm)^{-1}
    \frac{(\hbar c)^2}{A (m_N c^2)} K^2
    \ifproofpre{\\&\phantom{=}\quad}{}
    +
    \frac12
    (\hwcm)
    \frac{A (m_N c^2)}{(\hbar c)^2} R^2
    -
    \frac{3}{2},
    \end{aligned}
\end{equation}
where $\cveccm^\dagger$ and $\cveccm$ are the center-of-mass ladder operators
(see Sec.~F.3 of Ref.~\cite{caprio2020:intrinsic} for definitions),
$K^2=\abs{\kveccm}^2$ is the squared magnitude of the center-of-mass momentum
vector or, more precisely, wave vector, where $\pveccm=\hbar \kveccm$, and
$R^2=\abs{\xveccm}^2$ is the squared magnitude of the center-of-mass coordinate
vector~\cite{caprio2020:intrinsic}.  Taking the expectation value of the
expression in~(\ref{eqn:Ncm-hwcm}), we see that $\tbracket{\Ncm(\hwcm)}$ depends
on the many-body wave function only through the two expectation values
$\tbracket{K^2}$ and $\tbracket{R^2}$, which must then be taken in linear
combination, weighted by the appropriate numerical coefficients
from~(\ref{eqn:Ncm-hwcm}).  These expectation values are readily evaluated
within standard NCCI many-body codes, since $R^2$ and $K^2$ are simply scalar
two-body operators, like the Hamiltonian itself.

Then $\tbracket{\Ncm(\hwcm)}$ vanishes if and only if the wave function has pure
factorized harmonic-oscillator $0s$ center-of-mass motion, corresponding to the
given oscillator length.  A nonvanishing $\tbracket{\Ncm}$ measures, or at least
places a limit upon, the deviation from such pure factorized $0s$
motion.\footnote{In general, the many-body state $\tket{\Psi}$ may be decomposed
  into components with different eigenvalues of $\Ncm$:
\begin{math}
  \tket{\Psi}
  = \alpha_0 \tket{\Psi_{\Ncm=0}}
  + \alpha_1 \tket{\Psi_{\Ncm=1}}
  + \alpha_2 \tket{\Psi_{\Ncm=2}}
  +\cdots.
\end{math}
Then $\tbracket{\Ncm}=\sum_\nu \alpha_\nu^2 \nu $, which vanishes if and only if
$\tket{\Psi}=\tket{\Psi_{\Ncm=0}}$.  This is simply the variational principle
for the nonnegative-definite operator $\Ncm$.}  In particular, the total
contribution to the norm from components with nonzero excitation of the
center-of-mass degree of freedom is $P(\Ncm>0)\leq\tbracket{\Ncm}$.

However, as emphasized in
Ref.~\cite{hagen2009:coupled-cluster-com,*hagen2010:coupled-cluster}, simply
evaluating $\tbracket{\Ncm(\hw)}$, with $\hwcm$ taken as the $\hw$ of the
underlying oscillator basis, will, in general, overestimate the center-of-mass
contamination.  Even if it so happens that the wave function obtained in an NCCI
calculation, in some natural-orbital basis, factorizes (or approximately
factorizes), with $0s$ oscillator motion for the center of mass, there
is no reason to expect that the oscillator parameter for this center-of-mass
motion will match that of the oscillator basis used in the original NCCI
calculation which yielded the reference state from which the natural orbitals
were derived.  Rather, we must search for the value of $\hwcm$
in~(\ref{eqn:Ncm-hwcm}) which minimizes $\tbracket{\Ncm(\hwcm)}$.  This value,
denoted by $\hwcmtilde$ (or simply $\hbar\tilde{\omega}$ in Ref.~\cite{hagen2009:coupled-cluster-com,*hagen2010:coupled-cluster}), is readily extracted from~(\ref{eqn:Ncm-hwcm}) in
analytic form, as
\begin{equation}
  \label{eqn:hwcmtilde}
  \hwcmtilde=\frac{(\hbar c)^2}{A(m_Nc^2)}
  \biggl(
    \frac{\tbracket{K^2}}{\tbracket{R^2}}
    \biggr)^{1/2},
\end{equation}
and the corresponding minimized measure of the center-of-mass contamination,
$\Ncmtilde\equiv\tbracket{\Ncm(\hwcmtilde)}$, is given by
\begin{equation}
  \label{eqn:Ncmtilde}
  \Ncmtilde = \bigl( \tbracket{K^2} \tbracket{R^2} \bigr)^{1/2} - \frac{3}{2}.
\end{equation}

With this in mind, let us now examine the center-of-mass motion for the
$\isotope[3]{He}$ ground state wave functions obtained in a natural-orbital
basis.  The values of $\tbracket{\Ncm(\hwcm)}$, as we sweep $\hwcm$
in~(\ref{eqn:Ncm-hwcm}), are shown in
Fig.~\ref{fig:3he-Ncm-hwcm-functional-dependence}.  Each curve is simply
determined analytically, by~(\ref{eqn:Ncm-hwcm}), taking the calculated
$\tbracket{K^2}$ and $\tbracket{R^2}$ for the corresponding wave function.  We
again (as in Fig.~\ref{fig:3he-orbitals-Nmax-evolution-Pr}) take $\hw=9\,\MeV$
[Fig.~\ref{fig:3he-Ncm-hwcm-functional-dependence}(a)], $15\,\MeV$
[Fig.~\ref{fig:3he-Ncm-hwcm-functional-dependence}(b)], and $25\,\MeV$
[Fig.~\ref{fig:3he-Ncm-hwcm-functional-dependence}(c)] as representative values
for the oscillator parameter of the underlying oscillator basis (namely, below,
near, and above the variational energy minimum, respectively).

For $\Nmax=0$ (thick gray line), in
Fig.~\ref{fig:3he-Ncm-hwcm-functional-dependence}, recall that the natural
orbitals are simply the original oscillator functions, with oscillator parameter
$\hw$, and calculations in the natural-orbital basis are simply calculations in
the oscillator basis.  The center-of-mass motion is thus pure $0s$
motion, with $\hwcmtilde=\hw$ (vertical dotted line), for which $\Ncmtilde=0$.  (In
fact, curves identical to that shown would be obtained for any of the
$\Nmax$-truncated oscillator-basis calculations with this same $\hw$.)

Then, for the calculations in a natural-orbital basis proper, with $\Nmax=4$
(dotted line) through $16$ (solid line), in
Fig.~\ref{fig:3he-Ncm-hwcm-functional-dependence}, there is no $\hwcm$ for which
$\tbracket{\Ncm}$ vanishes.  Rather, the location of the minimum in
$\tbracket{\Ncm}$, given by~(\ref{eqn:hwcmtilde}) and~(\ref{eqn:Ncmtilde}), is
marked by a dot.

For $\hw=9\,\MeV$ [Figs.~\ref{fig:3he-Ncm-hwcm-functional-dependence}(a)], there
is an initial discontinuity going from the oscillator basis to a natural-orbital
basis, where $\Ncmtilde$ jumps to $\gtrsim10^{-2}$ for $\Nmax=4$, then steadily
decreases again, converging to a value $\sim10^{-3}$.  The optimal $\hwcm$ for
recognizing this approximate factorization is $\hwcmtilde\approx9.4\,\MeV$,
slightly above the $\hw$ of the underlying oscillator basis ($\hw=9\,\MeV$).

Moving to the other side of the variational minimum in $\hw$, for $\hw=25\,\MeV$
[Figs.~\ref{fig:3he-Ncm-hwcm-functional-dependence}(c)], there is again an
initial discontinuity, with $\Ncmtilde\sim10^{-2}$ for $\Nmax=4$, and
converging towards $\gtrsim10^{-3}$.  Here
the optimal $\hwcm$ for recognizing
this approximate factorization is $\hwcmtilde\approx 19\,\MeV$, notably below
the $\hw$ of the underlying oscillator basis ($\hw=25\,\MeV$).

Finally, for $\hw=15\,\MeV$
[Figs.~\ref{fig:3he-Ncm-hwcm-functional-dependence}(b)], near the variational
energy mimimum, after $\Ncmtilde$ initially jumps to $\Ncmtilde\lesssim10^{-2}$
for $\Nmax=4$, it then immediately drops to $\Ncmtilde\approx10^{-3}$ for higher
$\Nmax$.  The location of the minimum drifts slightly downward, from the $\hw$
of the underlying oscillator basis ($\hw=15\,\MeV$), toward $\hwcmtilde\approx
12.5\,\MeV$.

Thus, in each case, regardless of the $\hw$ for the underlying oscillator basis,
a reasonably pure $0s$ center-of-mass motion spontaneously emerges for the
$\isotope[3]{He}$ ground state, as recognized when the appropriate choice
$\hwcmtilde$ of oscillator parameter is used in measuring the center-of-mass
motion, implying also a high degree of center-of-mass factorization.
Furthermore, in each case, this $\hwcmtilde$ for which $0s$ motion is most closely
realized differs from the $\hw$ of the underlying oscillator basis.

To more systematically map out the behaviors we have just seen, $\hwcmtilde$ and
$\Ncmtilde$ are shown as functions of the underlying oscillator basis $\hw$ in
Fig.~\ref{fig:3he-minNcm-hw-scan}.  For large $\Nmax$, the dependence of
$\hwcmtilde$ on $\hw$ [Fig.~\ref{fig:3he-minNcm-hw-scan}(a)] is nearly linear, but
of shallower slope than the reference line $\hwcmtilde=\hw$ (dashed diagonal
line).  The oscillator parameter $\hwcmtilde$ for the center-of-mass motion
matches that of the underlying oscillator basis for the natural orbitals in the
vicinity of $\hw=10\,\MeV$ to $12.5\,\MeV$.  In this range of $\hw$, at high
$\Nmax$, one also observes that the purest $0s$ center-of-mass motion is obtained
[Fig.~\ref{fig:3he-minNcm-hw-scan}(b)], with $\Ncmtilde\lesssim10^{-4}$.

A rough intuitive understanding of the center-of-mass motion, in particular, the
behavior of the preferred $\hwcmtilde$ observed in
Fig.~\ref{fig:3he-minNcm-hw-scan}(a), follows from the $\hw$-dependence noted
above for the natural orbitals themselves (Sec.~\ref{sec:results-3he:orbitals}).
Recall the tendency, observed in Fig.~\ref{fig:3he-orbitals-Nmax-evolution-Pr},
for natural orbitals obtained from a low-$\hw$ underlying oscillator basis
[Fig.~\ref{fig:3he-orbitals-Nmax-evolution-Pr}(a)] to still resemble oscillator
orbitals, but of a somewhat higher $\hw$, closer to $\hw\approx15\,\MeV$, and
for natural orbitals obtained from a high-$\hw$ underlying oscillator basis
[Fig.~\ref{fig:3he-orbitals-Nmax-evolution-Pr}(c)] to resemble oscillator
orbitals of a somewhat lower $\hw$, again closer to $\hw\approx15\,\MeV$.  To
the extent that the low-lying natural orbitals resemble oscillator
orbitals of some $\hw$, then a (nominally $\Nmax$-truncated) calculation in such
a natural-orbital basis may be expected to have similar properties to an
($\Nmax$-truncated) calculation in an oscillator basis of this same $\hw$.  It
is thus perhaps not surprising that $\hwcmtilde$ of the center-of-mass wave
function follows the same overall trend as the ``effective'' $\hw$ of the
natural orbitals.

However, we must always keep in mind that $\tbracket{\Ncm}$ is, strictly, only a
measure of center-of-mass contamination, relative to harmonic-oscillator $0s$ motion, and therefore
only incidentally provides an upper bound on the breakdown of center-of-mass
factorization.  Nonzero $\tbracket{\Ncm}$ could reflect that factorization is
broken, but it could also simply mean that we have factorization which is of a
more difficult form to recognize, since the center-of-mass motion is not simply
described by a $0s$ oscillator wave function.

Furthermore, for the present many-body calculations in the natural-orbital
basis, recall that we have included no Lawson center-of-mass term
(Sec.~\ref{sec:methods}) in the Hamiltonian.  For now, we are thus identifying
the center-of-mass motion which emerges spontaneously when we diagonalize a
translationally-invariant intrinsic Hamiltonian, restricted to the particular
truncated many-body space of these calculations.  Starting from this baseline,
one may then explore the effect of including a Lawson term, which is expected to
refine the center-of-mass motion, at some cost to the convergence of the
intrinsic motion (see Ref.~\cite{constantinou2017:diss} for initial examples of
such calculations).  Here one might more naturally choose an $\hwcm$ parameter
for the Lawson term which reinforces the center-of-mass motion as it already
spontaneously emerges in the natural-orbital basis ($\hwcm=\hwcmtilde$) rather
than simply matching the oscillator parameter the underlying oscillator basis
($\hwcm=\hw$).
 
\section{Natural orbitals and halo structure: \boldmath$\isotope[6]{He}$}
\label{sec:results-6he}

\subsection{Convergence of observables}
\label{sec:results-6he:obs}

\begin{figure}[tp]
    \begin{center}
        \includegraphics[width=\ifproofpre{0.9}{0.5}\hsize]{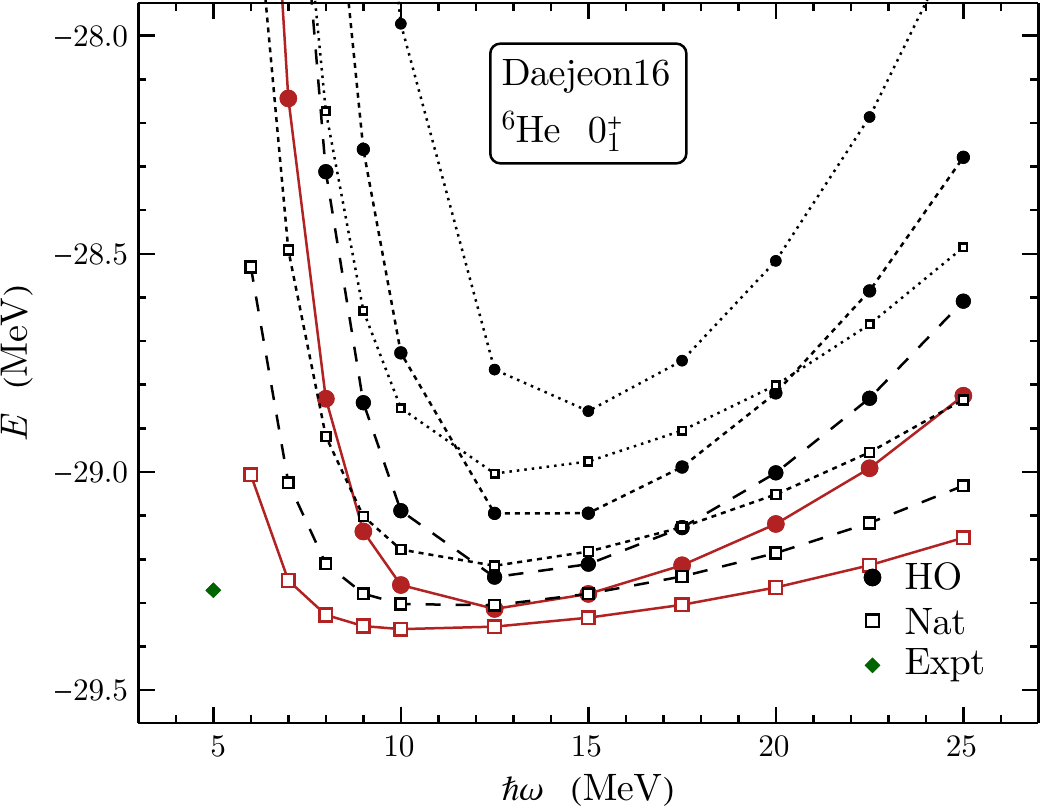}\\
    \end{center}
    \caption{The $\isotope[6]{He}$ ground-state energy, as calculated in
      oscillator (solid circles) and natural orbital (open squares)
      bases.  Calculated values are shown as functions of the basis parameter
      $\hw$, for successive even value of $\Nmax$, from $\Nmax=8$ (dotted lines)
      to $14$ (solid lines, highlighted).
      The experimental binding energy~\cite{wang2021:ame2020} is also shown (filled diamond).
    }
    \label{fig:6he-convergence-energy}
\end{figure}

\begin{figure}[tp]
    \begin{center}
      \includegraphics[width=\ifproofpre{0.9}{0.5}\hsize]{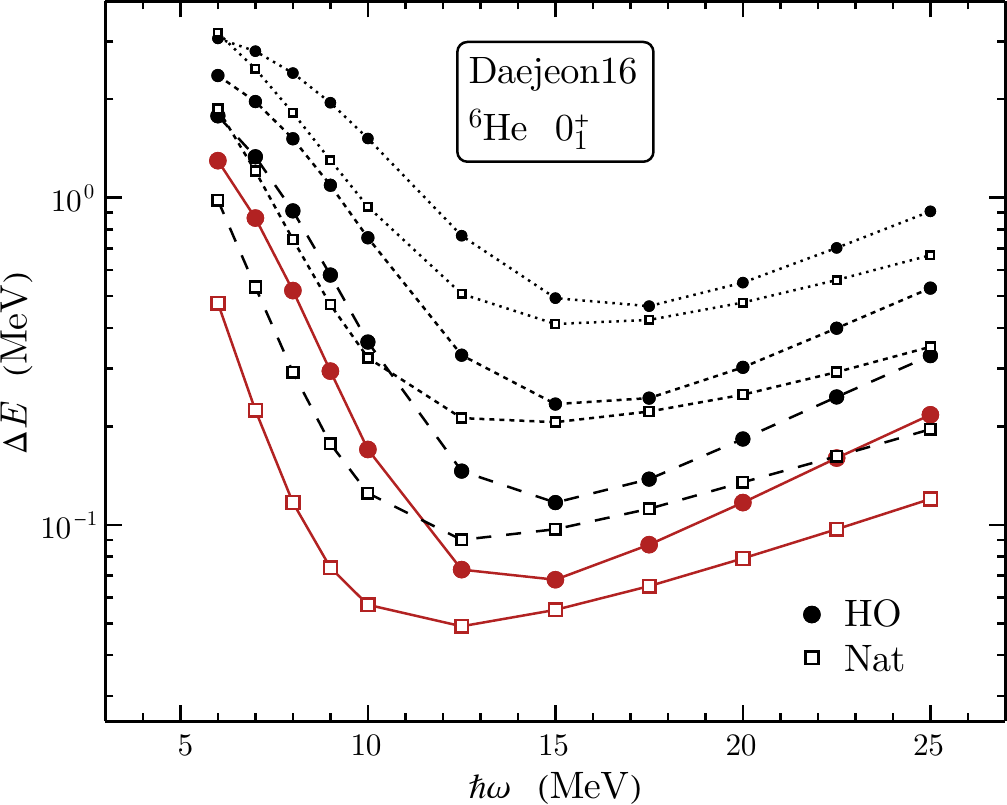}
    \end{center}
    \caption{Differences of calculated $\isotope[6]{He}$ ground-state energies
      obtained for successive $\Nmax$, as obtained for oscillator (solid
      circles) and natural orbital (open squares) bases, shown on a logarithmic
      scale.  }
    \label{fig:6he-convergence-energy-log-diff}
\end{figure}

\begin{figure}[tp]
    \begin{center}
        \includegraphics[width=\ifproofpre{0.9}{0.5}\hsize]{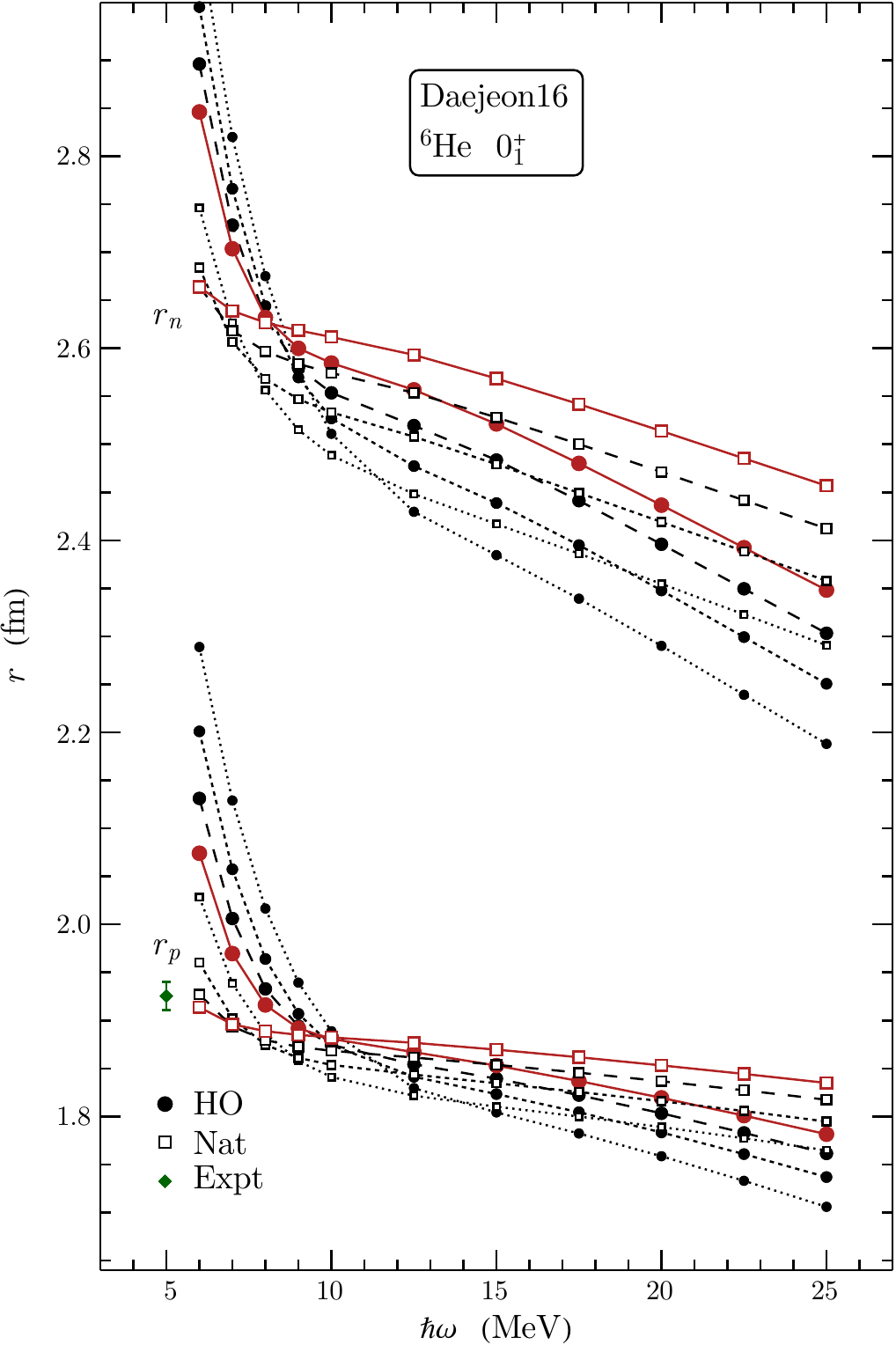}\\
    \end{center}
    \caption{The $\isotope[6]{He}$ ground-state point-proton and point-neutron r.m.s.\ radii, as
      calculated in oscillator (solid circles) and natural orbital (open
      squares) bases.  Calculated values are shown as functions of the basis
      parameter $\hw$, for successive even value of $\Nmax$, from $\Nmax=8$
      (dotted lines) to $14$ (solid lines, highlighted).
      The value deduced from the experimental charge radius~\cite{angeli2013:charge-radii} is also shown (filled diamond).
        }
    \label{fig:6he-convergence-rp-rn}
\end{figure}

For a halo nucleus, such as $\isotope[6]{He}$, the connection between natural
orbitals and the single-particle structure of the nucleus should be particularly
revealing. The natural orbitals occupied by halo nucleons may be expected to
reflect the large-distance behavior which generates the halo.  The ground state
of $\isotope[6]{He}$ is understood to be clusterized, consisting of a
$\isotope[4]{He}$ (or $\alpha$) core plus two weakly-bound neutrons.  This leads
to a spatially-extended neutron
distribution~\cite{jonson2004:light-dripline,tanihata2013:halo-expt}, with
possible correlations in the motion of the halo
neutrons~\cite{quaglioni2013:ncsm-rgm-cluster-6he,saaf2014:core-n-n-6he,romeroredondo2016:6he-correlations,robin2021:6he-entanglement}.
The weak binding is reflected in a small two-neutron separation energy
($\approx0.97\,\MeV$), while the extended spatial structure is reflected in a
marked increase in r.m.s.\ radius observables from $\isotope[4]{He}$ to
$\isotope[6]{He}$.  Having already explored the basic properties of NCCI
calculations in a natural-orbital basis for $\isotope[3]{He}$
(Sec.~\ref{sec:results-3he}), we will take these as a baseline for comparison
for $\isotope[6]{He}$. Let us first consider the calculated energy and radius
observables for $\isotope[6]{He}$, with a natural-orbital basis, then (in
the subsequent Sec.~\ref{sec:results-3he:orbitals} below) the radial wave functions of the orbitals
themselves.

The $\isotope[6]{He}$ ground state energy is shown in
Fig.~\ref{fig:6he-convergence-energy}, as calculated with oscillator
(solid circles) and natural-orbital (open squares) bases.  Here we consider
truncations through $\Nmax=14$, again with the Daejeon16 interaction.
The experimental binding energy~\cite{wang2021:ame2020} is shown for comparison (filled diamond).

The energy obtained with natural orbitals, in
Fig.~\ref{fig:6he-convergence-energy}, is consistently lower than that
obtained in the underlying oscillator basis, and is thus, by the variational
principle, closer to the true energy in the full many-body space.  In the vicinity of the variational minimum, the energy obtained with natural orbitals is
approximately ``one step'' in $\Nmax$ ahead of that obtained with oscillator
orbitals.  This relation strictly holds for the highest $\Nmax$ shown in
Fig.~\ref{fig:6he-convergence-energy} (\textit{i.e.}, the
energy obtained in the natural-orbital basis for $\Nmax=12$ lies below that
obtained in the oscillator basis for $\Nmax=14$).  The $\hw$
dependence of the calculated energy is, again, much reduced in the
natural-orbital basis, so the improvement of the natural-orbital results over
the oscillator-basis results becomes more marked as we move away from the
variational energy minimum and towards the extreme values of $\hw$ shown in
Fig.~\ref{fig:6he-convergence-energy}.

Whereas for $\isotope[3]{He}$ we could benchmark the calculated energies against
an effectively converged value obtained at much higher $\Nmax$, as in
Fig.~\ref{fig:3he-convergence-energy}(c), we no longer have this luxury for
$\isotope[6]{He}$, where the growth in dimension with $\Nmax$ is much more rapid
(Fig.~\ref{fig:dimension}).  We must simply compare the calculations obtained
with oscillator and natural-orbital bases, and for different $\Nmax$, against
each other.

The overall scale of the change in calculated energy with $\Nmax$ for
$\isotope[6]{He}$ is much larger than for $\isotope[3]{He}$.  In the vicinity of
the variational energy minimum, the change in calculated energy with each step
in $\Nmax$ is $\lesssim0.1\,\MeV$ (Fig.~\ref{fig:6he-convergence-energy}),
compared to steps of $\approx0.001\,\MeV$ for comparable $\Nmax$ in
$\isotope[3]{He}$ (Fig.~\ref{fig:3he-convergence-energy}).  This difference
might be taken to reflect the greater complication in reproducing a higher-$A$
system in general, as well as the challenging halo structure of
$\isotope[6]{He}$ in particular.

However, in judging convergence, what is important is not only the size of the
change between values calculated with successive $\Nmax$, but how this change
decreases with $\Nmax$.  A convenient baseline against which to compare the
convergence of the ground state energy is the hypothesis of
exponential convergence with respect to $\Nmax$,
\begin{equation}
  \label{eqn:exp-convergence}
  E(\Nmax)=E_{\infty}+a\exp(-c\Nmax),
\end{equation}
where $E_{\infty}$ is then the full-space
value~\cite{bogner2008:ncsm-converg-2N}.  The calculated values approach the
full-space value in a geometric progression with successive steps in $\Nmax$.
For exponential convergence, the residual $\delta E(\Nmax)\equiv
E(\Nmax)-E_{\infty}$ of the calculated energy relative to the full-space value,
considered above for $\isotope[3]{He}$ (Sec.~\ref{sec:results-3he:obs}), is
given by $\delta E(\Nmax)=a\exp(-c\Nmax)$, and thus decreases by a constant factor
$e^{-2c}$ with each (even) step in $\Nmax$. On a logarithmic plot of the
residual, as we considered for $\isotope[3]{He}$ in
Fig.~\ref{fig:3he-convergence-energy}(c), this appears as equally spaced steps
with respect to $\Nmax$, as was indeed approximately noted for $\isotope[3]{He}$
(Sec.~\ref{sec:results-3he:obs}).

For $\isotope[6]{He}$, we have no converged value with respect to which to take
residuals, and thus cannot generate a logarithmic plot of residuals as in
Fig.~\ref{fig:3he-convergence-energy}(c).  Nonetheless, we can still compare
successive calculated values of the energy, for successive truncations $\Nmax$,
and consider their difference $\Delta E(\Nmax)=E(\Nmax)-E(\Nmax-2)$.  For
exponential convergence, the ratio of successive steps
\begin{equation}
  \label{eqn:step-ratio}
  \eta(\Nmax)\equiv \frac{E(\Nmax)-E(\Nmax-2)}{E(\Nmax-2)-E(\Nmax-4)},
\end{equation}
is simply a constant $\eta=e^{-2c}$, independent of $\Nmax$.  That is,
$\eta=0.5$ corresponds to a step size in $E$ which is halved with each
successive step in $\Nmax$, and a smaller value of $\eta$ corresponds to a more
rapid exponential decay towards the full-space value.  Such differences which
decrease by a constant ratio again appear, on a logarithmic plot, to move
downward by equal increments with each step in $\Nmax$.

We thus consider a plot of $\log\abs{\Delta E}$ for the $\isotope[6]{He}$ ground
state energy, in Fig.~\ref{fig:6he-convergence-energy-log-diff}.  The overall
convergence behavior is qualitatively similar for calculations in oscillator
(solid circles) and natural orbital (open squares) bases.  The spacing between
curves for successive $\Nmax$ is roughly uniform with $\Nmax$, but decreases
gradually for higher $\Nmax$, \textit{i.e.}, the convergence ``slows down''
relative to exponential convergence.  For the energies calculated in either
basis, the step size $\Delta E$ decreases by a factor of $\sim0.4\mathrm{-}0.6$
with each step in $\Nmax$.  At low $\Nmax$, $\Delta E$ in either basis is
roughly comparable.  However, for high $\Nmax$, the curves representing $\Delta
E$ for the calculations in the natural-orbital basis lie approximately one step
in $\Nmax$ ahead of those for the oscillator basis.  Near the variational
minimum in energy ($\hw\approx 15\,\MeV$), this is consistent with the
observation from above, that the natural-orbital basis improves on the best
oscillator-basis energy by about one step in $\Nmax$.  But this observation
holds uniformly over a wide range extending to higher $\hw$, as well (at lower
$\hw$, the $\Delta E$ obtained with the natural orbital basis falls off much
more sharply with $\Nmax$).

We now consider the r.m.s.\ radii, which provide measures of the halo structure.
The calculated values of both $r_p$ and
$r_n$, for the $\isotope[6]{He}$ ground state, are shown in
Fig.~\ref{fig:6he-convergence-rp-rn}.
Note that the point-proton r.m.s.\ radius $r_p$, the point-neutron
r.m.s.\ radius $r_n$, and the matter (or total point-nucleon) radius $r_m$ form
a redundant set of observables, related by $Ar_m^2=Zr_p^2+Nr_n^2$.  It is thus worth briefly reviewing the physical
significance of these observables, in the context of
$\isotope[6]{He}$~\cite{lu2013:laser-neutron-rich,caprio2014:cshalo}.

Although $r_p$ does not
\textit{directly} measure neutron halo structure, it is nonetheless
\textit{indirectly} sensitive to this structure, and it is accessible to electromagnetic measurement, through its simple relation
to the charge radius.  It is important to keep in mind that $r_p$, as calculated
here and as accessed in experiment, is defined relative to the common center of
mass of the protons and neutrons (see, \textit{e.g.},
Refs.~\cite{bacca2012:6he-hyperspherical,tanihata2013:halo-expt,caprio2014:cshalo,caprio2020:intrinsic}).
In the cluster halo description of $\isotope[6]{He}$, the $\alpha$ recoils against
the halo neutrons, which consequently displaces the center of mass of the
$\alpha$ (and thus of the protons) relative to this common center of mass.  This
induces an increase in $r_p$ going from $\isotope[4]{He}$ to $\isotope[6]{He}$.
(There may also be contributions from modifications to the structure of the
$\alpha$ particle itself, or ``core
polarization''~\cite{lu2013:laser-neutron-rich}.)  Experimentally, the increase
in $r_p$ from $1.462(6)\,\fm$ for $\isotope[4]{He}$ to $1.934(9)\,\fm$ for
$\isotope[6]{He}$~\cite{wang2004:6he-radius-laser,brodeur2012:6he-8he-mass,lu2013:laser-neutron-rich}
is taken as a principal indicator of halo structure in $\isotope[6]{He}$.

Then, both $r_n$ and $r_m$ include direct contributions from the halo neutrons.
While $r_n$ is more selectively a measure of the neutron distribution, it is
$r_m$ which is extracted from nuclear reaction cross section or proton-nucleus
elastic scattering measurements.  The results thereby obtained for the
$\isotope{He}$ isotopes are model-dependent and contradictory (see Sec.~III\,A
of Ref.~\cite{caprio2014:cshalo} for an overview).  They variously suggest
$r_m\approx2.3\,\fm$--$2.7\,\fm$ in $\isotope[6]{He}$, corresponding to an
increase relative to $\isotope[4]{He}$ of $\approx50\text{--}90\%$.  Subject to
these uncertainties, the increased matter radius in $\isotope[6]{He}$ is again
taken as an indicator of halo structure.

For the calculated $r_p$ (lower curves in Fig.~\ref{fig:6he-convergence-rp-rn}), the results
obtained in the natural-orbital basis yield reduced $\hw$ dependence relative to
those obtained in the oscillator basis, much as already seen for
$\isotope[3]{He}$ (Fig.~\ref{fig:3he-convergence-hw-radius}). At the extremes in
$\hw$ shown in Fig.~\ref{fig:6he-convergence-rp-rn}, the calculations in the
natural-orbital basis thus lie several steps in $\Nmax$ ``ahead'' of the
oscillator-basis calculations.
Again, we show the value of $r_p$ deduced from the experimental $r_c$~\cite{angeli2013:charge-radii}
for comparison (filled diamond).

For $\isotope[6]{He}$, the curves of radius \textit{vs.}\ $\hw$ exhibit
recognizable crossing points regardless of which basis is used.  Recall
(Sec.~\ref{sec:results-3he:obs}) that these crossing points have been suggested
as a heuristic estimator of the full-space value.  These crossing points are
displaced in $\hw$ relative to each other~--- from $\hw\approx10\,\MeV$ for the
oscillator basis down to $\hw\approx7\,\MeV$ for the natural-orbital basis~---
but occur at comparable values for the observable ($r_p\approx1.9\,\fm$),
consistent with the experimental value noted above.

Naturally, given the halo structure, the calculated values for $r_n$ (upper
curves in Fig.~\ref{fig:6he-convergence-rp-rn}) are larger than for $r_p$, the
$\hw$ dependence is stronger, and the changes in calculated value with each step
in $\Nmax$ is larger.  Again, crossing points are obtained for calculations in
both the oscillator and natural-orbital bases, shifted towards lower $\hw$
(longer oscillator length) than for $r_p$, namely $\hw\approx9\,\MeV$ for the
oscillator basis, and approaching $\hw\approx6\,\MeV$ for the natural-orbital
basis.  This shift is perhaps not surprising given the larger radial extent of
the structure being described.  These two crossing points again occur at
comparable values for $r_n$, in the range $r_n\approx2.6\,\fm$--$2.7\,\fm$.  (In
conjunction with the above value for $r_p$, this suggests
$r_m\approx2.4\,\fm$--$2.5\,\fm$.)  The highest $\Nmax$ curves for the natural
orbital calculations develop a flat ``shoulder'', varying by $\lesssim0.05\,\fm$
over several steps in $\hw$.
This range of calculated $r_n$ values is compatible with the range $r_n \approx 2.5\,\fm - 3.0\,\fm$
suggested by the range of experimental matter radii (discussed above) in conjuction with $r_p$.

The transformation to a natural-orbital basis clearly does not definitively
solve the problem of convergence for the r.m.s.\ radius observables.
Nonetheless, it does contribute to taming the convergence behavior for these
observables.

\subsection{Natural orbitals}
\label{sec:results-6he:orbitals}

\begin{figure*}[tp]
    \begin{center}
        \includegraphics[width=\ifproofpre{0.9}{0.9}\hsize]{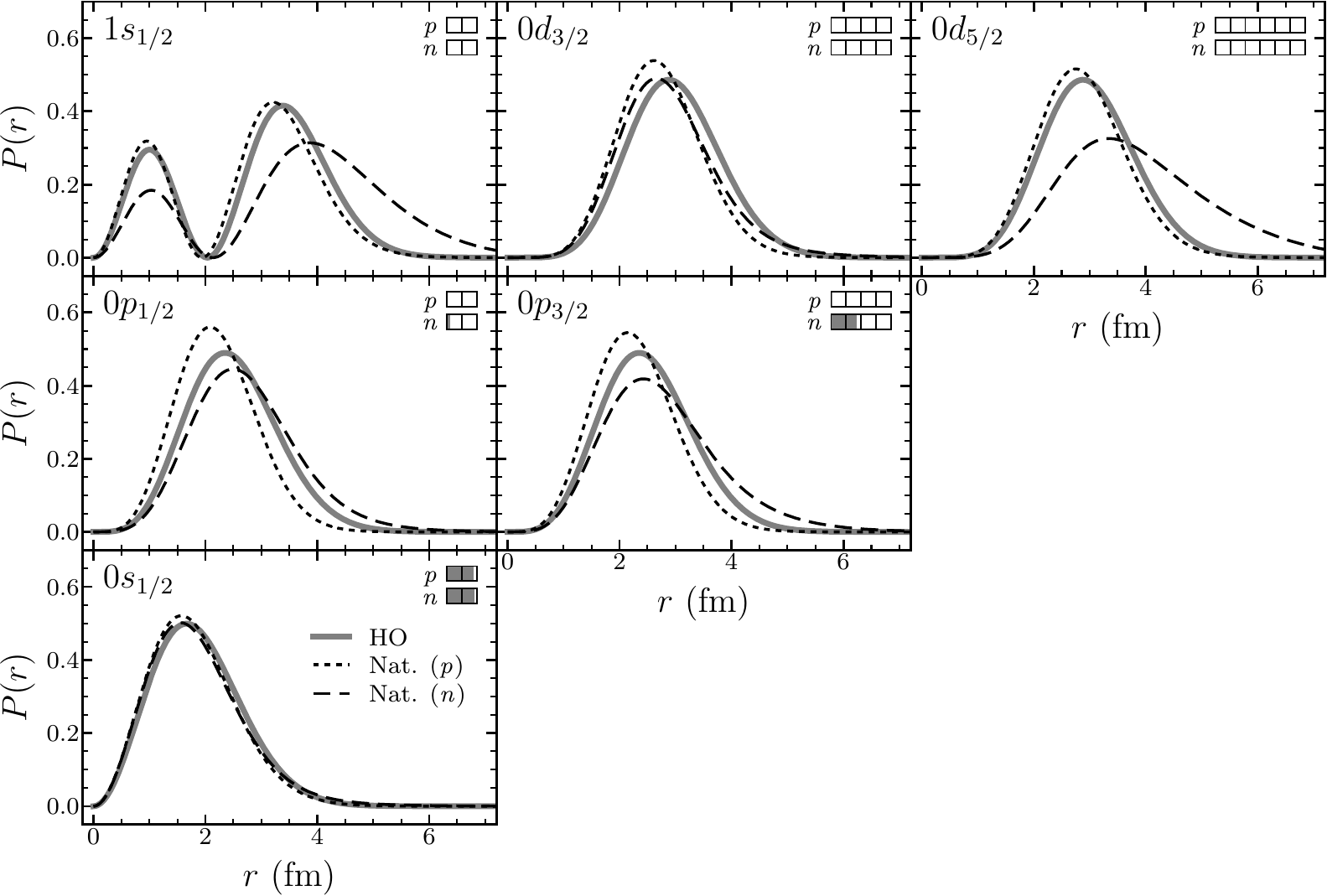}\\
    \end{center}
    \caption{Radial wave functions for the $\isotope[6]{He}$ $s$-, $p$-, and
      $sd$-shell natural orbitals, for both protons (short dashed lines) and
      neutrons (long dashed lines), plotted as the radial probability density
      $P(r)$.  These are obtained from the underlying oscillator-basis
      calculation near the variational minimum ($\hw=15\,\MeV$) and at high
      $\Nmax$ ($\Nmax=14$).  The corresponding oscillator radial functions for
      $\hw=15\,\MeV$ (thick gray lines) are shown for comparison.  The mean
      occupation $n_a$ for each natural orbital, from the corresponding
      eigenvalue of the scalar density matrix, is indicated by the filling of
      the bar at top right (upper bar for protons, lower bar for neutrons).
    }
    \label{fig:6he-orbitals-Pr}
\end{figure*}
\begin{figure*}[tp]
    \begin{center}
        \includegraphics[width=\ifproofpre{0.9}{0.9}\hsize]{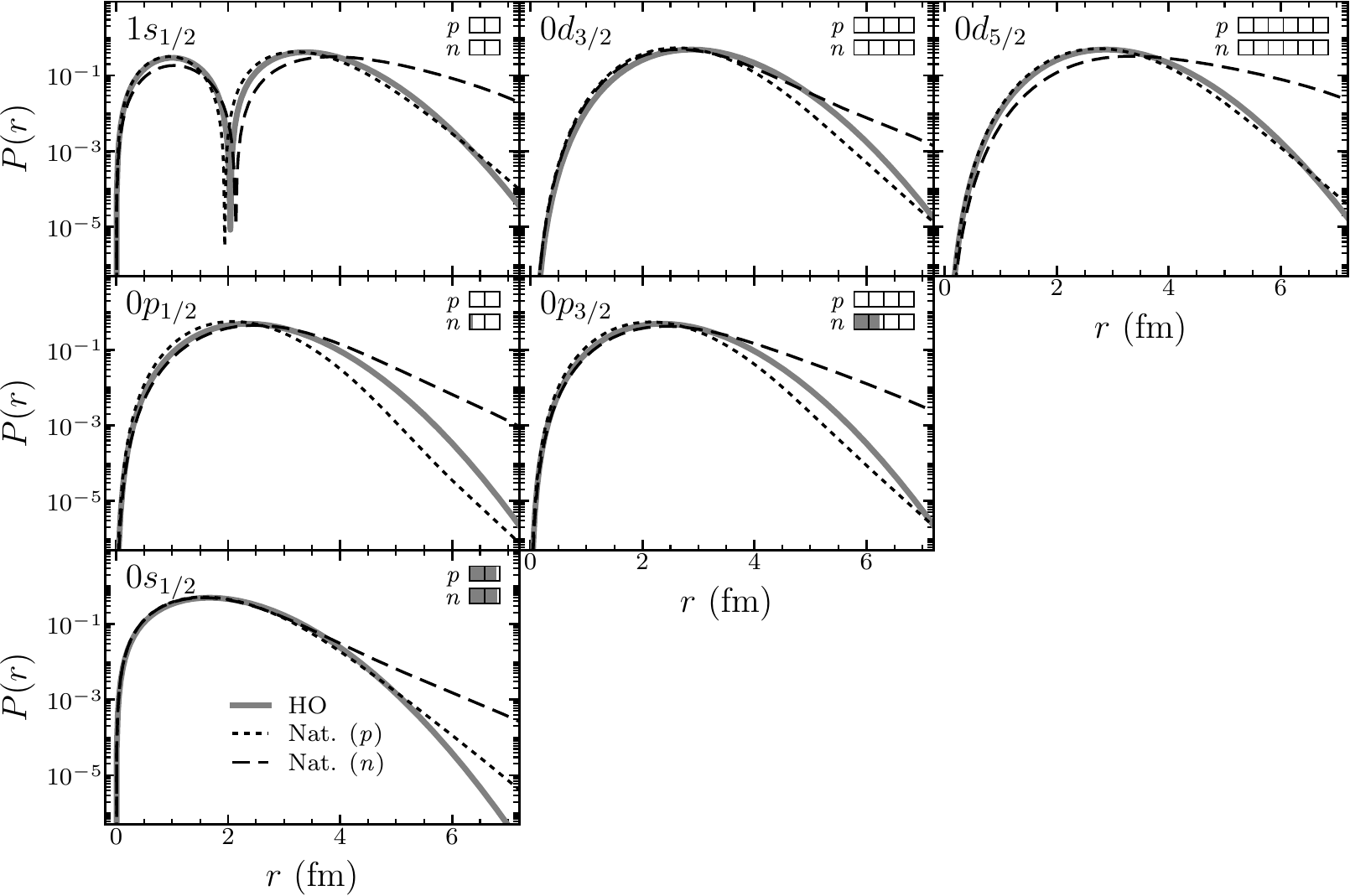}\\
    \end{center}
    \caption{Radial wave functions for the $\isotope[6]{He}$ $s$-, $p$-, and
      $sd$-shell natural orbitals, plotted as the radial probability density
      $P(r)$, as in Fig.~\ref{fig:6he-orbitals-Pr}, but now on a logarithmic
      scale.
    }
    \label{fig:6he-orbitals-logPr}
\end{figure*}

\begin{figure}[tp]
    \begin{center}
        \includegraphics[width=\ifproofpre{0.9}{0.5}\hsize]{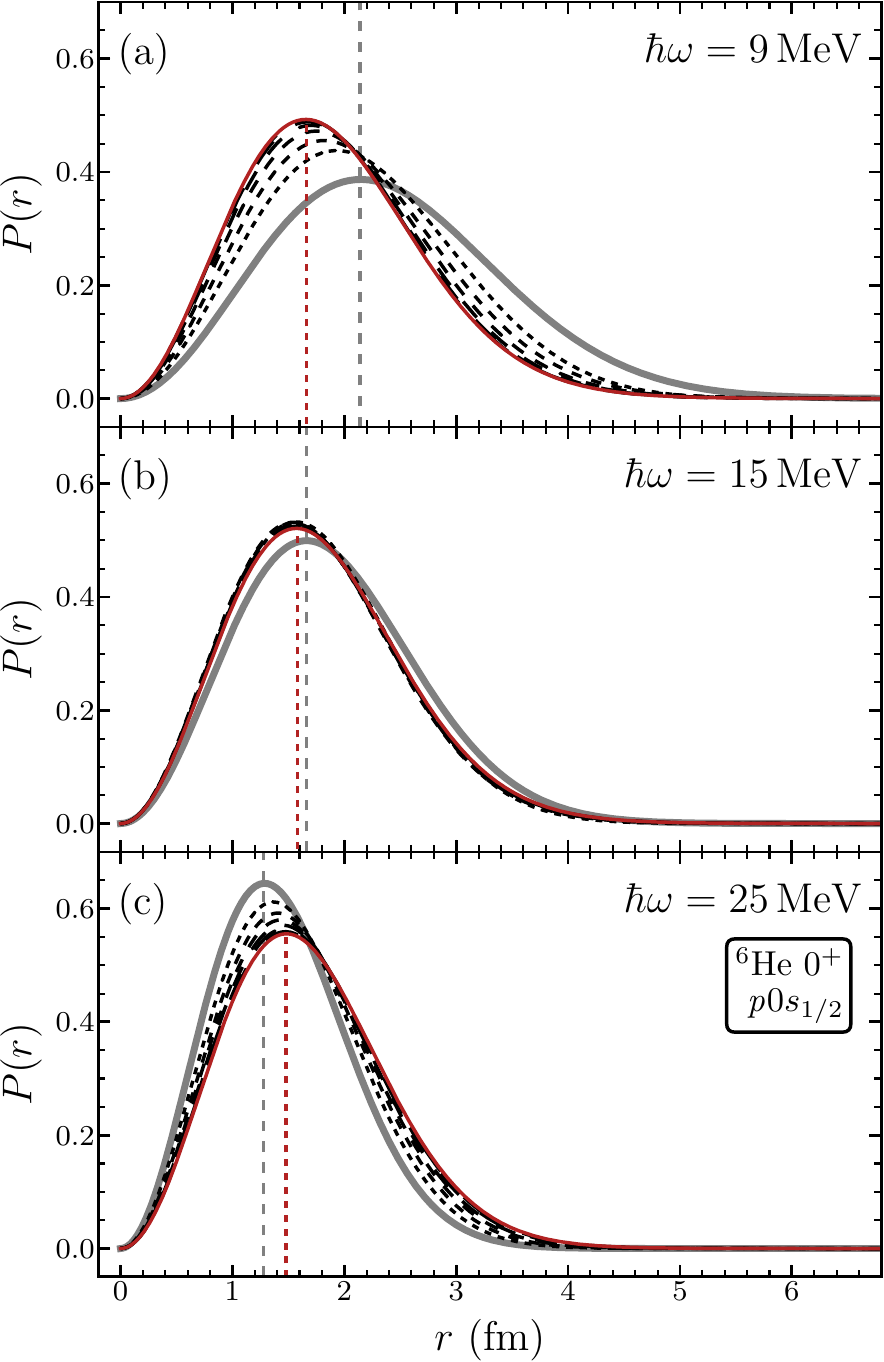}\\
    \end{center}
    \caption{Radial wave functions obtained for the $\isotope[6]{He}$ proton
      $0s_{1/2}$ natural orbital, from different underlying oscillator-basis
      calculations, plotted as the radial probability density $P(r)$.  Results
      are shown as obtained from underlying oscillator-basis calculations with
      (a)~$\hw=9\,\MeV$, (b)~$\hw=15\,\MeV$, and (c)~$\hw=25\,\MeV$.  Radial
      wave functions are shown for $\Nmax=2$ (dotted lines) through $\Nmax=14$
      (solid lines, highlighted), with the oscillator $0s$ function for the
      given $\hw$ (thick gray lines) shown for comparison. The locations of the
      peaks of the underlying harmonic-oscillator orbital and $\Nmax=14$ natural
      orbital are marked with dashed vertical lines.}
    \label{fig:6he-orbitals-ps1-Nmax-evolution-Pr}
\end{figure}

\begin{figure}[tp]
    \begin{center}
        \includegraphics[width=\ifproofpre{0.9}{0.5}\hsize]{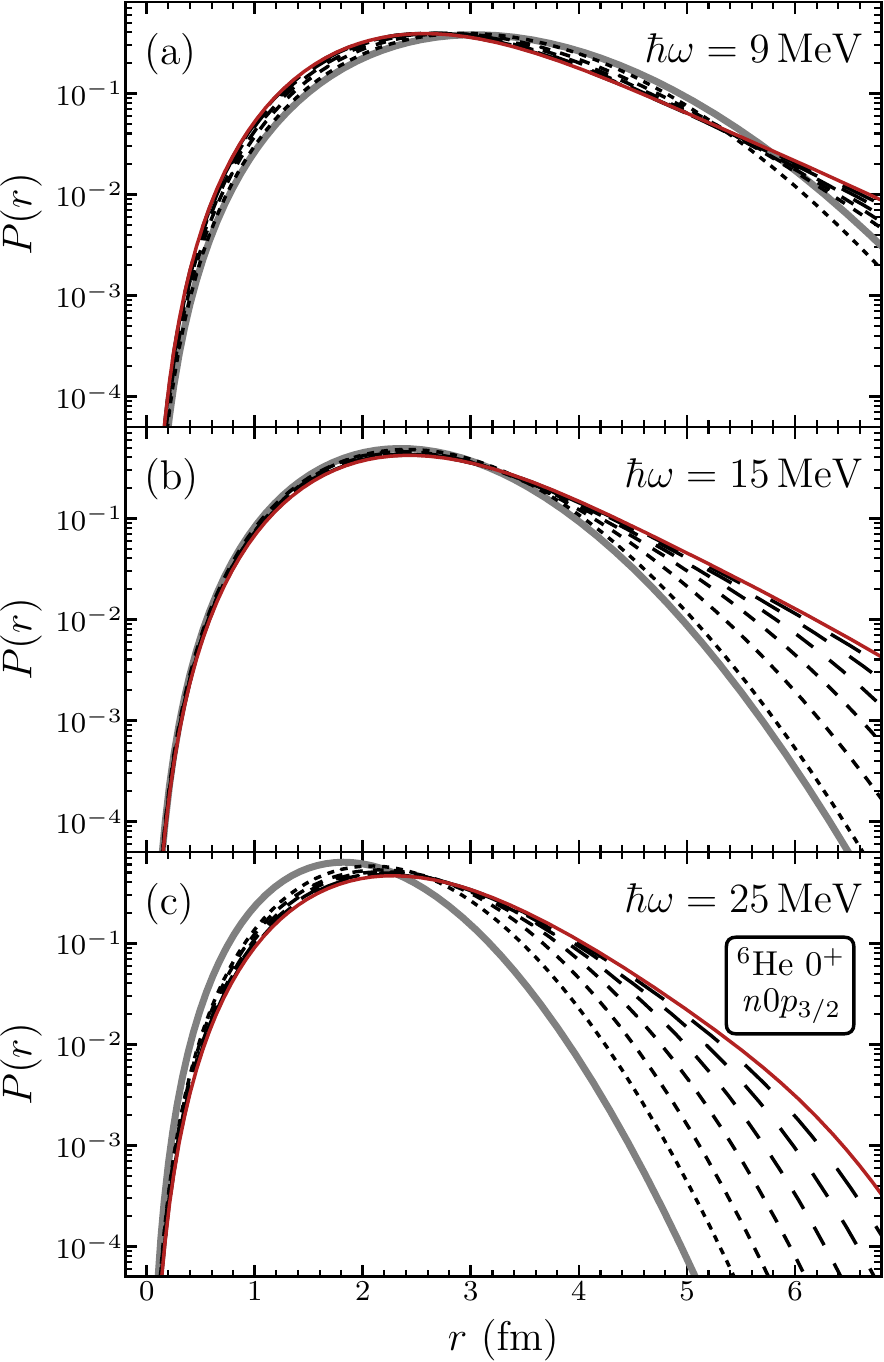}\\
    \end{center}
    \caption{Radial wave functions obtained for the $\isotope[6]{He}$ neutron
      $0p_{3/2}$ natural orbital, from different underlying oscillator-basis
      calculations, plotted as the radial probability density $P(r)$, on a logarithmic scale.  Results are shown as obtained from underlying
      oscillator-basis calculations with (a)~$\hw=9\,\MeV$,
      (b)~$\hw=15\,\MeV$, and (c)~$\hw=25\,\MeV$.  Radial wave functions are
      shown for $\Nmax=2$ (dotted lines) through $\Nmax=14$ (solid lines, highlighted),
      with the oscillator $0s$ function for the given $\hw$ (thick gray lines) shown
      for comparison.
    }
    \label{fig:6he-orbitals-np3-Nmax-evolution-logPr}
\end{figure}

\begin{figure}[tp]
    \begin{center}
        \includegraphics[width=\ifproofpre{0.9}{0.5}\hsize]{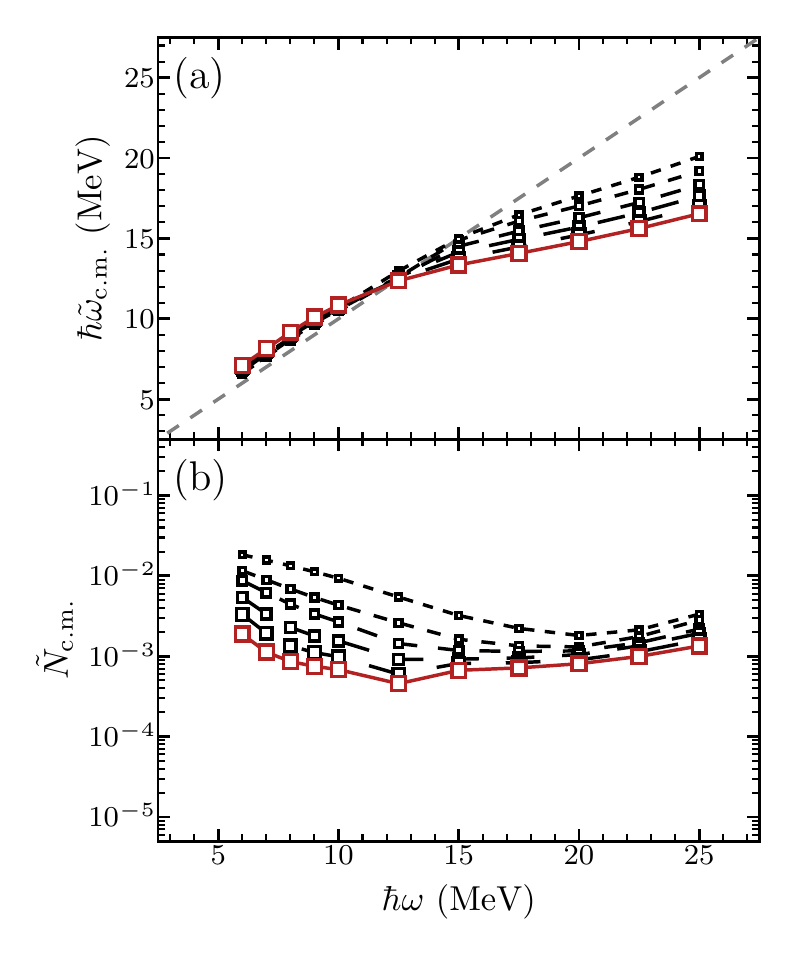}\\
    \end{center}
    \caption{Dependence of the approximate $0s$ center-of-mass motion of the
      calculated $\isotope[6]{He}$ ground state (and its degree of
      contamination) on the $\hw$ of the underlying oscillator basis, in
      calculations with a natural-orbital basis, as measured by (a)~$\hwcmtilde$
      and (b)~$\Ncmtilde$.  Results are shown for calculations with $\Nmax=4$
      (dotted lines) through $\Nmax=14$ (solid lines, highlighted).
        }
    \label{fig:6he-minNcm-hw-scan}
\end{figure}

Turning to the natural orbitals themselves, the radial wave functions for the
lowest natural orbitals are shown in Fig.~\ref{fig:6he-orbitals-Pr}, for a
high-$\Nmax$ calculation ($\Nmax=14$) with $\hw$ near the variational energy
minimum ($\hw=15\,\MeV$). The orbitals shown again correspond, by their $nlj$
labels, to the oscillator orbitals through the $sd$ shell, as in the analogous
figure above for $\isotope[3]{He}$ (Fig.~\ref{fig:3he-orbitals-Pr}).  Radial
functions are shown for both protons (short dashed lines) and neutrons (long
dashed lines), and the oscillator radial functions are again shown for
comparison (thick gray lines).

The mean occupations of these natural orbitals, indicated by the bars at top
right in each panel of Fig.~\ref{fig:6he-orbitals-Pr}, are not far from what
would be expected in a traditional shell model picture.  The $s$ shell is almost
filled, with an occupation of $1.81$ for protons and $1.86$ for neutrons.  Most
of the remaining occupation, out of a total occupation of $2$ for the protons and
$4$ for the neutrons, lies in the $p$ shell.  For the neutrons, in particular,
the $0p_{3/2}$ orbital, which would be the expected orbital for the two valence
neutrons in an extreme noninteracting shell model picture, naturally enough has
a mean occupation of $1.81$, while the $0p_{1/2}$ orbital accounts for a mean
occupation of $0.21$.  By contrast, the $1s_{1/2}$ orbital has a mean occupation
of $<0.05$.  Thus, the halo neutrons are decisively $p$-shell nucleons.  The
occupations for the low-lying natural orbitals are higher than for the
corresponding oscillator orbitals in the underlying calculation, but only
marginally so: the increase in occupation is by $\approx0.16$ for the neutron
$p_{3/2}$ orbital, but only at the level of $\approx0.01$ for the remaining
$s$-shell and $p$-shell orbitals, for both protons and neutrons.

The $0s_{1/2}$ natural orbitals appear virtually unchanged, in
Fig.~\ref{fig:6he-orbitals-Pr}, relative to the underlying oscillator orbital,
for both protons and neutrons.  This is consistent with an unmodified $\alpha$
``core''.  However, to examine the large-distance behavior, we turn to
logarithmic plots, shown in Fig.~\ref{fig:6he-orbitals-logPr}.  Intriguingly,
while both the proton and neutron natural orbitals have linear tails on the
logarithmic plot, indicating exponential fall-off, the decay constants differ,
with a slower fall-off (longer tail) for the neutron orbital.

The $0p_{3/2}$ orbital is of course of special interest, as the orbital
``occupied'' by the halo neutrons.  The peak of the probability distribution, in
the central region (Fig.~\ref{fig:6he-orbitals-Pr}), shifts only marginally
outward in the radial coordinate, on the scale of $\approx0.1\,\fm$.  But the
tail is noticeably extended even viewed on a linear scale.  This is confirmed as
a shallow exponential fall-off when viewed on a logarithmic scale
(Fig.~\ref{fig:6he-orbitals-logPr}).  In contrast, the peak for the
``unoccupied'' proton $0p_{3/2}$ orbital moves to smaller radius, by a
comparable amount, and the tail similarly is retracted
(Fig.~\ref{fig:6he-orbitals-Pr}), with a much steeper exponential fall-off
(Fig.~\ref{fig:6he-orbitals-logPr}).  Similar observations hold for the
$0p_{1/2}$ orbital, which, as noted above, is partially occupied by the valence
neutrons.

The $sd$-shell orbitals are notionally ``unoccupied'' orbitals for both the
protons and neutrons.  The mean occupations of these orbitals are each $\lesssim
0.05$.  The proton orbitals move radially inward, relative to the oscillator
orbital, both in terms of peak location and tail extent
(Fig.~\ref{fig:6he-orbitals-Pr}).  For the neutrons, the behavior is less
consistent. The second peak of the $1s_{1/2}$ orbital, as well as the peak of
the $0d_{5/2}$ orbital, both move markedly outwards, by $\approx1\,\fm$, and the
tails of these orbitals are even more exaggeratedly extended than for the
neutron $p$-shell orbitals.  Yet the neutron $0d_{3/2}$ orbital has a behavior
which closely resembles that of the corresponding proton orbital, in the central
region at least.  Asymptotically, the proton orbitals have similar exponential
tails, with faster decay than the neutron orbitals
(Fig.~\ref{fig:6he-orbitals-logPr}).

For a ``core'' orbital, the proton $0s_{1/2}$ orbital, we explore the dependence
on the $\hw$ and $\Nmax$ of the reference calculation in
Fig.~\ref{fig:6he-orbitals-ps1-Nmax-evolution-Pr}.  The sensitivity of the
natural orbital to the spectator $0s$ motion of the center of mass degree of
freedom is similar to that already seen for this same orbital in
$\isotope[3]{He}$, discussed in Sec.~\ref{sec:results-3he:orbitals}.  Once
again, convergence is rapidly reached with increasing $\Nmax$ for the reference
oscillator-basis calculation, while the shape of this converged natural orbital
is dependent upon the $\hw$ of the underlying oscillator-basis calculation,
which determines the $\hwcm$ of the center-of-mass zero-point motion.  In
$\isotope[6]{He}$ (Fig.~\ref{fig:6he-orbitals-ps1-Nmax-evolution-Pr}), the peak
location for the natural orbital depends on the $\hw$ of the reference
calculation less strongly than for $\isotope[3]{He}$
(Fig.~\ref{fig:3he-orbitals-Nmax-evolution-Pr}), especially at low $\hw$.

Then, for the principal ``halo'' orbital, the neutron $0p_{3/2}$ orbital, the
$\hw$ and $\Nmax$ dependence is similarly explored in
Fig.~\ref{fig:6he-orbitals-np3-Nmax-evolution-logPr}, now on a logarithmic
scale.  (The peak location has a similar dependence to that noted above for the
proton $0s_{1/2}$ orbital.)  The $\hw=9\,\MeV$ oscillator basis
[Fig.~\ref{fig:6he-orbitals-np3-Nmax-evolution-logPr}(a)], with its
comparatively long oscillator length, provides the best support in the tail
region, and thus the fastest realization of a region of exponential decay
(again, indicated by a straight line on the logarithmic plot).  In contrast,
the $\hw=25\,\MeV$ basis
[Fig.~\ref{fig:6he-orbitals-np3-Nmax-evolution-logPr}(c)] yields the slowest
grow-in of the exponential tail.

Finally, there is the question of the center-of-mass motion which emerges in
these calculations for $\isotope[6]{He}$ in a natural-orbital basis.  We apply
the same diagnostics for $\isotope[6]{He}$, shown in
Fig.~\ref{fig:6he-minNcm-hw-scan}, as considered earlier for $\isotope[3]{He}$
in Sec.~\ref{sec:results-3he:Ncm}.  That is, starting from the natural orbitals
obtained from a reference oscillator basis calculation of given $\hw$, we carry
out the many-body calculation for $\isotope[6]{He}$, then evaluate the
center-of-mass $\tbracket{R^2}$ and $\tbracket{K^2}$ observables.  From these,
we deduce the ``optimal'' value of the $\hwcm$ parameter for center-of-mass
motion, $\hwcmtilde$, such that the expectation value $\tbracket{\Ncm}$ of the center-of-mass
number operator assumes its minimum value $\Ncmtilde$.

Comparing the $\isotope[6]{He}$ results for the center-of-mass diagnostics
(Fig.~\ref{fig:6he-minNcm-hw-scan}) to the $\isotope[3]{He}$ results
(Fig.~\ref{fig:3he-minNcm-hw-scan}), a few features stand out.  The dependence
of both $\hwcmtilde$ [Fig.~\ref{fig:6he-minNcm-hw-scan}(a)] and $\Ncmtilde$
[Fig.~\ref{fig:6he-minNcm-hw-scan}(b)] on the reference basis parameters $\hw$
and $\Nmax$ is generally smoother for $\isotope[6]{He}$ than for $\isotope[3]{He}$.  The
zig-zagging irregularities of Fig.~\ref{fig:3he-minNcm-hw-scan} are no longer in
evidence.

The oscillator parameter $\hwcmtilde$ for the center-of-mass motion
[Fig.~\ref{fig:6he-minNcm-hw-scan}(a)] again matches that of the underlying
oscillator basis for the natural orbitals in the vicinity of $\hw=10\MeV$ to
$12.5\,\MeV$. However, the $\hw$ dependence of $\hwcmtilde$ is less steep, above
this point, for $\isotope[6]{He}$ than for $\isotope[3]{He}$ and, indeed, is
continuing to become shallower with increasing $\Nmax$.

Furthermore, $\Ncmtilde$ [Fig.~\ref{fig:6he-minNcm-hw-scan}(b)] is comparatively
independent of $\hw$.  By $\Nmax=14$, $\Ncmtilde$ has decreased to
$\approx10^{-3}$ over most of the $\hw$ range shown, and it continues to
decrease with increasing $\Nmax$.  Thus, in short, for practical purposes, a
near-pure harmonic-oscillator $0s$ center-of-mass motion is uniformly obtained in the many-body
calculations for $\isotope[6]{He}$ in the natural-orbital basis.

\section{Conclusion}
\label{sec:conclusion}

The nuclear many-body system is highly correlated, and thus inherently requires
many antisymmetrized product states (Slater determinants) for its accurate
description.  No choice of single-particle states can completely obviate the
need for superposing antisymmetrized product states in representing a correlated
system.  Nonetheless, a judicious choice of single-particle basis can accelerate
the convergence of the description of the many-body wave function in a
configuration-interaction basis.

Natural orbitals, obtained by diagonalizing the (scalar) one-body density
matrix, address this aim in a well-defined sense, by maximizing the occupation
of low-lying orbitals, or minimizing the depletion of the Fermi sea, in the
expansion of a specific many-body reference state.  After outlining the
procedure for constructing and using natural orbitals within the NCCI framework
(Sec.~\ref{sec:methods}), we have examined in detail the properties both of the
orbitals themselves and the consequent many-body calculations in a natural
orbital basis, first for the simple testbed case of $\isotope[3]{He}$
(Sec.~\ref{sec:results-3he}), then for the halo nucleus $\isotope[6]{He}$
(Sec.~\ref{sec:results-6he}).

There are several noteworthy factors (Sec.~\ref{sec:methods}) limiting what we might expect to
accomplish, in practice, with the natural orbital basis in NCCI calculations.
The natural orbitals are only known to limited accuracy, as they are obtained
from a reference many-body state which is only an approximation to the true
solution of the many-body problem (as it would be obtained in an untruncated
space).  This reference state is represented in terms of orbitals from a
truncated single-particle space, which limits the portion of the single-particle
space which the natural orbitals can sample.  Moreover, the many-body space in
which the reference state is calculated is then subjected to a nontrivial
truncation (\textit{e.g.}, by $\Nmax$), which may be expected to further
restrict the fidelity of the reference state found therein and, specifically,
the representation of high-lying orbitals in the scalar density obtained from
this reference state.

Furthermore, even if the reference state could be found exactly, and its natural
orbitals deduced exactly, one-body densities obtained from a laboratory-frame
reference state are not uniquely defined by the intrinsic structure.  Rather,
they reflect some spectator center-of-mass motion arbitrarily superposed on this
intrinsic structure.  The natural orbitals obtained from these densities are
then used in a many-body calculation which, although intended simply to
reproduce the intrinsic structure of interest, in practice must yield some possibly
complicated combination of intrinsic and center-of-mass motion.

Nonetheless, changing to a many-body basis constructed from natural orbitals
does permit an NCCI calculation to probe portions of the many-body space which
were not accessible in the original reference calculation.  If the initial
calculation is in an $\Nmax$-truncated harmonic oscillator basis, as here, then
the calculation in a natural orbital basis brings in highly-excited oscillator
configurations which were beyond the limit of the initial calculation.

We find that the transformation from harmonic oscillator orbitals to natural
orbitals serves in part to simply accomplish a dilation of the harmonic
oscillator basis, from the length scale (or $\hw$) of the underlying basis, to a
more optimal length scale (or $\hw$).  This observation (\textit{e.g.},
Fig.~\ref{fig:6he-orbitals-ps1-Nmax-evolution-Pr}) already explains the relative
insensitivity of calculated energies and other observables in a natural orbital
basis to the $\hw$ of the underlying oscillator basis: as one varies $\hw$, the
transformation to natural orbitals simply undoes this variation.  Such dilation,
in itself, merely recovers the results of a harmonic oscillator basis chosen
with optimal length scale, rather than improving on it.

More substantially, though, the transformation to natural orbitals provides
genuine modifications to the shape of the radial wave functions.  Notably, the
artificial Gaussian fall-off of the oscillator functions is modified to more
closely resemble the exponential fall-off physically expected from the finite
range of the nuclear interaction (\textit{e.g.},
Fig.~\ref{fig:6he-orbitals-logPr}).  These differences can account
for the improvements over the results obtained, even with an optimal choice of
$\hw$, using the harmonic oscillator basis.

In the present work, where we retain the convenient but simpleminded ``nominal
$\Nmax$'' truncation scheme (Sec.~\ref{sec:methods:ncci}) for the many-body
basis generated from the natural orbitals, we find improvements by about one
step in $\Nmax$ over the oscillator-basis calculations.  (Although illustrated
here for the Daejeon16 interaction, similar results are found with other
interactions, \textit{e.g.}, in the preliminary
study~\cite{constantinou2017:natorb-natowitz16,constantinou2017:diss} with
JISP16.)  While this improvement is incremental, it is nonetheless welcome.  The
computational cost of a second calculation (with natural orbitals) in a space of
the same dimension as the underlying reference calculation (with oscillator
orbitals) is typically far less than that of performing a new calculation in a
space of higher $\Nmax$, which typically entails an order-of-magnitude increase
in dimension (Fig.~\ref{fig:dimension}), and correspondingly larger increase in
memory demands and computational load~\cite{maris2013:ncci-chiral-ccp12}.

However, the present exploration is also intended to provide a baseline for
understanding more sophisticated many-body calculations based on natural
orbitals derived from NCCI reference calculations.  Here we reiterate that the
eigenvalues of the density matrix provide information on the ``importance'' of
orbitals, which could ostensibly be used to good effect in defining a weighting
scheme for the many-body truncation.  Moreover, natural orbitals provide a
reasonable starting point~\cite{tichai2019:natorb-mbpt} for hybrid many-body
calculations which incorporate truncated configuration-interaction bases for
portions of the calculation, \textit{e.g.}, in-medium
NCSM~\cite{gebrerufael2017:im-ncsm} and perturbatively-improved
NCSM~\cite{tichai2018:ncsm-perturbative} calculations.

\begin{acknowledgments}
We thank Guillaume Hupin for valuable discussions on the formulation of the
nuclear natural orbital problem, Mitch~A.\ McNanna for carrying out informative
preliminary studies in one dimension, and Jakub Herko, Anna~E.\ McCoy,
Charlotte~M.\ Wood, and Zhou Zhou for comments on the manuscript.  This material is based upon
work supported by the U.S.~Department of Energy, Office of Science, under Award
Numbers DE-FG02-95ER-40934, DE-FG02-91ER-40608, DE-SC0018223 (SciDAC4/NUCLEI),
and DE-FG02-87ER40371.  This research used computational resources of the
University of Notre Dame Center for Research Computing and of the National
Energy Research Scientific Computing Center (NERSC), a U.S.~Department of
Energy, Office of Science, user facility supported under
Contract~DE-AC02-05CH11231.
\end{acknowledgments}

\bibliographystyle{apsrev4-2}

\begin{thebibliography}{98}\makeatletter
\providecommand \@ifxundefined [1]{\@ifx{#1\undefined}
}\providecommand \@ifnum [1]{\ifnum #1\expandafter \@firstoftwo
 \else \expandafter \@secondoftwo
 \fi
}\providecommand \@ifx [1]{\ifx #1\expandafter \@firstoftwo
 \else \expandafter \@secondoftwo
 \fi
}\providecommand \natexlab [1]{#1}\providecommand \enquote  [1]{``#1''}\providecommand \bibnamefont  [1]{#1}\providecommand \bibfnamefont [1]{#1}\providecommand \citenamefont [1]{#1}\providecommand \href@noop [0]{\@secondoftwo}\providecommand \href [0]{\begingroup \@sanitize@url \@href}\providecommand \@href[1]{\@@startlink{#1}\@@href}\providecommand \@@href[1]{\endgroup#1\@@endlink}\providecommand \@sanitize@url [0]{\catcode `\\12\catcode `\$12\catcode
  `\&12\catcode `\#12\catcode `\^12\catcode `\_12\catcode `\%12\relax}\providecommand \@@startlink[1]{}\providecommand \@@endlink[0]{}\providecommand \url  [0]{\begingroup\@sanitize@url \@url }\providecommand \@url [1]{\endgroup\@href {#1}{\urlprefix }}\providecommand \urlprefix  [0]{URL }\providecommand \Eprint [0]{\href }\providecommand \doibase [0]{https://doi.org/}\providecommand \selectlanguage [0]{\@gobble}\providecommand \bibinfo  [0]{\@secondoftwo}\providecommand \bibfield  [0]{\@secondoftwo}\providecommand \translation [1]{[#1]}\providecommand \BibitemOpen [0]{}\providecommand \bibitemStop [0]{}\providecommand \bibitemNoStop [0]{.\EOS\space}\providecommand \EOS [0]{\spacefactor3000\relax}\providecommand \BibitemShut  [1]{\csname bibitem#1\endcsname}\let\auto@bib@innerbib\@empty
\bibitem [{\citenamefont {Navr\'{a}til}\ \emph {et~al.}(2000)\citenamefont
  {Navr\'{a}til}, \citenamefont {Vary},\ and\ \citenamefont
  {Barrett}}]{navratil2000:12c-ab-initio}\BibitemOpen
  \bibfield  {author} {\bibinfo {author} {\bibfnamefont {P.}~\bibnamefont
  {Navr\'{a}til}}, \bibinfo {author} {\bibfnamefont {J.~P.}\ \bibnamefont
  {Vary}},\ and\ \bibinfo {author} {\bibfnamefont {B.~R.}\ \bibnamefont
  {Barrett}},\ }\bibfield  {title} {\bibinfo {title} {Properties of
  $^{12}\mathrm{C}$ in the \textit{ab initio} nuclear shell model},\ }\href
  {https://doi.org/10.1103/PhysRevLett.84.5728} {\bibfield  {journal} {\bibinfo
   {journal} {Phys. Rev. Lett.}\ }\textbf {\bibinfo {volume} {84}},\ \bibinfo
  {pages} {5728} (\bibinfo {year} {2000})}\BibitemShut {NoStop}\bibitem [{\citenamefont {Pieper}\ \emph {et~al.}(2004)\citenamefont {Pieper},
  \citenamefont {Wiringa},\ and\ \citenamefont
  {Carlson}}]{pieper2004:gfmc-a6-8}\BibitemOpen
  \bibfield  {author} {\bibinfo {author} {\bibfnamefont {S.~C.}\ \bibnamefont
  {Pieper}}, \bibinfo {author} {\bibfnamefont {R.~B.}\ \bibnamefont
  {Wiringa}},\ and\ \bibinfo {author} {\bibfnamefont {J.}~\bibnamefont
  {Carlson}},\ }\bibfield  {title} {\bibinfo {title} {Quantum {M}onte {C}arlo
  calculations of excited states in {$A=6$--$8$} nuclei},\ }\href
  {https://doi.org/10.1103/PhysRevC.70.054325} {\bibfield  {journal} {\bibinfo
  {journal} {Phys. Rev. C}\ }\textbf {\bibinfo {volume} {70}},\ \bibinfo
  {pages} {054325} (\bibinfo {year} {2004})}\BibitemShut {NoStop}\bibitem [{\citenamefont {Neff}\ and\ \citenamefont
  {Feldmeier}(2004)}]{neff2004:cluster-fmd}\BibitemOpen
  \bibfield  {author} {\bibinfo {author} {\bibfnamefont {T.}~\bibnamefont
  {Neff}}\ and\ \bibinfo {author} {\bibfnamefont {H.}~\bibnamefont
  {Feldmeier}},\ }\bibfield  {title} {\bibinfo {title} {Cluster structures
  within fermionic molecular dynamics},\ }\href
  {https://doi.org/10.1016/j.nuclphysa.2004.04.061} {\bibfield  {journal}
  {\bibinfo  {journal} {Nucl. Phys. A}\ }\textbf {\bibinfo {volume} {738}},\
  \bibinfo {pages} {357} (\bibinfo {year} {2004})}\BibitemShut {NoStop}\bibitem [{\citenamefont {Hagen}\ \emph {et~al.}(2007)\citenamefont {Hagen},
  \citenamefont {Dean}, \citenamefont {Hjorth-Jensen}, \citenamefont
  {Papenbrock},\ and\ \citenamefont
  {Schwenk}}]{hagen2007:coupled-cluster-benchmark}\BibitemOpen
  \bibfield  {author} {\bibinfo {author} {\bibfnamefont {G.}~\bibnamefont
  {Hagen}}, \bibinfo {author} {\bibfnamefont {D.~J.}\ \bibnamefont {Dean}},
  \bibinfo {author} {\bibfnamefont {M.}~\bibnamefont {Hjorth-Jensen}}, \bibinfo
  {author} {\bibfnamefont {T.}~\bibnamefont {Papenbrock}},\ and\ \bibinfo
  {author} {\bibfnamefont {A.}~\bibnamefont {Schwenk}},\ }\bibfield  {title}
  {\bibinfo {title} {Benchmark calculations for $\isotope[3]{H}$,
  $\isotope[4]{He}$, $\isotope[16]{O}$, and $\isotope[40]{Ca}$ with \textit{ab
  initio} coupled-cluster theory},\ }\href
  {https://doi.org/10.1103/PhysRevC.76.044305} {\bibfield  {journal} {\bibinfo
  {journal} {Phys. Rev. C}\ }\textbf {\bibinfo {volume} {76}},\ \bibinfo
  {pages} {044305} (\bibinfo {year} {2007})}\BibitemShut {NoStop}\bibitem [{\citenamefont {Quaglioni}\ and\ \citenamefont
  {Navr{\'a}til}(2009)}]{quaglioni2009:ncsm-rgm}\BibitemOpen
  \bibfield  {author} {\bibinfo {author} {\bibfnamefont {S.}~\bibnamefont
  {Quaglioni}}\ and\ \bibinfo {author} {\bibfnamefont {P.}~\bibnamefont
  {Navr{\'a}til}},\ }\bibfield  {title} {\bibinfo {title} {\textit{Ab initio}
  many-body calculations of nucleon-nucleus scattering},\ }\href
  {https://doi.org/10.1103/PhysRevC.79.044606} {\bibfield  {journal} {\bibinfo
  {journal} {Phys. Rev. C}\ }\textbf {\bibinfo {volume} {79}},\ \bibinfo
  {pages} {044606} (\bibinfo {year} {2009})}\BibitemShut {NoStop}\bibitem [{\citenamefont {Bacca}\ \emph {et~al.}(2012)\citenamefont {Bacca},
  \citenamefont {Barnea},\ and\ \citenamefont
  {Schwenk}}]{bacca2012:6he-hyperspherical}\BibitemOpen
  \bibfield  {author} {\bibinfo {author} {\bibfnamefont {S.}~\bibnamefont
  {Bacca}}, \bibinfo {author} {\bibfnamefont {N.}~\bibnamefont {Barnea}},\ and\
  \bibinfo {author} {\bibfnamefont {A.}~\bibnamefont {Schwenk}},\ }\bibfield
  {title} {\bibinfo {title} {Matter and charge radii of $\isotope[6]{He}$ in
  the hyperspherical-harmonics approach},\ }\href
  {https://doi.org/10.1103/PhysRevC.86.034321} {\bibfield  {journal} {\bibinfo
  {journal} {Phys. Rev. C}\ }\textbf {\bibinfo {volume} {86}},\ \bibinfo
  {pages} {034321} (\bibinfo {year} {2012})}\BibitemShut {NoStop}\bibitem [{\citenamefont {Shimizu}\ \emph {et~al.}(2012)\citenamefont
  {Shimizu}, \citenamefont {Abe}, \citenamefont {Tsunoda}, \citenamefont
  {Utsuno}, \citenamefont {Yoshida}, \citenamefont {Mizusaki}, \citenamefont
  {Honma},\ and\ \citenamefont {Otsuka}}]{shimizu2012:mcsm}\BibitemOpen
  \bibfield  {author} {\bibinfo {author} {\bibfnamefont {N.}~\bibnamefont
  {Shimizu}}, \bibinfo {author} {\bibfnamefont {T.}~\bibnamefont {Abe}},
  \bibinfo {author} {\bibfnamefont {Y.}~\bibnamefont {Tsunoda}}, \bibinfo
  {author} {\bibfnamefont {Y.}~\bibnamefont {Utsuno}}, \bibinfo {author}
  {\bibfnamefont {T.}~\bibnamefont {Yoshida}}, \bibinfo {author} {\bibfnamefont
  {T.}~\bibnamefont {Mizusaki}}, \bibinfo {author} {\bibfnamefont
  {M.}~\bibnamefont {Honma}},\ and\ \bibinfo {author} {\bibfnamefont
  {T.}~\bibnamefont {Otsuka}},\ }\bibfield  {title} {\bibinfo {title}
  {New-generation {M}onte {C}arlo shell model for the {K} computer era},\
  }\href {https://doi.org/10.1093/ptep/pts012} {\bibfield  {journal} {\bibinfo
  {journal} {Prog. Exp. Theor. Phys.}\ }\textbf {\bibinfo {volume} {2012}},\
  \bibinfo {pages} {01A205} (\bibinfo {year} {2012})}\BibitemShut {NoStop}\bibitem [{\citenamefont {Dytrych}\ \emph {et~al.}(2013)\citenamefont
  {Dytrych}, \citenamefont {Launey}, \citenamefont {Draayer}, \citenamefont
  {Maris}, \citenamefont {Vary}, \citenamefont {Saule}, \citenamefont
  {Catalyurek}, \citenamefont {Sosonkina}, \citenamefont {Langr},\ and\
  \citenamefont {Caprio}}]{dytrych2013:su3ncsm}\BibitemOpen
  \bibfield  {author} {\bibinfo {author} {\bibfnamefont {T.}~\bibnamefont
  {Dytrych}}, \bibinfo {author} {\bibfnamefont {K.~D.}\ \bibnamefont {Launey}},
  \bibinfo {author} {\bibfnamefont {J.~P.}\ \bibnamefont {Draayer}}, \bibinfo
  {author} {\bibfnamefont {P.}~\bibnamefont {Maris}}, \bibinfo {author}
  {\bibfnamefont {J.~P.}\ \bibnamefont {Vary}}, \bibinfo {author}
  {\bibfnamefont {E.}~\bibnamefont {Saule}}, \bibinfo {author} {\bibfnamefont
  {U.}~\bibnamefont {Catalyurek}}, \bibinfo {author} {\bibfnamefont
  {M.}~\bibnamefont {Sosonkina}}, \bibinfo {author} {\bibfnamefont
  {D.}~\bibnamefont {Langr}},\ and\ \bibinfo {author} {\bibfnamefont {M.~A.}\
  \bibnamefont {Caprio}},\ }\bibfield  {title} {\bibinfo {title} {Collective
  modes in light nuclei from first principles},\ }\href
  {https://doi.org/10.1103/PhysRevLett.111.252501} {\bibfield  {journal}
  {\bibinfo  {journal} {Phys. Rev. Lett.}\ }\textbf {\bibinfo {volume} {111}},\
  \bibinfo {pages} {252501} (\bibinfo {year} {2013})}\BibitemShut {NoStop}\bibitem [{\citenamefont {Barrett}\ \emph {et~al.}(2013)\citenamefont
  {Barrett}, \citenamefont {Navr\'{a}til},\ and\ \citenamefont
  {Vary}}]{barrett2013:ncsm}\BibitemOpen
  \bibfield  {author} {\bibinfo {author} {\bibfnamefont {B.~R.}\ \bibnamefont
  {Barrett}}, \bibinfo {author} {\bibfnamefont {P.}~\bibnamefont
  {Navr\'{a}til}},\ and\ \bibinfo {author} {\bibfnamefont {J.~P.}\ \bibnamefont
  {Vary}},\ }\bibfield  {title} {\bibinfo {title} {\textit{Ab initio} no core
  shell model},\ }\href {https://doi.org/10.1016/j.ppnp.2012.10.003} {\bibfield
   {journal} {\bibinfo  {journal} {Prog. Part. Nucl. Phys.}\ }\textbf {\bibinfo
  {volume} {69}},\ \bibinfo {pages} {131} (\bibinfo {year} {2013})}\BibitemShut
  {NoStop}\bibitem [{\citenamefont {Baroni}\ \emph {et~al.}(2013)\citenamefont {Baroni},
  \citenamefont {Navr{\'a}til},\ and\ \citenamefont
  {Quaglioni}}]{baroni2013:7he-ncsmc}\BibitemOpen
  \bibfield  {author} {\bibinfo {author} {\bibfnamefont {S.}~\bibnamefont
  {Baroni}}, \bibinfo {author} {\bibfnamefont {P.}~\bibnamefont
  {Navr{\'a}til}},\ and\ \bibinfo {author} {\bibfnamefont {S.}~\bibnamefont
  {Quaglioni}},\ }\bibfield  {title} {\bibinfo {title} {Unified \textit{ab
  initio} approach to bound and unbound states: {N}o-core shell model with
  continuum and its application to $\isotope[7]{He}$},\ }\href
  {https://doi.org/10.1103/PhysRevC.87.034326} {\bibfield  {journal} {\bibinfo
  {journal} {Phys. Rev. C}\ }\textbf {\bibinfo {volume} {87}},\ \bibinfo
  {pages} {034326} (\bibinfo {year} {2013})}\BibitemShut {NoStop}\bibitem [{\citenamefont {Wiringa}\ \emph {et~al.}(1995)\citenamefont
  {Wiringa}, \citenamefont {Stoks},\ and\ \citenamefont
  {Schiavilla}}]{wiringa1995:nn-av18}\BibitemOpen
  \bibfield  {author} {\bibinfo {author} {\bibfnamefont {R.~B.}\ \bibnamefont
  {Wiringa}}, \bibinfo {author} {\bibfnamefont {V.~G.~J.}\ \bibnamefont
  {Stoks}},\ and\ \bibinfo {author} {\bibfnamefont {R.}~\bibnamefont
  {Schiavilla}},\ }\bibfield  {title} {\bibinfo {title} {Accurate
  nucleon-nucleon potential with charge-independence breaking},\ }\href
  {https://doi.org/10.1103/PhysRevC.51.38} {\bibfield  {journal} {\bibinfo
  {journal} {Phys. Rev. C}\ }\textbf {\bibinfo {volume} {51}},\ \bibinfo
  {pages} {38} (\bibinfo {year} {1995})}\BibitemShut {NoStop}\bibitem [{\citenamefont {Entem}\ and\ \citenamefont
  {Machleidt}(2003)}]{entem2003:chiral-nn-potl}\BibitemOpen
  \bibfield  {author} {\bibinfo {author} {\bibfnamefont {D.~R.}\ \bibnamefont
  {Entem}}\ and\ \bibinfo {author} {\bibfnamefont {R.}~\bibnamefont
  {Machleidt}},\ }\bibfield  {title} {\bibinfo {title} {Accurate
  charge-dependent nucleon-nucleon potential at fourth order of chiral
  perturbation theory},\ }\href {https://doi.org/10.1103/PhysRevC.68.041001}
  {\bibfield  {journal} {\bibinfo  {journal} {Phys. Rev. C}\ }\textbf {\bibinfo
  {volume} {68}},\ \bibinfo {pages} {041001(R)} (\bibinfo {year}
  {2003})}\BibitemShut {NoStop}\bibitem [{\citenamefont {Shirokov}\ \emph {et~al.}(2007)\citenamefont
  {Shirokov}, \citenamefont {Vary}, \citenamefont {Mazur},\ and\ \citenamefont
  {Weber}}]{shirokov2007:nn-jisp16}\BibitemOpen
  \bibfield  {author} {\bibinfo {author} {\bibfnamefont {A.~M.}\ \bibnamefont
  {Shirokov}}, \bibinfo {author} {\bibfnamefont {J.~P.}\ \bibnamefont {Vary}},
  \bibinfo {author} {\bibfnamefont {A.~I.}\ \bibnamefont {Mazur}},\ and\
  \bibinfo {author} {\bibfnamefont {T.~A.}\ \bibnamefont {Weber}},\ }\bibfield
  {title} {\bibinfo {title} {Realistic nuclear {H}amiltonian: \textit{Ab exitu}
  approach},\ }\href {https://doi.org/10.1016/j.physletb.2006.10.066}
  {\bibfield  {journal} {\bibinfo  {journal} {Phys. Lett. B}\ }\textbf
  {\bibinfo {volume} {644}},\ \bibinfo {pages} {33} (\bibinfo {year}
  {2007})}\BibitemShut {NoStop}\bibitem [{\citenamefont {Epelbaum}\ \emph {et~al.}(2009)\citenamefont
  {Epelbaum}, \citenamefont {Hammer},\ and\ \citenamefont
  {Mei\ss{}ner}}]{epelbaum2009:nuclear-forces}\BibitemOpen
  \bibfield  {author} {\bibinfo {author} {\bibfnamefont {E.}~\bibnamefont
  {Epelbaum}}, \bibinfo {author} {\bibfnamefont {H.-W.}\ \bibnamefont
  {Hammer}},\ and\ \bibinfo {author} {\bibfnamefont {U.-G.}\ \bibnamefont
  {Mei\ss{}ner}},\ }\bibfield  {title} {\bibinfo {title} {Modern theory of
  nuclear forces},\ }\href {https://doi.org/10.1103/RevModPhys.81.1773}
  {\bibfield  {journal} {\bibinfo  {journal} {Rev. Mod. Phys.}\ }\textbf
  {\bibinfo {volume} {81}},\ \bibinfo {pages} {1773} (\bibinfo {year}
  {2009})}\BibitemShut {NoStop}\bibitem [{\citenamefont {Helgaker}\ \emph {et~al.}(2000)\citenamefont
  {Helgaker}, \citenamefont {J{\o}rgensen},\ and\ \citenamefont
  {Olsen}}]{helgaker2000:electron-structure}\BibitemOpen
  \bibfield  {author} {\bibinfo {author} {\bibfnamefont {T.}~\bibnamefont
  {Helgaker}}, \bibinfo {author} {\bibfnamefont {P.}~\bibnamefont
  {J{\o}rgensen}},\ and\ \bibinfo {author} {\bibfnamefont {J.}~\bibnamefont
  {Olsen}},\ }\href@noop {} {\emph {\bibinfo {title} {Molecular
  Electron-Structure Theory}}}\ (\bibinfo  {publisher} {Wiley},\ \bibinfo
  {address} {Chichester},\ \bibinfo {year} {2000})\BibitemShut {NoStop}\bibitem [{\citenamefont {Moshinsky}\ and\ \citenamefont
  {Smirnov}(1996)}]{moshinsky1996:oscillator}\BibitemOpen
  \bibfield  {author} {\bibinfo {author} {\bibfnamefont {M.}~\bibnamefont
  {Moshinsky}}\ and\ \bibinfo {author} {\bibfnamefont {Y.~F.}\ \bibnamefont
  {Smirnov}},\ }\href@noop {} {\emph {\bibinfo {title} {The Harmonic Oscillator
  in Modern Physics}}}\ (\bibinfo  {publisher} {Harwood Academic Publishers},\
  \bibinfo {address} {Amsterdam},\ \bibinfo {year} {1996})\BibitemShut
  {NoStop}\bibitem [{\citenamefont {Elliott}\ and\ \citenamefont
  {Skyrme}(1955)}]{elliott1955:com-shell}\BibitemOpen
  \bibfield  {author} {\bibinfo {author} {\bibfnamefont {J.~P.}\ \bibnamefont
  {Elliott}}\ and\ \bibinfo {author} {\bibfnamefont {T.~H.~R.}\ \bibnamefont
  {Skyrme}},\ }\bibfield  {title} {\bibinfo {title} {Centre-of-mass effects in
  the nuclear shell-model},\ }\href {https://doi.org/10.1098/rspa.1955.0239}
  {\bibfield  {journal} {\bibinfo  {journal} {Proc. R. Soc. London A}\ }\textbf
  {\bibinfo {volume} {232}},\ \bibinfo {pages} {561} (\bibinfo {year}
  {1955})}\BibitemShut {NoStop}\bibitem [{\citenamefont {Caprio}\ \emph {et~al.}(2020)\citenamefont {Caprio},
  \citenamefont {McCoy},\ and\ \citenamefont {Fasano}}]{caprio2020:intrinsic}\BibitemOpen
  \bibfield  {author} {\bibinfo {author} {\bibfnamefont {M.~A.}\ \bibnamefont
  {Caprio}}, \bibinfo {author} {\bibfnamefont {A.~E.}\ \bibnamefont {McCoy}},\
  and\ \bibinfo {author} {\bibfnamefont {P.~J.}\ \bibnamefont {Fasano}},\
  }\bibfield  {title} {\bibinfo {title} {Intrinsic operators for the
  translationally-invariant many-body problem},\ }\href
  {https://doi.org/10.1088/1361-6471/ab9d38} {\bibfield  {journal} {\bibinfo
  {journal} {J. Phys. G}\ }\textbf {\bibinfo {volume} {47}},\ \bibinfo {pages}
  {122001} (\bibinfo {year} {2020})}\BibitemShut {NoStop}\bibitem [{\citenamefont {Davies}\ \emph {et~al.}(1966)\citenamefont {Davies},
  \citenamefont {Krieger},\ and\ \citenamefont
  {Baranger}}]{davies1966:hartree-fock}\BibitemOpen
  \bibfield  {author} {\bibinfo {author} {\bibfnamefont {K.~T.~R.}\
  \bibnamefont {Davies}}, \bibinfo {author} {\bibfnamefont {S.~J.}\
  \bibnamefont {Krieger}},\ and\ \bibinfo {author} {\bibfnamefont
  {M.}~\bibnamefont {Baranger}},\ }\bibfield  {title} {\bibinfo {title} {A
  study of the {H}artree-{F}ock approximation as applied to finite nuclei},\
  }\href {https://doi.org/10.1016/0029-5582(66)91013-3} {\bibfield  {journal}
  {\bibinfo  {journal} {Nucl. Phys.}\ }\textbf {\bibinfo {volume} {84}},\
  \bibinfo {pages} {545} (\bibinfo {year} {1966})}\BibitemShut {NoStop}\bibitem [{\citenamefont {Caprio}\ \emph {et~al.}(2012)\citenamefont {Caprio},
  \citenamefont {Maris},\ and\ \citenamefont {Vary}}]{caprio2012:csbasis}\BibitemOpen
  \bibfield  {author} {\bibinfo {author} {\bibfnamefont {M.~A.}\ \bibnamefont
  {Caprio}}, \bibinfo {author} {\bibfnamefont {P.}~\bibnamefont {Maris}},\ and\
  \bibinfo {author} {\bibfnamefont {J.~P.}\ \bibnamefont {Vary}},\ }\bibfield
  {title} {\bibinfo {title} {The {C}oulomb-{S}turmian basis for the nuclear
  many-body problem},\ }\href {https://doi.org/10.1103/PhysRevC.86.034312}
  {\bibfield  {journal} {\bibinfo  {journal} {Phys. Rev. C}\ }\textbf {\bibinfo
  {volume} {86}},\ \bibinfo {pages} {034312} (\bibinfo {year}
  {2012})}\BibitemShut {NoStop}\bibitem [{\citenamefont {Caprio}\ \emph {et~al.}(2014)\citenamefont {Caprio},
  \citenamefont {Maris},\ and\ \citenamefont {Vary}}]{caprio2014:cshalo}\BibitemOpen
  \bibfield  {author} {\bibinfo {author} {\bibfnamefont {M.~A.}\ \bibnamefont
  {Caprio}}, \bibinfo {author} {\bibfnamefont {P.}~\bibnamefont {Maris}},\ and\
  \bibinfo {author} {\bibfnamefont {J.~P.}\ \bibnamefont {Vary}},\ }\bibfield
  {title} {\bibinfo {title} {Halo nuclei $\isotope[6]{He}$ and
  $\isotope[8]{He}$ with the {C}oulomb-{S}turmian basis},\ }\href
  {https://doi.org/10.1103/PhysRevC.90.034305} {\bibfield  {journal} {\bibinfo
  {journal} {Phys. Rev. C}\ }\textbf {\bibinfo {volume} {90}},\ \bibinfo
  {pages} {034305} (\bibinfo {year} {2014})}\BibitemShut {NoStop}\bibitem [{\citenamefont
  {L{\"o}wdin}(1955)}]{loewdin1955:natural-orbitals-part1}\BibitemOpen
  \bibfield  {author} {\bibinfo {author} {\bibfnamefont {P.-O.}\ \bibnamefont
  {L{\"o}wdin}},\ }\bibfield  {title} {\bibinfo {title} {Quantum theory of
  many-particle systems. {I}. {P}hysical interpretations by means of density
  matrices, natural spin-orbitals, and convergence problems in the method of
  configurational interaction},\ }\href
  {https://doi.org/10.1103/PhysRev.97.1474} {\bibfield  {journal} {\bibinfo
  {journal} {Phys. Rev.}\ }\textbf {\bibinfo {volume} {97}},\ \bibinfo {pages}
  {1474} (\bibinfo {year} {1955})}\BibitemShut {NoStop}\bibitem [{\citenamefont {Shull}\ and\ \citenamefont
  {L{\"o}wdin}(1955{\natexlab{a}})}]{shull1955:natural-orbitals-helium}\BibitemOpen
  \bibfield  {author} {\bibinfo {author} {\bibfnamefont {H.}~\bibnamefont
  {Shull}}\ and\ \bibinfo {author} {\bibfnamefont {P.-O.}\ \bibnamefont
  {L{\"o}wdin}},\ }\bibfield  {title} {\bibinfo {title} {Natural spin orbitals
  for helium},\ }\href {https://doi.org/10.1063/1.1742383} {\bibfield
  {journal} {\bibinfo  {journal} {J. Chem. Phys.}\ }\textbf {\bibinfo {volume}
  {23}},\ \bibinfo {pages} {1565} (\bibinfo {year}
  {1955}{\natexlab{a}})}\BibitemShut {NoStop}\bibitem [{\citenamefont {L{\"o}wdin}\ and\ \citenamefont
  {Shull}(1956)}]{loewdin1956:natural-orbital}\BibitemOpen
  \bibfield  {author} {\bibinfo {author} {\bibfnamefont {P.-O.}\ \bibnamefont
  {L{\"o}wdin}}\ and\ \bibinfo {author} {\bibfnamefont {H.}~\bibnamefont
  {Shull}},\ }\bibfield  {title} {\bibinfo {title} {Natural orbitals in the
  quantum theory of two-electron systems},\ }\href
  {https://doi.org/10.1103/PhysRev.101.1730} {\bibfield  {journal} {\bibinfo
  {journal} {Phys. Rev.}\ }\textbf {\bibinfo {volume} {101}},\ \bibinfo {pages}
  {1730} (\bibinfo {year} {1956})}\BibitemShut {NoStop}\bibitem [{\citenamefont {Davidson}(1972)}]{davidson1972:natural-orbital}\BibitemOpen
  \bibfield  {author} {\bibinfo {author} {\bibfnamefont {E.~R.}\ \bibnamefont
  {Davidson}},\ }\bibfield  {title} {\bibinfo {title} {Properties and uses of
  natural orbitals},\ }\href {https://doi.org/10.1103/RevModPhys.44.451}
  {\bibfield  {journal} {\bibinfo  {journal} {Rev. Mod. Phys.}\ }\textbf
  {\bibinfo {volume} {44}},\ \bibinfo {pages} {451} (\bibinfo {year}
  {1972})}\BibitemShut {NoStop}\bibitem [{\citenamefont {Mahaux}\ and\ \citenamefont
  {Sartor}(1991)}]{mahaux1991:single-particle}\BibitemOpen
  \bibfield  {author} {\bibinfo {author} {\bibfnamefont {C.}~\bibnamefont
  {Mahaux}}\ and\ \bibinfo {author} {\bibfnamefont {R.}~\bibnamefont
  {Sartor}},\ }\bibinfo {title} {Single-particle motion in nuclei},\ in\ \href
  {https://doi.org/10.1007/978-1-4613-9910-0_1} {\emph {\bibinfo {booktitle}
  {Adv. Nucl. Phys.}}},\ Vol.~\bibinfo {volume} {20},\ \bibinfo {editor}
  {edited by\ \bibinfo {editor} {\bibfnamefont {J.~W.}\ \bibnamefont {Negele}}\
  and\ \bibinfo {editor} {\bibfnamefont {E.}~\bibnamefont {Vogt}}}\ (\bibinfo
  {publisher} {Springer},\ \bibinfo {address} {Boston},\ \bibinfo {year}
  {1991})\ p.~\bibinfo {pages} {1}\BibitemShut {NoStop}\bibitem [{\citenamefont {Lalazissis}\ \emph {et~al.}(1992)\citenamefont
  {Lalazissis}, \citenamefont {Massen},\ and\ \citenamefont
  {Panos}}]{lalazissis1992:natural-orbital-shell}\BibitemOpen
  \bibfield  {author} {\bibinfo {author} {\bibfnamefont {G.~A.}\ \bibnamefont
  {Lalazissis}}, \bibinfo {author} {\bibfnamefont {S.~E.}\ \bibnamefont
  {Massen}},\ and\ \bibinfo {author} {\bibfnamefont {C.~P.}\ \bibnamefont
  {Panos}},\ }\bibfield  {title} {\bibinfo {title} {Systematic study of the
  effect of short range correlations on the occupation numbers of the shell
  model orbits in light nuclei},\ }\href
  {https://doi.org/10.1103/PhysRevC.46.201} {\bibfield  {journal} {\bibinfo
  {journal} {Phys. Rev. C}\ }\textbf {\bibinfo {volume} {46}},\ \bibinfo
  {pages} {201} (\bibinfo {year} {1992})}\BibitemShut {NoStop}\bibitem [{\citenamefont {Stoitsov}\ \emph
  {et~al.}(1993{\natexlab{a}})\citenamefont {Stoitsov}, \citenamefont
  {Antonov},\ and\ \citenamefont
  {Dimitrova}}]{stoitsov1993:natural-orbital-nuclei}\BibitemOpen
  \bibfield  {author} {\bibinfo {author} {\bibfnamefont {M.~V.}\ \bibnamefont
  {Stoitsov}}, \bibinfo {author} {\bibfnamefont {A.~N.}\ \bibnamefont
  {Antonov}},\ and\ \bibinfo {author} {\bibfnamefont {S.~S.}\ \bibnamefont
  {Dimitrova}},\ }\bibfield  {title} {\bibinfo {title} {Natural orbital
  representation in nuclei},\ }\href {https://doi.org/10.1103/PhysRevC.47.R455}
  {\bibfield  {journal} {\bibinfo  {journal} {Phys. Rev. C}\ }\textbf {\bibinfo
  {volume} {47}},\ \bibinfo {pages} {R455} (\bibinfo {year}
  {1993}{\natexlab{a}})}\BibitemShut {NoStop}\bibitem [{\citenamefont {Stoitsov}\ \emph
  {et~al.}(1993{\natexlab{b}})\citenamefont {Stoitsov}, \citenamefont
  {Antonov},\ and\ \citenamefont
  {Dimitrova}}]{stoitsov1993:natural-orbital-correlation}\BibitemOpen
  \bibfield  {author} {\bibinfo {author} {\bibfnamefont {M.~V.}\ \bibnamefont
  {Stoitsov}}, \bibinfo {author} {\bibfnamefont {A.~N.}\ \bibnamefont
  {Antonov}},\ and\ \bibinfo {author} {\bibfnamefont {S.~S.}\ \bibnamefont
  {Dimitrova}},\ }\bibfield  {title} {\bibinfo {title} {Natural orbital
  representation and short-range correlations in nuclei},\ }\href
  {https://doi.org/10.1103/PhysRevC.48.74} {\bibfield  {journal} {\bibinfo
  {journal} {Phys. Rev. C}\ }\textbf {\bibinfo {volume} {48}},\ \bibinfo
  {pages} {74} (\bibinfo {year} {1993}{\natexlab{b}})}\BibitemShut {NoStop}\bibitem [{\citenamefont {Shin}\ \emph {et~al.}(2017)\citenamefont {Shin},
  \citenamefont {Kim}, \citenamefont {Maris}, \citenamefont {Vary},
  \citenamefont {Forss\'en}, \citenamefont {Rotureau},\ and\ \citenamefont
  {Michel}}]{shin2017:6li-gsm-ncfc}\BibitemOpen
  \bibfield  {author} {\bibinfo {author} {\bibfnamefont {I.~J.}\ \bibnamefont
  {Shin}}, \bibinfo {author} {\bibfnamefont {Y.}~\bibnamefont {Kim}}, \bibinfo
  {author} {\bibfnamefont {P.}~\bibnamefont {Maris}}, \bibinfo {author}
  {\bibfnamefont {J.~P.}\ \bibnamefont {Vary}}, \bibinfo {author}
  {\bibfnamefont {C.}~\bibnamefont {Forss\'en}}, \bibinfo {author}
  {\bibfnamefont {J.}~\bibnamefont {Rotureau}},\ and\ \bibinfo {author}
  {\bibfnamefont {N.}~\bibnamefont {Michel}},\ }\bibfield  {title} {\bibinfo
  {title} {\textit{Ab initio} no-core solutions for \isotope[6]{Li}},\ }\href
  {https://doi.org/10.1088/1361-6471/aa6cb7} {\bibfield  {journal} {\bibinfo
  {journal} {J. Phys. G}\ }\textbf {\bibinfo {volume} {44}},\ \bibinfo {pages}
  {075103} (\bibinfo {year} {2017})}\BibitemShut {NoStop}\bibitem [{\citenamefont {Jaganathen}\ \emph {et~al.}(2017)\citenamefont
  {Jaganathen}, \citenamefont {{Id Betan}}, \citenamefont {Michel},
  \citenamefont {Nazarewicz},\ and\ \citenamefont
  {P{\l}oszajczak}}]{jaganathen2017:gamow-shell-psd-quantified}\BibitemOpen
  \bibfield  {author} {\bibinfo {author} {\bibfnamefont {Y.}~\bibnamefont
  {Jaganathen}}, \bibinfo {author} {\bibfnamefont {R.~M.}\ \bibnamefont {{Id
  Betan}}}, \bibinfo {author} {\bibfnamefont {N.}~\bibnamefont {Michel}},
  \bibinfo {author} {\bibfnamefont {W.}~\bibnamefont {Nazarewicz}},\ and\
  \bibinfo {author} {\bibfnamefont {M.}~\bibnamefont {P{\l}oszajczak}},\
  }\bibfield  {title} {\bibinfo {title} {Quantified {G}amow shell model
  interaction for $psd$-shell nuclei},\ }\href
  {https://doi.org/10.1103/PhysRevC.96.054316} {\bibfield  {journal} {\bibinfo
  {journal} {Phys. Rev. C}\ }\textbf {\bibinfo {volume} {96}},\ \bibinfo
  {pages} {054316} (\bibinfo {year} {2017})}\BibitemShut {NoStop}\bibitem [{\citenamefont {L{\"o}wdin}(1960)}]{loewdin1960:natural-orbital}\BibitemOpen
  \bibfield  {author} {\bibinfo {author} {\bibfnamefont {P.-O.}\ \bibnamefont
  {L{\"o}wdin}},\ }\bibfield  {title} {\bibinfo {title} {Expansion theorems for
  the total wave function and extended {H}artree-{F}ock schemes},\ }\href
  {https://doi.org/10.1103/RevModPhys.32.328} {\bibfield  {journal} {\bibinfo
  {journal} {Rev. Mod. Phys.}\ }\textbf {\bibinfo {volume} {32}},\ \bibinfo
  {pages} {328} (\bibinfo {year} {1960})}\BibitemShut {NoStop}\bibitem [{\citenamefont {Kobe}(1969)}]{kobe1969:natural-orbital-variational}\BibitemOpen
  \bibfield  {author} {\bibinfo {author} {\bibfnamefont {D.~H.}\ \bibnamefont
  {Kobe}},\ }\bibfield  {title} {\bibinfo {title} {Natural orbitals,
  divergences, and variational principles},\ }\href
  {https://doi.org/10.1063/1.1671034} {\bibfield  {journal} {\bibinfo
  {journal} {J. Chem. Phys.}\ }\textbf {\bibinfo {volume} {50}},\ \bibinfo
  {pages} {5183} (\bibinfo {year} {1969})}\BibitemShut {NoStop}\bibitem [{\citenamefont {Constantinou}\ \emph {et~al.}(2017)\citenamefont
  {Constantinou}, \citenamefont {Caprio}, \citenamefont {Vary},\ and\
  \citenamefont {Maris}}]{constantinou2017:natorb-natowitz16}\BibitemOpen
  \bibfield  {author} {\bibinfo {author} {\bibfnamefont
  {{\mbox{Ch}}.}~\bibnamefont {Constantinou}}, \bibinfo {author} {\bibfnamefont
  {M.~A.}\ \bibnamefont {Caprio}}, \bibinfo {author} {\bibfnamefont {J.~P.}\
  \bibnamefont {Vary}},\ and\ \bibinfo {author} {\bibfnamefont
  {P.}~\bibnamefont {Maris}},\ }\bibfield  {title} {\bibinfo {title}
  {\textit{Ab initio} properties of the halo nucleus $\isotope[6]{He}$ in a
  natural orbital basis},\ }\href {https://doi.org/10.1007/s41365-017-0332-6}
  {\bibfield  {journal} {\bibinfo  {journal} {Nucl. Sci. Techniques}\ }\textbf
  {\bibinfo {volume} {28}},\ \bibinfo {pages} {179} (\bibinfo {year}
  {2017})}\BibitemShut {NoStop}\bibitem [{\citenamefont {Constantinou}(2017)}]{constantinou2017:diss}\BibitemOpen
  \bibfield  {author} {\bibinfo {author} {\bibfnamefont
  {{\mbox{Ch}}.}~\bibnamefont {Constantinou}},\ }\emph {\bibinfo {title}
  {Natural orbitals for the no-core configuration interaction approach}},\
  \href {https://curate.nd.edu/show/ff365427x19} {Ph.D. thesis},\ \bibinfo
  {school} {University of Notre Dame} (\bibinfo {year} {2017})\BibitemShut
  {NoStop}\bibitem [{\citenamefont {Puddu}(2018)}]{puddu2018:deuteron-natural-orbitals}\BibitemOpen
  \bibfield  {author} {\bibinfo {author} {\bibfnamefont {G.}~\bibnamefont
  {Puddu}},\ }\bibfield  {title} {\bibinfo {title} {Many-body calculations with
  deuteron based single-particle bases and their associated natural orbits},\
  }\href {https://doi.org/10.1088/1402-4896/aabbf3} {\bibfield  {journal}
  {\bibinfo  {journal} {Physica Scripta}\ }\textbf {\bibinfo {volume} {93}},\
  \bibinfo {pages} {065301} (\bibinfo {year} {2018})}\BibitemShut {NoStop}\bibitem [{\citenamefont {Tichai}\ \emph {et~al.}(2019)\citenamefont {Tichai},
  \citenamefont {M{\"u}ller}, \citenamefont {Vobig},\ and\ \citenamefont
  {Roth}}]{tichai2019:natorb-mbpt}\BibitemOpen
  \bibfield  {author} {\bibinfo {author} {\bibfnamefont {A.}~\bibnamefont
  {Tichai}}, \bibinfo {author} {\bibfnamefont {J.}~\bibnamefont {M{\"u}ller}},
  \bibinfo {author} {\bibfnamefont {K.}~\bibnamefont {Vobig}},\ and\ \bibinfo
  {author} {\bibfnamefont {R.}~\bibnamefont {Roth}},\ }\bibfield  {title}
  {\bibinfo {title} {Natural orbitals for \textit{ab initio} no-core shell
  model calculations},\ }\href {https://doi.org/10.1103/PhysRevC.99.034321}
  {\bibfield  {journal} {\bibinfo  {journal} {Phys. Rev. C}\ }\textbf {\bibinfo
  {volume} {99}},\ \bibinfo {pages} {034321} (\bibinfo {year}
  {2019})}\BibitemShut {NoStop}\bibitem [{\citenamefont {Hoppe}\ \emph {et~al.}(2021)\citenamefont {Hoppe},
  \citenamefont {Tichai}, \citenamefont {Heinz}, \citenamefont {Hebeler},\ and\
  \citenamefont {Schwenk}}]{hoppe2021:natorb-mbpt-imsrg}\BibitemOpen
  \bibfield  {author} {\bibinfo {author} {\bibfnamefont {J.}~\bibnamefont
  {Hoppe}}, \bibinfo {author} {\bibfnamefont {A.}~\bibnamefont {Tichai}},
  \bibinfo {author} {\bibfnamefont {M.}~\bibnamefont {Heinz}}, \bibinfo
  {author} {\bibfnamefont {K.}~\bibnamefont {Hebeler}},\ and\ \bibinfo {author}
  {\bibfnamefont {A.}~\bibnamefont {Schwenk}},\ }\bibfield  {title} {\bibinfo
  {title} {Natural orbitals for many-body expansion methods},\ }\href
  {https://doi.org/10.1103/PhysRevC.103.014321} {\bibfield  {journal} {\bibinfo
   {journal} {Phys. Rev. C}\ }\textbf {\bibinfo {volume} {103}},\ \bibinfo
  {pages} {014321} (\bibinfo {year} {2021})}\BibitemShut {NoStop}\bibitem [{\citenamefont {Robin}\ \emph {et~al.}(2021)\citenamefont {Robin},
  \citenamefont {Savage},\ and\ \citenamefont
  {Pillet}}]{robin2021:6he-entanglement}\BibitemOpen
  \bibfield  {author} {\bibinfo {author} {\bibfnamefont {C.}~\bibnamefont
  {Robin}}, \bibinfo {author} {\bibfnamefont {M.~J.}\ \bibnamefont {Savage}},\
  and\ \bibinfo {author} {\bibfnamefont {N.}~\bibnamefont {Pillet}},\
  }\bibfield  {title} {\bibinfo {title} {Entanglement rearrangement in
  self-consistent nuclear structure calculations},\ }\href
  {https://doi.org/10.1103/PhysRevC.103.034325} {\bibfield  {journal} {\bibinfo
   {journal} {Phys. Rev. C}\ }\textbf {\bibinfo {volume} {103}},\ \bibinfo
  {pages} {034325} (\bibinfo {year} {2021})}\BibitemShut {NoStop}\bibitem [{\citenamefont {Shirokov}\ \emph {et~al.}(2016)\citenamefont
  {Shirokov}, \citenamefont {Shin}, \citenamefont {Kim}, \citenamefont
  {Sosonkina}, \citenamefont {Maris},\ and\ \citenamefont
  {Vary}}]{shirokov2016:nn-daejeon16}\BibitemOpen
  \bibfield  {author} {\bibinfo {author} {\bibfnamefont {A.~M.}\ \bibnamefont
  {Shirokov}}, \bibinfo {author} {\bibfnamefont {I.~J.}\ \bibnamefont {Shin}},
  \bibinfo {author} {\bibfnamefont {Y.}~\bibnamefont {Kim}}, \bibinfo {author}
  {\bibfnamefont {M.}~\bibnamefont {Sosonkina}}, \bibinfo {author}
  {\bibfnamefont {P.}~\bibnamefont {Maris}},\ and\ \bibinfo {author}
  {\bibfnamefont {J.~P.}\ \bibnamefont {Vary}},\ }\bibfield  {title} {\bibinfo
  {title} {{N3LO} {$NN$} interaction adjusted to light nuclei in \textit{ab
  exitu} approach},\ }\href {https://doi.org/10.1016/j.physletb.2016.08.006}
  {\bibfield  {journal} {\bibinfo  {journal} {Phys. Lett. B}\ }\textbf
  {\bibinfo {volume} {761}},\ \bibinfo {pages} {87} (\bibinfo {year}
  {2016})}\BibitemShut {NoStop}\bibitem [{\citenamefont {Maris}\ \emph {et~al.}(2019)\citenamefont {Maris},
  \citenamefont {Shin},\ and\ \citenamefont
  {Vary}}]{maris2019:daejeon16-lenpic-pshell-ntse18}\BibitemOpen
  \bibfield  {author} {\bibinfo {author} {\bibfnamefont {P.}~\bibnamefont
  {Maris}}, \bibinfo {author} {\bibfnamefont {I.~J.}\ \bibnamefont {Shin}},\
  and\ \bibinfo {author} {\bibfnamefont {J.~P.}\ \bibnamefont {Vary}},\
  }\bibfield  {title} {\bibinfo {title} {\textit{Ab initio} structure of
  $p$-shell nuclei with chiral effective field theory and {D}aejeon16
  interactions},\ }in\ \href
  {http://www.ntse.khb.ru/files/uploads/2018/proceedings/Vary.pdf} {\emph
  {\bibinfo {booktitle} {Proceedings of the International Conference Nuclear
  Theory in the Supercomputing Era 2018}}},\ \bibinfo {editor} {edited by\
  \bibinfo {editor} {\bibfnamefont {A.~M.}\ \bibnamefont {Shirokov}}\ and\
  \bibinfo {editor} {\bibfnamefont {A.~I.}\ \bibnamefont {Mazur}}}\ (\bibinfo
  {publisher} {Pacific National University, Khabarovsk, Russia},\ \bibinfo
  {year} {2019})\ p.\ \bibinfo {pages} {168}\BibitemShut {NoStop}\bibitem [{\citenamefont {Ring}\ and\ \citenamefont
  {Schuck}(1980)}]{ring1980:nuclear-many-body}\BibitemOpen
  \bibfield  {author} {\bibinfo {author} {\bibfnamefont {P.}~\bibnamefont
  {Ring}}\ and\ \bibinfo {author} {\bibfnamefont {P.}~\bibnamefont {Schuck}},\
  }\href@noop {} {\emph {\bibinfo {title} {The Nuclear Many-Body Problem}}}\
  (\bibinfo  {publisher} {Springer-Verlag},\ \bibinfo {address} {New York},\
  \bibinfo {year} {1980})\BibitemShut {NoStop}\bibitem [{\citenamefont {Coleman}\ and\ \citenamefont
  {Yukalov}(2000)}]{coleman2000:reduced-density-matrices}\BibitemOpen
  \bibfield  {author} {\bibinfo {author} {\bibfnamefont {A.~J.}\ \bibnamefont
  {Coleman}}\ and\ \bibinfo {author} {\bibfnamefont {V.~I.}\ \bibnamefont
  {Yukalov}},\ }\href@noop {} {\emph {\bibinfo {title} {Reduced Density
  Matrices}}},\ \bibinfo {series} {Lecture Notes in Chemistry}, Vol.~\bibinfo
  {volume} {72}\ (\bibinfo  {publisher} {Springer},\ \bibinfo {address}
  {Berlin},\ \bibinfo {year} {2000})\BibitemShut {NoStop}\bibitem [{\citenamefont {McWeeny}\ and\ \citenamefont
  {Kutzelnigg}(1968)}]{mcweeny1968:symmetry-natorb-part1-adapted}\BibitemOpen
  \bibfield  {author} {\bibinfo {author} {\bibfnamefont {R.}~\bibnamefont
  {McWeeny}}\ and\ \bibinfo {author} {\bibfnamefont {W.}~\bibnamefont
  {Kutzelnigg}},\ }\bibfield  {title} {\bibinfo {title} {Symmetry properties of
  natural orbitals and geminals {I}. {C}onstruction of spin- and
  symmetry-adapted functions},\ }\href {https://doi.org/10.1002/qua.560020203}
  {\bibfield  {journal} {\bibinfo  {journal} {Int. J. Quantum. Chem.}\ }\textbf
  {\bibinfo {volume} {2}},\ \bibinfo {pages} {187} (\bibinfo {year}
  {1968})}\BibitemShut {NoStop}\bibitem [{\citenamefont {Suhonen}(2007)}]{suhonen2007:nucleons-nucleus}\BibitemOpen
  \bibfield  {author} {\bibinfo {author} {\bibfnamefont {J.}~\bibnamefont
  {Suhonen}},\ }\href {https://doi.org/10.1007/978-3-540-48861-3} {\emph
  {\bibinfo {title} {From Nucleons to Nucleus}}}\ (\bibinfo  {publisher}
  {Springer-Verlag},\ \bibinfo {address} {Berlin},\ \bibinfo {year}
  {2007})\BibitemShut {NoStop}\bibitem [{\citenamefont {Sakurai}(1994)}]{sakurai1994:qm}\BibitemOpen
  \bibfield  {author} {\bibinfo {author} {\bibfnamefont {J.~J.}\ \bibnamefont
  {Sakurai}},\ }\href@noop {} {\emph {\bibinfo {title} {Modern Quantum
  Mechanics}}},\ \bibinfo {edition} {rev.}\ ed.,\ edited by\ \bibinfo {editor}
  {\bibfnamefont {S.~F.}\ \bibnamefont {Tuan}}\ (\bibinfo  {publisher}
  {Addison-Wesley},\ \bibinfo {address} {Reading, Massachusetts},\ \bibinfo
  {year} {1994})\BibitemShut {NoStop}\bibitem [{\citenamefont {Dirac}(1930)}]{dirac1930:thomas-atom-exchange}\BibitemOpen
  \bibfield  {author} {\bibinfo {author} {\bibfnamefont {P.~A.~M.}\
  \bibnamefont {Dirac}},\ }\bibfield  {title} {\bibinfo {title} {Note on
  exchange phenomena in the {T}homas atom},\ }\href
  {https://doi.org/10.1017/S0305004100016108} {\bibfield  {journal} {\bibinfo
  {journal} {Math. Proc. Cambridge Phil. Soc.}\ }\textbf {\bibinfo {volume}
  {26}},\ \bibinfo {pages} {376} (\bibinfo {year} {1930})}\BibitemShut
  {NoStop}\bibitem [{\citenamefont
  {Fan}(1949)}]{fan1949:transformation-eigenvalues-part1}\BibitemOpen
  \bibfield  {author} {\bibinfo {author} {\bibfnamefont {K.}~\bibnamefont
  {Fan}},\ }\bibfield  {title} {\bibinfo {title} {On a theorem of {W}eyl
  concerning eigenvalues of linear transformations {I}},\ }\href
  {https://doi.org/10.1073/pnas.35.11.652} {\bibfield  {journal} {\bibinfo
  {journal} {Proc. Nat. Acad. Sci. USA}\ }\textbf {\bibinfo {volume} {35}},\
  \bibinfo {pages} {652} (\bibinfo {year} {1949})}\BibitemShut {NoStop}\bibitem [{\citenamefont {Whitehead}\ \emph {et~al.}(1977)\citenamefont
  {Whitehead}, \citenamefont {Watt}, \citenamefont {Cole},\ and\ \citenamefont
  {Morrison}}]{whitehead1977:shell-methods}\BibitemOpen
  \bibfield  {author} {\bibinfo {author} {\bibfnamefont {R.~R.}\ \bibnamefont
  {Whitehead}}, \bibinfo {author} {\bibfnamefont {A.}~\bibnamefont {Watt}},
  \bibinfo {author} {\bibfnamefont {B.~J.}\ \bibnamefont {Cole}},\ and\
  \bibinfo {author} {\bibfnamefont {I.}~\bibnamefont {Morrison}},\ }\bibfield
  {title} {\bibinfo {title} {Computational methods for shell-model
  calculations},\ }\href {https://doi.org/10.1007/978-1-4615-8234-2_2}
  {\bibfield  {journal} {\bibinfo  {journal} {Adv. Nucl. Phys.}\ }\textbf
  {\bibinfo {volume} {9}},\ \bibinfo {pages} {123} (\bibinfo {year}
  {1977})}\BibitemShut {NoStop}\bibitem [{\citenamefont {Edmonds}(1960)}]{edmonds1960:am}\BibitemOpen
  \bibfield  {author} {\bibinfo {author} {\bibfnamefont {A.~R.}\ \bibnamefont
  {Edmonds}},\ }\href@noop {} {\emph {\bibinfo {title} {Angular Momentum in
  Quantum Mechanics}}},\ \bibinfo {edition} {2nd}\ ed.,\ \bibinfo {series}
  {Investigations in Physics}\ No.~\bibinfo {number} {4}\ (\bibinfo
  {publisher} {Princeton University Press},\ \bibinfo {address} {Princeton, New
  Jersey},\ \bibinfo {year} {1960})\BibitemShut {NoStop}\bibitem [{\citenamefont {Rowe}\ and\ \citenamefont
  {Wood}(2010)}]{rowe2010:rowanwood}\BibitemOpen
  \bibfield  {author} {\bibinfo {author} {\bibfnamefont {D.~J.}\ \bibnamefont
  {Rowe}}\ and\ \bibinfo {author} {\bibfnamefont {J.~L.}\ \bibnamefont
  {Wood}},\ }\href {https://doi.org/10.1142/6209} {\emph {\bibinfo {title}
  {Fundamentals of Nuclear Models: Foundational Models}}}\ (\bibinfo
  {publisher} {World Scientific},\ \bibinfo {address} {Singapore},\ \bibinfo
  {year} {2010})\BibitemShut {NoStop}\bibitem [{\citenamefont {Bethe}\ and\ \citenamefont
  {Rose}(1937)}]{bethe1937:nuclear-kinetic-com}\BibitemOpen
  \bibfield  {author} {\bibinfo {author} {\bibfnamefont {H.~A.}\ \bibnamefont
  {Bethe}}\ and\ \bibinfo {author} {\bibfnamefont {M.~E.}\ \bibnamefont
  {Rose}},\ }\bibfield  {title} {\bibinfo {title} {Kinetic energy of nuclei in
  the {H}artree model},\ }\href {https://doi.org/10.1103/PhysRev.51.283}
  {\bibfield  {journal} {\bibinfo  {journal} {Phys. Rev.}\ }\textbf {\bibinfo
  {volume} {51}},\ \bibinfo {pages} {283} (\bibinfo {year} {1937})}\BibitemShut
  {NoStop}\bibitem [{\citenamefont {Brussaard}\ and\ \citenamefont
  {Glaudemans}(1977)}]{brussaard1977:shell}\BibitemOpen
  \bibfield  {author} {\bibinfo {author} {\bibfnamefont {P.~J.}\ \bibnamefont
  {Brussaard}}\ and\ \bibinfo {author} {\bibfnamefont {P.~W.~M.}\ \bibnamefont
  {Glaudemans}},\ }\href@noop {} {\emph {\bibinfo {title} {Shell-Model
  Applications in Nuclear Spectroscopy}}}\ (\bibinfo  {publisher}
  {North-Holland Publishing Company},\ \bibinfo {address} {Amsterdam},\
  \bibinfo {year} {1977})\BibitemShut {NoStop}\bibitem [{\citenamefont {Gloeckner}\ and\ \citenamefont
  {Lawson}(1974)}]{gloeckner1974:spurious-com}\BibitemOpen
  \bibfield  {author} {\bibinfo {author} {\bibfnamefont {D.~H.}\ \bibnamefont
  {Gloeckner}}\ and\ \bibinfo {author} {\bibfnamefont {R.~D.}\ \bibnamefont
  {Lawson}},\ }\bibfield  {title} {\bibinfo {title} {Spurious center-of-mass
  motion},\ }\href {https://doi.org/10.1016/0370-2693(74)90390-6} {\bibfield
  {journal} {\bibinfo  {journal} {Phys. Lett. B}\ }\textbf {\bibinfo {volume}
  {53}},\ \bibinfo {pages} {313} (\bibinfo {year} {1974})}\BibitemShut
  {NoStop}\bibitem [{\citenamefont {Lawson}(1980)}]{lawson1980:shell}\BibitemOpen
  \bibfield  {author} {\bibinfo {author} {\bibfnamefont {R.~D.}\ \bibnamefont
  {Lawson}},\ }\href@noop {} {\emph {\bibinfo {title} {Theory of the Nuclear
  Shell Model}}}\ (\bibinfo  {publisher} {Clarendon Press},\ \bibinfo {address}
  {Oxford},\ \bibinfo {year} {1980})\BibitemShut {NoStop}\bibitem [{\citenamefont {Lane}(1960)}]{lane1960:reduced-widths}\BibitemOpen
  \bibfield  {author} {\bibinfo {author} {\bibfnamefont {A.~M.}\ \bibnamefont
  {Lane}},\ }\bibfield  {title} {\bibinfo {title} {Reduced widths of individual
  nuclear energy levels},\ }\href {https://doi.org/10.1103/RevModPhys.32.519}
  {\bibfield  {journal} {\bibinfo  {journal} {Rev. Mod. Phys.}\ }\textbf
  {\bibinfo {volume} {32}},\ \bibinfo {pages} {519} (\bibinfo {year}
  {1960})}\BibitemShut {NoStop}\bibitem [{\citenamefont {Bogner}\ \emph {et~al.}(2008)\citenamefont {Bogner},
  \citenamefont {Furnstahl}, \citenamefont {Maris}, \citenamefont {Perry},
  \citenamefont {Schwenk},\ and\ \citenamefont
  {Vary}}]{bogner2008:ncsm-converg-2N}\BibitemOpen
  \bibfield  {author} {\bibinfo {author} {\bibfnamefont {S.~K.}\ \bibnamefont
  {Bogner}}, \bibinfo {author} {\bibfnamefont {R.~J.}\ \bibnamefont
  {Furnstahl}}, \bibinfo {author} {\bibfnamefont {P.}~\bibnamefont {Maris}},
  \bibinfo {author} {\bibfnamefont {R.~J.}\ \bibnamefont {Perry}}, \bibinfo
  {author} {\bibfnamefont {A.}~\bibnamefont {Schwenk}},\ and\ \bibinfo {author}
  {\bibfnamefont {J.}~\bibnamefont {Vary}},\ }\bibfield  {title} {\bibinfo
  {title} {Convergence in the no-core shell model with low-momentum two-nucleon
  interactions},\ }\href {https://doi.org/10.1016/j.nuclphysa.2007.12.008}
  {\bibfield  {journal} {\bibinfo  {journal} {Nucl. Phys. A}\ }\textbf
  {\bibinfo {volume} {801}},\ \bibinfo {pages} {21} (\bibinfo {year}
  {2008})}\BibitemShut {NoStop}\bibitem [{\citenamefont {Maris}\ and\ \citenamefont
  {Vary}(2013)}]{maris2013:ncsm-pshell}\BibitemOpen
  \bibfield  {author} {\bibinfo {author} {\bibfnamefont {P.}~\bibnamefont
  {Maris}}\ and\ \bibinfo {author} {\bibfnamefont {J.~P.}\ \bibnamefont
  {Vary}},\ }\bibfield  {title} {\bibinfo {title} {\textit{Ab initio} nuclear
  structure calculations of $p$-shell nuclei with {JISP16}},\ }\href
  {https://doi.org/10.1142/S0218301313300166} {\bibfield  {journal} {\bibinfo
  {journal} {Int. J. Mod. Phys. E}\ }\textbf {\bibinfo {volume} {22}},\
  \bibinfo {pages} {1330016} (\bibinfo {year} {2013})}\BibitemShut {NoStop}\bibitem [{\citenamefont {Caprio}\ \emph {et~al.}(2015)\citenamefont {Caprio},
  \citenamefont {Maris}, \citenamefont {Vary},\ and\ \citenamefont
  {Smith}}]{caprio2015:berotor-ijmpe}\BibitemOpen
  \bibfield  {author} {\bibinfo {author} {\bibfnamefont {M.~A.}\ \bibnamefont
  {Caprio}}, \bibinfo {author} {\bibfnamefont {P.}~\bibnamefont {Maris}},
  \bibinfo {author} {\bibfnamefont {J.~P.}\ \bibnamefont {Vary}},\ and\
  \bibinfo {author} {\bibfnamefont {R.}~\bibnamefont {Smith}},\ }\bibfield
  {title} {\bibinfo {title} {Collective rotation from \textit{ab initio}
  theory},\ }\href {https://doi.org/10.1142/S0218301315410025} {\bibfield
  {journal} {\bibinfo  {journal} {Int. J. Mod. Phys. E}\ }\textbf {\bibinfo
  {volume} {24}},\ \bibinfo {pages} {1541002} (\bibinfo {year}
  {2015})}\BibitemShut {NoStop}\bibitem [{\citenamefont {Caprio}\ \emph {et~al.}(2021)\citenamefont {Caprio},
  \citenamefont {Fasano}, \citenamefont {Maris},\ and\ \citenamefont
  {McCoy}}]{caprio2021:emratio}\BibitemOpen
  \bibfield  {author} {\bibinfo {author} {\bibfnamefont {M.~A.}\ \bibnamefont
  {Caprio}}, \bibinfo {author} {\bibfnamefont {P.~J.}\ \bibnamefont {Fasano}},
  \bibinfo {author} {\bibfnamefont {P.}~\bibnamefont {Maris}},\ and\ \bibinfo
  {author} {\bibfnamefont {A.~E.}\ \bibnamefont {McCoy}},\ }\bibfield  {title}
  {\bibinfo {title} {Quadrupole moments and proton-neutron structure in
  $p$-shell mirror nuclei},\ }\href
  {https://doi.org/10.1103/PhysRevC.104.034319} {\bibfield  {journal} {\bibinfo
   {journal} {Phys. Rev. C}\ }\textbf {\bibinfo {volume} {104}},\ \bibinfo
  {pages} {034319} (\bibinfo {year} {2021})}\BibitemShut {NoStop}\bibitem [{\citenamefont {Hagen}\ \emph {et~al.}(2009)\citenamefont {Hagen},
  \citenamefont {Papenbrock},\ and\ \citenamefont
  {Dean}}]{hagen2009:coupled-cluster-com}\BibitemOpen
  \bibfield  {author} {\bibinfo {author} {\bibfnamefont {G.}~\bibnamefont
  {Hagen}}, \bibinfo {author} {\bibfnamefont {T.}~\bibnamefont {Papenbrock}},\
  and\ \bibinfo {author} {\bibfnamefont {D.~J.}\ \bibnamefont {Dean}},\
  }\bibfield  {title} {\bibinfo {title} {Solution of the center-of-mass problem
  in nuclear structure calculations},\ }\href
  {https://doi.org/10.1103/PhysRevLett.102.062503} {\bibfield  {journal}
  {\bibinfo  {journal} {Phys. Rev. Lett.}\ }\textbf {\bibinfo {volume} {103}},\
  \bibinfo {pages} {062503} (\bibinfo {year} {2009})}\BibitemShut {NoStop}\bibitem [{\citenamefont {Hagen}\ \emph {et~al.}(2010)\citenamefont {Hagen},
  \citenamefont {Papenbrock}, \citenamefont {Dean},\ and\ \citenamefont
  {Hjorth-Jensen}}]{hagen2010:coupled-cluster}\BibitemOpen
  \bibfield  {author} {\bibinfo {author} {\bibfnamefont {G.}~\bibnamefont
  {Hagen}}, \bibinfo {author} {\bibfnamefont {T.}~\bibnamefont {Papenbrock}},
  \bibinfo {author} {\bibfnamefont {D.~J.}\ \bibnamefont {Dean}},\ and\
  \bibinfo {author} {\bibfnamefont {M.}~\bibnamefont {Hjorth-Jensen}},\
  }\bibfield  {title} {\bibinfo {title} {\textit{Ab initio} coupled-cluster
  approach to nuclear structure with modern nucleon-nucleon interactions},\
  }\href {https://doi.org/10.1103/PhysRevC.82.034330} {\bibfield  {journal}
  {\bibinfo  {journal} {Phys. Rev. C}\ }\textbf {\bibinfo {volume} {82}},\
  \bibinfo {pages} {034330} (\bibinfo {year} {2010})}\BibitemShut {NoStop}\bibitem [{\citenamefont {Roth}\ \emph {et~al.}(2009)\citenamefont {Roth},
  \citenamefont {Gour},\ and\ \citenamefont
  {Piecuch}}]{roth2009:it-ncsm-cc-cm-truncated}\BibitemOpen
  \bibfield  {author} {\bibinfo {author} {\bibfnamefont {R.}~\bibnamefont
  {Roth}}, \bibinfo {author} {\bibfnamefont {J.~R.}\ \bibnamefont {Gour}},\
  and\ \bibinfo {author} {\bibfnamefont {P.}~\bibnamefont {Piecuch}},\
  }\bibfield  {title} {\bibinfo {title} {Center-of-mass problem in truncated
  configuration interaction and coupled-cluster calculations},\ }\href
  {https://doi.org/10.1016/j.physletb.2009.07.071} {\bibfield  {journal}
  {\bibinfo  {journal} {Phys. Lett. B}\ }\textbf {\bibinfo {volume} {679}},\
  \bibinfo {pages} {334} (\bibinfo {year} {2009})}\BibitemShut {NoStop}\bibitem [{\citenamefont {Hergert}\ \emph {et~al.}(2016)\citenamefont
  {Hergert}, \citenamefont {Bogner}, \citenamefont {Morris}, \citenamefont
  {Schwenk},\ and\ \citenamefont {Tsukiyama}}]{hergert2016:imsrg}\BibitemOpen
  \bibfield  {author} {\bibinfo {author} {\bibfnamefont {H.}~\bibnamefont
  {Hergert}}, \bibinfo {author} {\bibfnamefont {S.~K.}\ \bibnamefont {Bogner}},
  \bibinfo {author} {\bibfnamefont {T.~D.}\ \bibnamefont {Morris}}, \bibinfo
  {author} {\bibfnamefont {A.}~\bibnamefont {Schwenk}},\ and\ \bibinfo {author}
  {\bibfnamefont {K.}~\bibnamefont {Tsukiyama}},\ }\bibfield  {title} {\bibinfo
  {title} {The in-medium similarity renormalization group: A novel \textit{ab
  initio} method for nuclei},\ }\href
  {https://doi.org/10.1016/j.physrep.2015.12.007} {\bibfield  {journal}
  {\bibinfo  {journal} {Phys. Rep.}\ }\textbf {\bibinfo {volume} {621}},\
  \bibinfo {pages} {165} (\bibinfo {year} {2016})}\BibitemShut {NoStop}\bibitem [{\citenamefont {Shull}\ and\ \citenamefont
  {L{\"o}wdin}(1955{\natexlab{b}})}]{shull1955-continuum}\BibitemOpen
  \bibfield  {author} {\bibinfo {author} {\bibfnamefont {H.}~\bibnamefont
  {Shull}}\ and\ \bibinfo {author} {\bibfnamefont {P.-O.}\ \bibnamefont
  {L{\"o}wdin}},\ }\bibfield  {title} {\bibinfo {title} {Role of the continuum
  in superposition of configurations},\ }\href
  {https://doi.org/10.1063/1.1742296} {\bibfield  {journal} {\bibinfo
  {journal} {J. Chem. Phys.}\ }\textbf {\bibinfo {volume} {23}},\ \bibinfo
  {pages} {1362} (\bibinfo {year} {1955}{\natexlab{b}})}\BibitemShut {NoStop}\bibitem [{\citenamefont {Weniger}(1985)}]{weniger1985:fourier-plane-wave}\BibitemOpen
  \bibfield  {author} {\bibinfo {author} {\bibfnamefont {E.~J.}\ \bibnamefont
  {Weniger}},\ }\bibfield  {title} {\bibinfo {title} {Weakly convergent
  expansions of a plane wave and their use in {F}ourier integrals},\ }\href
  {https://doi.org/10.1063/1.526970} {\bibfield  {journal} {\bibinfo  {journal}
  {J. Math. Phys.}\ }\textbf {\bibinfo {volume} {26}},\ \bibinfo {pages} {276}
  (\bibinfo {year} {1985})}\BibitemShut {NoStop}\bibitem [{\citenamefont {McCoy}\ and\ \citenamefont
  {Caprio}(2016)}]{mccoy2016:lgalg}\BibitemOpen
  \bibfield  {author} {\bibinfo {author} {\bibfnamefont {A.~E.}\ \bibnamefont
  {McCoy}}\ and\ \bibinfo {author} {\bibfnamefont {M.~A.}\ \bibnamefont
  {Caprio}},\ }\bibfield  {title} {\bibinfo {title} {Algebraic evaluation of
  matrix elements in the {L}aguerre function basis},\ }\href
  {https://doi.org/10.1063/1.4941327} {\bibfield  {journal} {\bibinfo
  {journal} {J. Math. Phys.}\ }\textbf {\bibinfo {volume} {57}},\ \bibinfo
  {pages} {021708} (\bibinfo {year} {2016})}\BibitemShut {NoStop}\bibitem [{\citenamefont {Negele}\ and\ \citenamefont
  {Orland}(1988)}]{negele1988:many-particle}\BibitemOpen
  \bibfield  {author} {\bibinfo {author} {\bibfnamefont {J.~W.}\ \bibnamefont
  {Negele}}\ and\ \bibinfo {author} {\bibfnamefont {H.}~\bibnamefont
  {Orland}},\ }\href@noop {} {\emph {\bibinfo {title} {Quantum Many-Particle
  Systems}}}\ (\bibinfo  {publisher} {Addison-Wesley},\ \bibinfo {address}
  {Redwood City, CA},\ \bibinfo {year} {1988})\BibitemShut {NoStop}\bibitem [{\citenamefont {Abe}\ \emph {et~al.}(2012)\citenamefont {Abe},
  \citenamefont {Maris}, \citenamefont {Otsuka}, \citenamefont {Shimizu},
  \citenamefont {Utsuno},\ and\ \citenamefont {Vary}}]{abe2012:fci-mcsm-ncfc}\BibitemOpen
  \bibfield  {author} {\bibinfo {author} {\bibfnamefont {T.}~\bibnamefont
  {Abe}}, \bibinfo {author} {\bibfnamefont {P.}~\bibnamefont {Maris}}, \bibinfo
  {author} {\bibfnamefont {T.}~\bibnamefont {Otsuka}}, \bibinfo {author}
  {\bibfnamefont {N.}~\bibnamefont {Shimizu}}, \bibinfo {author} {\bibfnamefont
  {Y.}~\bibnamefont {Utsuno}},\ and\ \bibinfo {author} {\bibfnamefont {J.~P.}\
  \bibnamefont {Vary}},\ }\bibfield  {title} {\bibinfo {title} {Benchmarks of
  the full configuration interaction, {M}onte {C}arlo shell model, and no-core
  full configuration methods},\ }\href
  {https://doi.org/10.1103/PhysRevC.86.054301} {\bibfield  {journal} {\bibinfo
  {journal} {Phys. Rev. C}\ }\textbf {\bibinfo {volume} {86}},\ \bibinfo
  {pages} {054301} (\bibinfo {year} {2012})}\BibitemShut {NoStop}\bibitem [{\citenamefont {Dytrych}\ \emph {et~al.}(2008)\citenamefont
  {Dytrych}, \citenamefont {Sviratcheva}, \citenamefont {Draayer},
  \citenamefont {Bahri},\ and\ \citenamefont {Vary}}]{dytrych2008:sp-ncsm}\BibitemOpen
  \bibfield  {author} {\bibinfo {author} {\bibfnamefont {T.}~\bibnamefont
  {Dytrych}}, \bibinfo {author} {\bibfnamefont {K.~D.}\ \bibnamefont
  {Sviratcheva}}, \bibinfo {author} {\bibfnamefont {J.~P.}\ \bibnamefont
  {Draayer}}, \bibinfo {author} {\bibfnamefont {C.}~\bibnamefont {Bahri}},\
  and\ \bibinfo {author} {\bibfnamefont {J.~P.}\ \bibnamefont {Vary}},\
  }\bibfield  {title} {\bibinfo {title} {\textit{Ab initio} symplectic no-core
  shell model},\ }\href {https://doi.org/10.1088/0954-3899/35/12/123101}
  {\bibfield  {journal} {\bibinfo  {journal} {J. Phys. G}\ }\textbf {\bibinfo
  {volume} {35}},\ \bibinfo {pages} {123101} (\bibinfo {year}
  {2008})}\BibitemShut {NoStop}\bibitem [{\citenamefont {Vary}\ \emph {et~al.}(2018)\citenamefont {Vary},
  \citenamefont {Maris}, \citenamefont {Fasano},\ and\ \citenamefont
  {Caprio}}]{vary2018:gentrunc-ostuka17}\BibitemOpen
  \bibfield  {author} {\bibinfo {author} {\bibfnamefont {J.~P.}\ \bibnamefont
  {Vary}}, \bibinfo {author} {\bibfnamefont {P.}~\bibnamefont {Maris}},
  \bibinfo {author} {\bibfnamefont {P.~J.}\ \bibnamefont {Fasano}},\ and\
  \bibinfo {author} {\bibfnamefont {M.~A.}\ \bibnamefont {Caprio}},\ }\bibfield
   {title} {\bibinfo {title} {Perspectives on nuclear structure and scattering
  with the \textit{ab initio} no-core shell model},\ }\href
  {https://doi.org/10.7566/JPSCP.23.012001} {\bibfield  {journal} {\bibinfo
  {journal} {JPS Conf. Proc.}\ }\textbf {\bibinfo {volume} {23}},\ \bibinfo
  {pages} {012001} (\bibinfo {year} {2018})}\BibitemShut {NoStop}\bibitem [{\citenamefont {Roth}\ and\ \citenamefont
  {Navr\'atil}(2007)}]{roth2007:it-ncsm-40ca}\BibitemOpen
  \bibfield  {author} {\bibinfo {author} {\bibfnamefont {R.}~\bibnamefont
  {Roth}}\ and\ \bibinfo {author} {\bibfnamefont {P.}~\bibnamefont
  {Navr\'atil}},\ }\bibfield  {title} {\bibinfo {title} {\textit{Ab initio}
  study of $\isotope[40]{Ca}$ with an importance-truncated no-core shell
  model},\ }\href {https://doi.org/10.1103/PhysRevLett.99.092501} {\bibfield
  {journal} {\bibinfo  {journal} {Phys. Rev. Lett.}\ }\textbf {\bibinfo
  {volume} {99}},\ \bibinfo {pages} {092501} (\bibinfo {year}
  {2007})}\BibitemShut {NoStop}\bibitem [{\citenamefont {Lanczos}(1950)}]{lanczos1950:algorithm}\BibitemOpen
  \bibfield  {author} {\bibinfo {author} {\bibfnamefont {C.}~\bibnamefont
  {Lanczos}},\ }\bibfield  {title} {\bibinfo {title} {An iteration method for
  the solution of the eigenvalue problem of linear differential and integral
  operators},\ }\href {https://doi.org/10.6028/jres.045.026} {\bibfield
  {journal} {\bibinfo  {journal} {J. Res. Natl. Bur. Stand. (U. S.)}\ }\textbf
  {\bibinfo {volume} {45}},\ \bibinfo {pages} {255} (\bibinfo {year}
  {1950})}\BibitemShut {NoStop}\bibitem [{\citenamefont {Goldberger}\ and\ \citenamefont
  {Watson}(1964)}]{goldberger1964:collision}\BibitemOpen
  \bibfield  {author} {\bibinfo {author} {\bibfnamefont {M.~L.}\ \bibnamefont
  {Goldberger}}\ and\ \bibinfo {author} {\bibfnamefont {K.~M.}\ \bibnamefont
  {Watson}},\ }\href@noop {} {\emph {\bibinfo {title} {Collision Theory}}}\
  (\bibinfo  {publisher} {Wiley},\ \bibinfo {address} {New York},\ \bibinfo
  {year} {1964})\BibitemShut {NoStop}\bibitem [{\citenamefont {Navr\'atil}(2021)}]{navratil2021:trinv-obme}\BibitemOpen
  \bibfield  {author} {\bibinfo {author} {\bibfnamefont {P.}~\bibnamefont
  {Navr\'atil}},\ }\bibfield  {title} {\bibinfo {title} {Translationally
  invariant matrix elements of general one-body operators},\ }\href
  {https://doi.org/10.1103/PhysRevC.104.064322} {\bibfield  {journal} {\bibinfo
   {journal} {Phys. Rev. C}\ }\textbf {\bibinfo {volume} {104}},\ \bibinfo
  {pages} {064322} (\bibinfo {year} {2021})},\ \Eprint
  {https://arxiv.org/abs/2109.04017} {arXiv:2109.04017 [nucl-th]} \BibitemShut
  {NoStop}\bibitem [{\citenamefont {Hagen}\ \emph {et~al.}(2006)\citenamefont {Hagen},
  \citenamefont {Hjorth-Jensen},\ and\ \citenamefont
  {Michel}}]{hagen2006:gdm-realistic}\BibitemOpen
  \bibfield  {author} {\bibinfo {author} {\bibfnamefont {G.}~\bibnamefont
  {Hagen}}, \bibinfo {author} {\bibfnamefont {M.}~\bibnamefont
  {Hjorth-Jensen}},\ and\ \bibinfo {author} {\bibfnamefont {N.}~\bibnamefont
  {Michel}},\ }\bibfield  {title} {\bibinfo {title} {{G}amow shell model and
  realistic nucleon-nucleon interactions},\ }\href
  {https://doi.org/10.1103/PhysRevC.73.064307} {\bibfield  {journal} {\bibinfo
  {journal} {Phys. Rev. C}\ }\textbf {\bibinfo {volume} {73}},\ \bibinfo
  {pages} {064307} (\bibinfo {year} {2006})}\BibitemShut {NoStop}\bibitem [{\citenamefont {Wang}\ \emph {et~al.}(2021)\citenamefont {Wang},
  \citenamefont {Huang}, \citenamefont {Kondev}, \citenamefont {Audi},\ and\
  \citenamefont {Naimi}}]{wang2021:ame2020}\BibitemOpen
  \bibfield  {author} {\bibinfo {author} {\bibfnamefont {M.}~\bibnamefont
  {Wang}}, \bibinfo {author} {\bibfnamefont {W.}~\bibnamefont {Huang}},
  \bibinfo {author} {\bibfnamefont {F.}~\bibnamefont {Kondev}}, \bibinfo
  {author} {\bibfnamefont {G.}~\bibnamefont {Audi}},\ and\ \bibinfo {author}
  {\bibfnamefont {S.}~\bibnamefont {Naimi}},\ }\bibfield  {title} {\bibinfo
  {title} {The {AME} 2020 atomic mass evaluation ({II}). {T}ables, graphs and
  references},\ }\href {https://doi.org/10.1088/1674-1137/abddaf} {\bibfield
  {journal} {\bibinfo  {journal} {Chin. Phys. C}\ }\textbf {\bibinfo {volume}
  {45}},\ \bibinfo {pages} {030003} (\bibinfo {year} {2021})}\BibitemShut
  {NoStop}\bibitem [{\citenamefont {Bogner}\ \emph {et~al.}(2007)\citenamefont {Bogner},
  \citenamefont {Furnstahl},\ and\ \citenamefont
  {Perry}}]{bogner2007:srg-nucleon}\BibitemOpen
  \bibfield  {author} {\bibinfo {author} {\bibfnamefont {S.~K.}\ \bibnamefont
  {Bogner}}, \bibinfo {author} {\bibfnamefont {R.~J.}\ \bibnamefont
  {Furnstahl}},\ and\ \bibinfo {author} {\bibfnamefont {R.~J.}\ \bibnamefont
  {Perry}},\ }\bibfield  {title} {\bibinfo {title} {Similarity renormalization
  group for nucleon-nucleon interactions},\ }\href
  {https://doi.org/10.1103/PhysRevC.75.061001} {\bibfield  {journal} {\bibinfo
  {journal} {Phys. Rev. C}\ }\textbf {\bibinfo {volume} {75}},\ \bibinfo
  {pages} {061001(R)} (\bibinfo {year} {2007})}\BibitemShut {NoStop}\bibitem [{\citenamefont {Aktulga}\ \emph {et~al.}(2013)\citenamefont
  {Aktulga}, \citenamefont {Yang}, \citenamefont {Ng}, \citenamefont {Maris},\
  and\ \citenamefont {Vary}}]{aktulga2013:mfdn-scalability}\BibitemOpen
  \bibfield  {author} {\bibinfo {author} {\bibfnamefont {H.~M.}\ \bibnamefont
  {Aktulga}}, \bibinfo {author} {\bibfnamefont {C.}~\bibnamefont {Yang}},
  \bibinfo {author} {\bibfnamefont {E.~G.}\ \bibnamefont {Ng}}, \bibinfo
  {author} {\bibfnamefont {P.}~\bibnamefont {Maris}},\ and\ \bibinfo {author}
  {\bibfnamefont {J.~P.}\ \bibnamefont {Vary}},\ }\bibfield  {title} {\bibinfo
  {title} {Improving the scalability of symmetric iterative eigensolver for
  multi-core platforms},\ }\href {https://doi.org/10.1002/cpe.3129} {\bibfield
  {journal} {\bibinfo  {journal} {Concurrency Computat.: Pract. Exper.}\
  }\textbf {\bibinfo {volume} {26}},\ \bibinfo {pages} {2631} (\bibinfo {year}
  {2013})}\BibitemShut {NoStop}\bibitem [{\citenamefont {Shao}\ \emph {et~al.}(2018)\citenamefont {Shao},
  \citenamefont {Aktulga}, \citenamefont {Yang}, \citenamefont {Ng},
  \citenamefont {Maris},\ and\ \citenamefont
  {Vary}}]{shao2018:ncci-preconditioned}\BibitemOpen
  \bibfield  {author} {\bibinfo {author} {\bibfnamefont {M.}~\bibnamefont
  {Shao}}, \bibinfo {author} {\bibfnamefont {H.~M.}\ \bibnamefont {Aktulga}},
  \bibinfo {author} {\bibfnamefont {C.}~\bibnamefont {Yang}}, \bibinfo {author}
  {\bibfnamefont {E.~G.}\ \bibnamefont {Ng}}, \bibinfo {author} {\bibfnamefont
  {P.}~\bibnamefont {Maris}},\ and\ \bibinfo {author} {\bibfnamefont {J.~P.}\
  \bibnamefont {Vary}},\ }\bibfield  {title} {\bibinfo {title} {Accelerating
  nuclear configuration interaction calculations through a preconditioned block
  iterative eigensolver},\ }\href {https://doi.org/10.1016/j.cpc.2017.09.004}
  {\bibfield  {journal} {\bibinfo  {journal} {Comput. Phys. Commun.}\ }\textbf
  {\bibinfo {volume} {222}},\ \bibinfo {pages} {1} (\bibinfo {year}
  {2018})}\BibitemShut {NoStop}\bibitem [{\citenamefont {Caprio}\ and\ \citenamefont {Fasano}()}]{code-shell}\BibitemOpen
  \bibfield  {author} {\bibinfo {author} {\bibfnamefont {M.~A.}\ \bibnamefont
  {Caprio}}\ and\ \bibinfo {author} {\bibfnamefont {P.~J.}\ \bibnamefont
  {Fasano}},\ }\href {http://github.com/nd-nuclear-theory/shell} {\bibinfo
  {title} {\textup{computer code library \texttt{shell}}}}\BibitemShut
  {NoStop}\bibitem [{\citenamefont {Forssen}\ \emph {et~al.}(2008)\citenamefont
  {Forssen}, \citenamefont {Vary}, \citenamefont {Caurier},\ and\ \citenamefont
  {Navratil}}]{forssen2008:ncsm-sequences}\BibitemOpen
  \bibfield  {author} {\bibinfo {author} {\bibfnamefont {C.}~\bibnamefont
  {Forssen}}, \bibinfo {author} {\bibfnamefont {J.~P.}\ \bibnamefont {Vary}},
  \bibinfo {author} {\bibfnamefont {E.}~\bibnamefont {Caurier}},\ and\ \bibinfo
  {author} {\bibfnamefont {P.}~\bibnamefont {Navratil}},\ }\bibfield  {title}
  {\bibinfo {title} {Converging sequences in the ab-initio no-core shell
  model},\ }\href {https://doi.org/10.1103/PhysRevC.77.024301} {\bibfield
  {journal} {\bibinfo  {journal} {Phys. Rev. C}\ }\textbf {\bibinfo {volume}
  {77}},\ \bibinfo {pages} {024301} (\bibinfo {year} {2008})}\BibitemShut
  {NoStop}\bibitem [{\citenamefont {Maris}\ \emph {et~al.}(2009)\citenamefont {Maris},
  \citenamefont {Vary},\ and\ \citenamefont {Shirokov}}]{maris2009:ncfc}\BibitemOpen
  \bibfield  {author} {\bibinfo {author} {\bibfnamefont {P.}~\bibnamefont
  {Maris}}, \bibinfo {author} {\bibfnamefont {J.~P.}\ \bibnamefont {Vary}},\
  and\ \bibinfo {author} {\bibfnamefont {A.~M.}\ \bibnamefont {Shirokov}},\
  }\bibfield  {title} {\bibinfo {title} {\textit{Ab initio} no-core full
  configuration calculations of light nuclei},\ }\href
  {https://doi.org/10.1103/PhysRevC.79.014308} {\bibfield  {journal} {\bibinfo
  {journal} {Phys. Rev. C}\ }\textbf {\bibinfo {volume} {79}},\ \bibinfo
  {pages} {014308} (\bibinfo {year} {2009})}\BibitemShut {NoStop}\bibitem [{\citenamefont {Angeli}\ and\ \citenamefont
  {Marinova}(2013)}]{angeli2013:charge-radii}\BibitemOpen
  \bibfield  {author} {\bibinfo {author} {\bibfnamefont {I.}~\bibnamefont
  {Angeli}}\ and\ \bibinfo {author} {\bibfnamefont {K.~P.}\ \bibnamefont
  {Marinova}},\ }\bibfield  {title} {\bibinfo {title} {Table of experimental
  nuclear ground state charge radii: An update},\ }\href
  {https://doi.org/10.1016/j.adt.2011.12.006} {\bibfield  {journal} {\bibinfo
  {journal} {At. Data Nucl. Data Tables}\ }\textbf {\bibinfo {volume} {99}},\
  \bibinfo {pages} {69} (\bibinfo {year} {2013})}\BibitemShut {NoStop}\bibitem [{\citenamefont {Friar}\ \emph {et~al.}(1997)\citenamefont {Friar},
  \citenamefont {Martorell},\ and\ \citenamefont
  {Sprung}}]{friar1997:charge-radius-correction}\BibitemOpen
  \bibfield  {author} {\bibinfo {author} {\bibfnamefont {J.~L.}\ \bibnamefont
  {Friar}}, \bibinfo {author} {\bibfnamefont {J.}~\bibnamefont {Martorell}},\
  and\ \bibinfo {author} {\bibfnamefont {D.~W.~L.}\ \bibnamefont {Sprung}},\
  }\bibfield  {title} {\bibinfo {title} {Nuclear sizes and the isotope shift},\
  }\href {https://doi.org/10.1103/PhysRevA.56.4579} {\bibfield  {journal}
  {\bibinfo  {journal} {Phys. Rev. A}\ }\textbf {\bibinfo {volume} {56}},\
  \bibinfo {pages} {4579} (\bibinfo {year} {1997})}\BibitemShut {NoStop}\bibitem [{\citenamefont {Lu}\ \emph {et~al.}(2013)\citenamefont {Lu},
  \citenamefont {Mueller}, \citenamefont {Drake}, \citenamefont
  {N{\"o}rtersh{\"a}user}, \citenamefont {Pieper},\ and\ \citenamefont
  {Yan}}]{lu2013:laser-neutron-rich}\BibitemOpen
  \bibfield  {author} {\bibinfo {author} {\bibfnamefont {Z.-T.}\ \bibnamefont
  {Lu}}, \bibinfo {author} {\bibfnamefont {P.}~\bibnamefont {Mueller}},
  \bibinfo {author} {\bibfnamefont {G.~W.~F.}\ \bibnamefont {Drake}}, \bibinfo
  {author} {\bibfnamefont {W.}~\bibnamefont {N{\"o}rtersh{\"a}user}}, \bibinfo
  {author} {\bibfnamefont {S.~C.}\ \bibnamefont {Pieper}},\ and\ \bibinfo
  {author} {\bibfnamefont {Z.-C.}\ \bibnamefont {Yan}},\ }\bibfield  {title}
  {\bibinfo {title} {Laser probing of neutron-rich nuclei in light atoms},\
  }\href {https://doi.org/10.1103/RevModPhys.85.1383} {\bibfield  {journal}
  {\bibinfo  {journal} {Rev. Mod. Phys.}\ }\textbf {\bibinfo {volume} {85}},\
  \bibinfo {pages} {1383} (\bibinfo {year} {2013})}\BibitemShut {NoStop}\bibitem [{\citenamefont {Nogga}\ \emph {et~al.}(2006)\citenamefont {Nogga},
  \citenamefont {Navr\'atil}, \citenamefont {Barrett},\ and\ \citenamefont
  {Vary}}]{nogga2006:7li-ncsm-chiral}\BibitemOpen
  \bibfield  {author} {\bibinfo {author} {\bibfnamefont {A.}~\bibnamefont
  {Nogga}}, \bibinfo {author} {\bibfnamefont {P.}~\bibnamefont {Navr\'atil}},
  \bibinfo {author} {\bibfnamefont {B.~R.}\ \bibnamefont {Barrett}},\ and\
  \bibinfo {author} {\bibfnamefont {J.~P.}\ \bibnamefont {Vary}},\ }\bibfield
  {title} {\bibinfo {title} {Spectra and binding energy predictions of chiral
  interactions for $^{7}\mathrm{Li}$},\ }\href
  {https://doi.org/10.1103/PhysRevC.73.064002} {\bibfield  {journal} {\bibinfo
  {journal} {Phys. Rev. C}\ }\textbf {\bibinfo {volume} {73}},\ \bibinfo
  {pages} {064002} (\bibinfo {year} {2006})}\BibitemShut {NoStop}\bibitem [{\citenamefont {Cockrell}\ \emph {et~al.}(2012)\citenamefont
  {Cockrell}, \citenamefont {Vary},\ and\ \citenamefont
  {Maris}}]{cockrell2012:li-ncfc}\BibitemOpen
  \bibfield  {author} {\bibinfo {author} {\bibfnamefont {C.}~\bibnamefont
  {Cockrell}}, \bibinfo {author} {\bibfnamefont {J.~P.}\ \bibnamefont {Vary}},\
  and\ \bibinfo {author} {\bibfnamefont {P.}~\bibnamefont {Maris}},\ }\bibfield
   {title} {\bibinfo {title} {Lithium isotopes within the \textit{ab initio}
  no-core full configuration approach},\ }\href
  {https://doi.org/10.1103/PhysRevC.86.034325} {\bibfield  {journal} {\bibinfo
  {journal} {Phys. Rev. C}\ }\textbf {\bibinfo {volume} {86}},\ \bibinfo
  {pages} {034325} (\bibinfo {year} {2012})}\BibitemShut {NoStop}\bibitem [{\citenamefont {Jonson}(2004)}]{jonson2004:light-dripline}\BibitemOpen
  \bibfield  {author} {\bibinfo {author} {\bibfnamefont {B.}~\bibnamefont
  {Jonson}},\ }\bibfield  {title} {\bibinfo {title} {Light dripline nuclei},\
  }\href {https://doi.org/10.1016/j.physrep.2003.07.004} {\bibfield  {journal}
  {\bibinfo  {journal} {Phys. Rep.}\ }\textbf {\bibinfo {volume} {389}},\
  \bibinfo {pages} {1} (\bibinfo {year} {2004})}\BibitemShut {NoStop}\bibitem [{\citenamefont {Tanihata}\ \emph {et~al.}(2013)\citenamefont
  {Tanihata}, \citenamefont {Savajols},\ and\ \citenamefont
  {Kanungo}}]{tanihata2013:halo-expt}\BibitemOpen
  \bibfield  {author} {\bibinfo {author} {\bibfnamefont {I.}~\bibnamefont
  {Tanihata}}, \bibinfo {author} {\bibfnamefont {H.}~\bibnamefont {Savajols}},\
  and\ \bibinfo {author} {\bibfnamefont {R.}~\bibnamefont {Kanungo}},\
  }\bibfield  {title} {\bibinfo {title} {Recent experimental progress in
  nuclear halo structure studies},\ }\href
  {https://doi.org/10.1016/j.ppnp.2012.07.001} {\bibfield  {journal} {\bibinfo
  {journal} {Prog. Part. Nucl. Phys.}\ }\textbf {\bibinfo {volume} {68}},\
  \bibinfo {pages} {215} (\bibinfo {year} {2013})}\BibitemShut {NoStop}\bibitem [{\citenamefont {Quaglioni}\ \emph {et~al.}(2013)\citenamefont
  {Quaglioni}, \citenamefont {Romero-Redondo},\ and\ \citenamefont
  {Navr{\'a}til}}]{quaglioni2013:ncsm-rgm-cluster-6he}\BibitemOpen
  \bibfield  {author} {\bibinfo {author} {\bibfnamefont {S.}~\bibnamefont
  {Quaglioni}}, \bibinfo {author} {\bibfnamefont {C.}~\bibnamefont
  {Romero-Redondo}},\ and\ \bibinfo {author} {\bibfnamefont {P.}~\bibnamefont
  {Navr{\'a}til}},\ }\bibfield  {title} {\bibinfo {title} {Three-cluster
  dynamics within an \textit{ab initio} framework},\ }\href
  {https://doi.org/10.1103/PhysRevC.88.034320} {\bibfield  {journal} {\bibinfo
  {journal} {Phys. Rev. C}\ }\textbf {\bibinfo {volume} {88}},\ \bibinfo
  {pages} {034320} (\bibinfo {year} {2013})}\BibitemShut {NoStop}\bibitem [{\citenamefont {S{\"a}{\"a}f}\ and\ \citenamefont
  {Forss\'en}(2014)}]{saaf2014:core-n-n-6he}\BibitemOpen
  \bibfield  {author} {\bibinfo {author} {\bibfnamefont {D.}~\bibnamefont
  {S{\"a}{\"a}f}}\ and\ \bibinfo {author} {\bibfnamefont {C.}~\bibnamefont
  {Forss\'en}},\ }\bibfield  {title} {\bibinfo {title} {Microscopic description
  of translationally invariant $\mathrm{core} + n + n$ overlap functions},\
  }\href {https://doi.org/10.1103/PhysRevC.89.011303} {\bibfield  {journal}
  {\bibinfo  {journal} {Phys. Rev. C}\ }\textbf {\bibinfo {volume} {89}},\
  \bibinfo {pages} {011303(R)} (\bibinfo {year} {2014})}\BibitemShut {NoStop}\bibitem [{\citenamefont {Romero-Redondo}\ \emph {et~al.}(2016)\citenamefont
  {Romero-Redondo}, \citenamefont {Quaglioni}, \citenamefont {Navr\'{a}til},\
  and\ \citenamefont {Hupin}}]{romeroredondo2016:6he-correlations}\BibitemOpen
  \bibfield  {author} {\bibinfo {author} {\bibfnamefont {C.}~\bibnamefont
  {Romero-Redondo}}, \bibinfo {author} {\bibfnamefont {S.}~\bibnamefont
  {Quaglioni}}, \bibinfo {author} {\bibfnamefont {P.}~\bibnamefont
  {Navr\'{a}til}},\ and\ \bibinfo {author} {\bibfnamefont {G.}~\bibnamefont
  {Hupin}},\ }\bibfield  {title} {\bibinfo {title} {How many-body correlations
  and $\alpha$ clustering shape $\isotope[6]{He}$},\ }\href
  {https://doi.org/10.1103/PhysRevLett.117.222501} {\bibfield  {journal}
  {\bibinfo  {journal} {Phys. Rev. Lett.}\ }\textbf {\bibinfo {volume} {117}},\
  \bibinfo {pages} {222501} (\bibinfo {year} {2016})}\BibitemShut {NoStop}\bibitem [{\citenamefont {Wang}\ \emph {et~al.}(2004)\citenamefont {Wang},
  \citenamefont {Mueller}, \citenamefont {Bailey}, \citenamefont {Drake},
  \citenamefont {Greene}, \citenamefont {Henderson}, \citenamefont {Holt},
  \citenamefont {Janssens}, \citenamefont {Jiang}, \citenamefont {Lu},
  \citenamefont {O'Connor}, \citenamefont {Pardo}, \citenamefont {Rehm},
  \citenamefont {Schiffer},\ and\ \citenamefont
  {Tang}}]{wang2004:6he-radius-laser}\BibitemOpen
  \bibfield  {author} {\bibinfo {author} {\bibfnamefont {L.-B.}\ \bibnamefont
  {Wang}}, \bibinfo {author} {\bibfnamefont {P.}~\bibnamefont {Mueller}},
  \bibinfo {author} {\bibfnamefont {K.}~\bibnamefont {Bailey}}, \bibinfo
  {author} {\bibfnamefont {G.~W.~F.}\ \bibnamefont {Drake}}, \bibinfo {author}
  {\bibfnamefont {J.~P.}\ \bibnamefont {Greene}}, \bibinfo {author}
  {\bibfnamefont {D.}~\bibnamefont {Henderson}}, \bibinfo {author}
  {\bibfnamefont {R.~J.}\ \bibnamefont {Holt}}, \bibinfo {author}
  {\bibfnamefont {R.~V.~F.}\ \bibnamefont {Janssens}}, \bibinfo {author}
  {\bibfnamefont {C.~L.}\ \bibnamefont {Jiang}}, \bibinfo {author}
  {\bibfnamefont {Z.-T.}\ \bibnamefont {Lu}}, \bibinfo {author} {\bibfnamefont
  {T.~P.}\ \bibnamefont {O'Connor}}, \bibinfo {author} {\bibfnamefont {R.~C.}\
  \bibnamefont {Pardo}}, \bibinfo {author} {\bibfnamefont {K.~E.}\ \bibnamefont
  {Rehm}}, \bibinfo {author} {\bibfnamefont {J.~P.}\ \bibnamefont {Schiffer}},\
  and\ \bibinfo {author} {\bibfnamefont {X.~D.}\ \bibnamefont {Tang}},\
  }\bibfield  {title} {\bibinfo {title} {Laser spectroscopic determination of
  the $\isotope[6]{He}$ nuclear charge radius},\ }\href
  {https://doi.org/10.1103/PhysRevLett.93.142501} {\bibfield  {journal}
  {\bibinfo  {journal} {Phys. Rev. Lett.}\ }\textbf {\bibinfo {volume} {93}},\
  \bibinfo {pages} {142501} (\bibinfo {year} {2004})}\BibitemShut {NoStop}\bibitem [{\citenamefont {Brodeur}\ \emph {et~al.}(2012)\citenamefont
  {Brodeur}, \citenamefont {Brunner}, \citenamefont {Champagne}, \citenamefont
  {Ettenauer}, \citenamefont {Smith}, \citenamefont {Lapierre}, \citenamefont
  {Ringle}, \citenamefont {Ryjkov}, \citenamefont {Bacca}, \citenamefont
  {Delheij}, \citenamefont {Drake}, \citenamefont {Lunney}, \citenamefont
  {Schwenk},\ and\ \citenamefont {Dilling}}]{brodeur2012:6he-8he-mass}\BibitemOpen
  \bibfield  {author} {\bibinfo {author} {\bibfnamefont {M.}~\bibnamefont
  {Brodeur}}, \bibinfo {author} {\bibfnamefont {T.}~\bibnamefont {Brunner}},
  \bibinfo {author} {\bibfnamefont {C.}~\bibnamefont {Champagne}}, \bibinfo
  {author} {\bibfnamefont {S.}~\bibnamefont {Ettenauer}}, \bibinfo {author}
  {\bibfnamefont {M.~J.}\ \bibnamefont {Smith}}, \bibinfo {author}
  {\bibfnamefont {A.}~\bibnamefont {Lapierre}}, \bibinfo {author}
  {\bibfnamefont {R.}~\bibnamefont {Ringle}}, \bibinfo {author} {\bibfnamefont
  {V.~L.}\ \bibnamefont {Ryjkov}}, \bibinfo {author} {\bibfnamefont
  {S.}~\bibnamefont {Bacca}}, \bibinfo {author} {\bibfnamefont
  {P.}~\bibnamefont {Delheij}}, \bibinfo {author} {\bibfnamefont {G.~W.~F.}\
  \bibnamefont {Drake}}, \bibinfo {author} {\bibfnamefont {D.}~\bibnamefont
  {Lunney}}, \bibinfo {author} {\bibfnamefont {A.}~\bibnamefont {Schwenk}},\
  and\ \bibinfo {author} {\bibfnamefont {J.}~\bibnamefont {Dilling}},\
  }\bibfield  {title} {\bibinfo {title} {First direct mass measurement of the
  two-neutron halo nucleus $\isotope[6]{He}$ and improved mass for the
  four-neutron halo $\isotope[8]{He}$},\ }\href
  {https://doi.org/10.1103/PhysRevLett.108.052504} {\bibfield  {journal}
  {\bibinfo  {journal} {Phys. Rev. Lett.}\ }\textbf {\bibinfo {volume} {108}},\
  \bibinfo {pages} {052504} (\bibinfo {year} {2012})}\BibitemShut {NoStop}\bibitem [{\citenamefont {Maris}\ \emph {et~al.}(2013)\citenamefont {Maris},
  \citenamefont {Aktulga}, \citenamefont {Binder}, \citenamefont {Calci},
  \citenamefont {{\c C}ataly{\"u}rek}, \citenamefont {Langhammer},
  \citenamefont {Ng}, \citenamefont {Saule}, \citenamefont {Roth},
  \citenamefont {Vary},\ and\ \citenamefont
  {Yang}}]{maris2013:ncci-chiral-ccp12}\BibitemOpen
  \bibfield  {author} {\bibinfo {author} {\bibfnamefont {P.}~\bibnamefont
  {Maris}}, \bibinfo {author} {\bibfnamefont {H.~M.}\ \bibnamefont {Aktulga}},
  \bibinfo {author} {\bibfnamefont {S.}~\bibnamefont {Binder}}, \bibinfo
  {author} {\bibfnamefont {A.}~\bibnamefont {Calci}}, \bibinfo {author}
  {\bibfnamefont {{\"U}.~V.}\ \bibnamefont {{\c C}ataly{\"u}rek}}, \bibinfo
  {author} {\bibfnamefont {J.}~\bibnamefont {Langhammer}}, \bibinfo {author}
  {\bibfnamefont {E.}~\bibnamefont {Ng}}, \bibinfo {author} {\bibfnamefont
  {E.}~\bibnamefont {Saule}}, \bibinfo {author} {\bibfnamefont
  {R.}~\bibnamefont {Roth}}, \bibinfo {author} {\bibfnamefont {J.~P.}\
  \bibnamefont {Vary}},\ and\ \bibinfo {author} {\bibfnamefont
  {C.}~\bibnamefont {Yang}},\ }\bibfield  {title} {\bibinfo {title} {No core
  {CI} calculations for light nuclei with chiral 2- and 3-body forces},\ }\href
  {https://doi.org/10.1088/1742-6596/454/1/012063} {\bibfield  {journal}
  {\bibinfo  {journal} {J. Phys. Conf. Ser.}\ }\textbf {\bibinfo {volume}
  {454}},\ \bibinfo {pages} {012063} (\bibinfo {year} {2013})}\BibitemShut
  {NoStop}\bibitem [{\citenamefont {Gebrerufael}\ \emph {et~al.}(2017)\citenamefont
  {Gebrerufael}, \citenamefont {Vobig}, \citenamefont {Hergert},\ and\
  \citenamefont {Roth}}]{gebrerufael2017:im-ncsm}\BibitemOpen
  \bibfield  {author} {\bibinfo {author} {\bibfnamefont {E.}~\bibnamefont
  {Gebrerufael}}, \bibinfo {author} {\bibfnamefont {K.}~\bibnamefont {Vobig}},
  \bibinfo {author} {\bibfnamefont {H.}~\bibnamefont {Hergert}},\ and\ \bibinfo
  {author} {\bibfnamefont {R.}~\bibnamefont {Roth}},\ }\bibfield  {title}
  {\bibinfo {title} {\textit{Ab initio} description of open-shell nuclei:
  {M}erging no-core shell model and in-medium similarity renormalization
  group},\ }\href {https://doi.org/10.1103/PhysRevLett.118.152503} {\bibfield
  {journal} {\bibinfo  {journal} {Phys. Rev. Lett.}\ }\textbf {\bibinfo
  {volume} {118}},\ \bibinfo {pages} {152503} (\bibinfo {year}
  {2017})}\BibitemShut {NoStop}\bibitem [{\citenamefont {Tichai}\ \emph {et~al.}(2018)\citenamefont {Tichai},
  \citenamefont {Gebrerufael}, \citenamefont {Vobig},\ and\ \citenamefont
  {Roth}}]{tichai2018:ncsm-perturbative}\BibitemOpen
  \bibfield  {author} {\bibinfo {author} {\bibfnamefont {A.}~\bibnamefont
  {Tichai}}, \bibinfo {author} {\bibfnamefont {E.}~\bibnamefont {Gebrerufael}},
  \bibinfo {author} {\bibfnamefont {K.}~\bibnamefont {Vobig}},\ and\ \bibinfo
  {author} {\bibfnamefont {R.}~\bibnamefont {Roth}},\ }\bibfield  {title}
  {\bibinfo {title} {Open-shell nuclei from no-core shell model with
  perturbative improvement},\ }\href
  {https://doi.org/10.1016/j.physletb.2018.10.029} {\bibfield  {journal}
  {\bibinfo  {journal} {Phys. Lett. B}\ }\textbf {\bibinfo {volume} {786}},\
  \bibinfo {pages} {448} (\bibinfo {year} {2018})}\BibitemShut {NoStop}\end{thebibliography}
\nocite{control:title-on}

\end{document}